\documentclass[journal]{IEEEtran}

\usepackage{xcolor,soul,framed, colortbl}

\colorlet{shadecolor}{yellow}

\usepackage[pdftex]{graphicx}
\graphicspath{{../pdf/}{../jpeg/}}
\DeclareGraphicsExtensions{.pdf,.jpeg,.png}

\usepackage[cmex10]{amsmath}
\usepackage{array}
\usepackage{mdwmath}
\usepackage{mdwtab}
\usepackage{eqparbox}
\usepackage{url}
\usepackage{changes}

\usepackage{enumitem,textcomp,amsfonts,multicol,verbatim,epsfig,amssymb,bbding,cases,bm,multirow,dsfont,bbm,tabularx,makecell,booktabs,lscape,chngcntr,mwe,adjustbox}
\usepackage{acronym}
\usepackage{longtable}
\usepackage{supertabular}
\usepackage{lipsum}
\usepackage{hyperref}

\hypersetup{
 colorlinks=true,
 linkcolor=black,
 citecolor=black,
 filecolor=black,
 urlcolor=black,
 }

\usepackage{nomencl}
\makenomenclature
\begin{document}
\bstctlcite{IEEEexample:BSTcontrol}
    \title{
    Navigating the Dual-Use Nature and Security Implications of Reconfigurable Intelligent Surfaces in Next-Generation Wireless Systems}
  \author{Hetong~Wang,~Tiejun~Lv,~\IEEEmembership{Senior Member, IEEE}, Yashuai~Cao, Weicai~Li,~\IEEEmembership{Graduate Student Member, IEEE}, Jie~Zeng,~\IEEEmembership{Senior Member, IEEE}, Pingmu~Huang, and Muhammad Khurram Khan,~\IEEEmembership{Senior Member, IEEE}

      % <-this % stops a space

\thanks{Manuscript received 17 January 2025; revised 1 July 2025, and 16 September 2025; accepted 10 October 2025. This paper was supported in part by the National Natural Science Foundation of China under No. 62271068, the Beijing Natural Science Foundation under Grant No. L222046, and the Deputyship
for Research and Innovation, Ministry of Education in Saudi
Arabia (IFKSU-HCRA-3-2). (\emph{corresponding author: Tiejun Lv}.)}
\thanks{Hetong Wang and Tiejun Lv are with the School of Information and Communication Engineering, Beijing University of Posts and Telecommunications (BUPT), Beijing 100876, China (e-mail: \{htwang\_61, lvtiejun\}@bupt.edu.cn).}
\thanks{Yashuai Cao is with the School of Intelligence Science and Technology, University of Science and Technology Beijing, Beijing 100083, China (e-mail: caoys@ustb.edu.cn).}
\thanks{Weicai Li is with the School of Information Communication Engineering, Beijing Information Science and Technology University, Beijing, China (e-mail: liweicai@bupt.edu.cn).}
\thanks{Jie Zeng is with the School of Cyberspace Science and Technology, Beijing Institute of Technology, Beijing 100081, China, and the Beijing National Research Center for Information Science and Technology, Tsinghua University, Beijing 100084, China (e-mail: zengjie@bit.edu.cn).}
\thanks{Pingmu Huang is with the School of Artificial Intelligence, Beijing University of Posts and Telecommunications (BUPT), Beijing 100876, China (e-mail: pmhuang@bupt.edu.cn).}
\thanks{Muhammad Khurram Khan is with the Center of Excellence in Information Assurance, King Saud University, Riyadh 11653, Saudi Arabia (e-mail: mkhurram@ksu.edu.sa).}
}

% The paper headers
%\markboth{IEEE TRANSACTIONS ON MICROWAVE THEORY AND TECHNIQUES, VOL.~60, NO.~12, DECEMBER~2012
%}{Roberg \MakeLowercase{\textit{et al.}}: High-Efficiency Diode and Transistor Rectifiers}

% ====================================================================
\maketitle

% === ABSTRACT ====================================================================
% =================================================================================
\begin{abstract}
Reconfigurable intelligent surface (RIS) technology offers significant promise in enhancing wireless communication systems, but its dual-use potential also introduces substantial security risks. This survey explores the security implications of RIS in next-generation wireless networks. We first highlight the dual-use nature of RIS, demonstrating how its communication-enhancing capabilities can be exploited by adversaries to compromise legitimate users. We identify a new class of security vulnerabilities termed ``passive-active hybrid attacks,'' where RIS, despite passively handling signals, can be reconfigured to actively engage in malicious activities, enabling various RIS-assisted attacks, such as eavesdropping, man-in-the-middle (MITM), replay, reflection jamming, and side-channel attacks. Furthermore, we reveal how adversaries can exploit the openness of wireless channels to introduce adversarial perturbations in artificial intelligence-driven RIS networks, disrupting communication terminals and causing misclassifications or errors in RIS reflection predictions. Despite these risks, RIS technology also plays a critical role in enhancing security and privacy across radio frequency (RF) and visible light communication (VLC) systems. By synthesizing current insights and highlighting emerging threats, we provide actionable insights into cross-layer collaboration, advanced adversarial defenses, and the balance between security and cost. This survey provides a comprehensive overview of RIS technology's security landscape and underscores the urgent need for robust security frameworks in the development of future wireless systems.
\end{abstract}

% === Abbreviations ====================================================================

% For peer review papers, you can put extra information on the cover
% page as needed:
% \ifCLASSOPTIONpeerreview
% \begin{center} \bfseries EDICS Category: 3-BBND \end{center}
% \fi
%
% For peerreview papers, this IEEEtran command inserts a page break and
% creates the second title. It will be ignored for other modes.
\IEEEpeerreviewmaketitle
% ====================================================================

% === Abbreviations =============================================================
\printnomenclature
\nomenclature[A]{AN}{Artificial noise}
\nomenclature[B]{BS}{Base station}
\nomenclature[C]{CSI}{Channel state information}
\nomenclature[C]{CEE}{Channel estimation error}
\nomenclature[D]{DC}{Direct-current}
\nomenclature[D]{DoF}{Degree of freedom}
\nomenclature[E]{EM}{Electromagnetic}
\nomenclature[E]{EE}{Energy efficiency}
\nomenclature[E]{ES}{Energy splitting}
\nomenclature[E]{Eve}{Eavesdropper}
\nomenclature[H]{H mode}{Energy harvesting mode}
\nomenclature[L]{LoS}{Line-of-sight}
\nomenclature[M]{MIMO}{Multiple input multiple output}
\nomenclature[M]{MS}{Mode switching}
\nomenclature[P]{PNSC}{Probability of non-zero secrecy capacity}
\nomenclature[Q]{QoS}{Quality of service}
\nomenclature[R]{RF}{Radio frequency}
\nomenclature[R]{R mode}{Reflection mode}
\nomenclature[R]{RIS}{Reconfigurable intelligent surface}
\nomenclature[I]{IP}{Internet protocol}
\nomenclature[S]{SE}{Spectral efficiency}
\nomenclature[S]{SINR}{Signal-to-interference-plus-noise ratio}
\nomenclature[S]{SISO}{Single input single output}
\nomenclature[S]{SIMO}{Single input multiple output}
\nomenclature[S]{SOP}{Secrecy outage probability}
\nomenclature[S]{S mode}{Signal relay mode}
\nomenclature[S]{SR}{Secrecy rate}
\nomenclature[S]{STAR-RIS}{Simultaneous transmitting and reflecting RIS}
\nomenclature[T]{TS}{Time switching}
\nomenclature[T]{T mode}{Transmission mode}
\nomenclature[T]{T\&R mode}{Transmission and reflection mode}
\nomenclature[U]{UAV}{Unmanned aerial vehicle}
\nomenclature[U]{ULA}{Uniform linear array}
\nomenclature[T]{THz}{Terahertz}
\nomenclature[M]{mmWave}{Millimeter-wave}
\nomenclature[1]{2D}{Two-dimensional}
\nomenclature[A]{AAUC}{Angle aware user cooperation}
\nomenclature[A]{AF}{Amplify-and-forward}
\nomenclature[B]{B5G}{Beyond 5G}
\nomenclature[C]{CCP}{Charnes-Coopers}
\nomenclature[C]{CDF}{Cumulative distribution function}
\nomenclature[C]{CR}{Cognitive radio}
\nomenclature[D]{DL}{Deep learning}
\nomenclature[D]{DNN}{Deep neural network}
\nomenclature[D]{DF}{Decode-and-forward}
\nomenclature[D]{D2D}{Device-to-device}
\nomenclature[D]{DTX}{D2D transmitter}
\nomenclature[D]{DRX}{D2D receiver}
\nomenclature[E]{EH}{Energy harvesting}
\nomenclature[E]{ER}{Energy receiver}
\nomenclature[F]{FD}{Full-duplex}
\nomenclature[F]{FPP-SCA}{Feasible point pursuit-successive convex approximation}
\nomenclature[F]{FD-CJ}{Full-duplex cooperative jamming}
\nomenclature[G]{GA}{Genetic algorithm}
\nomenclature[G]{Gbps}{Gigabits-per-second}
\nomenclature[G]{GDA}{Gradient-descent-ascent}
\nomenclature[I]{IBCD}{Inexact block coordinate descent}
\nomenclature[I]{ID}{Information decoder}
\nomenclature[I]{IM/DD}{Intensity modulation/direction detection}
\nomenclature[K]{KKT}{Karush-Kuhn-Tucker}
\nomenclature[K]{KM}{Kuhn-Munkres}
\nomenclature[L]{LED}{Light emitting diode}
\nomenclature[M]{MISO}{Multiple input single output}
\nomenclature[M]{MM}{Majorization-minimization}
\nomenclature[N]{NOMA}{Non-orthogonal multiple access}
\nomenclature[N]{NE}{Nash equilibrium}
\nomenclature[P]{PBF}{Passive beamforming}
\nomenclature[P]{PS}{Power splitting}
\nomenclature[P]{PU}{Primary user}
\nomenclature[S]{SCA}{Successive convex approximation}
\nomenclature[S]{SDR}{Semidefinite relaxation}
\nomenclature[S]{SOCP}{Second-order cone program}
\nomenclature[S]{SU}{Secondary user}
\nomenclature[S]{SWIPT}{Simultaneous wireless information and power transfer}
\nomenclature[V]{VLC}{Visible light communication}
\nomenclature[1]{5G}{Fifth-generation}
\nomenclature[1]{6G}{Six-generation}
\nomenclature[D]{DDPG}{Deep deterministic policy gradient}
\nomenclature[P]{PD}{Photo-detector}
\nomenclature[P]{PSO-II}{Particle swarm optimization-initialization intervention}
\nomenclature[R]{RSF}{Reflected spot finding}
\nomenclature[A]{ACA}{Active channel aging}
\nomenclature[A]{AI}{Artificial intelligence}
\nomenclature[B]{BCD}{Block coordinate descent}
\nomenclature[D]{DT}{Data transmission}
\nomenclature[F]{FDMA}{Frequency division multiple access}
\nomenclature[F]{FP}{Fractional programming}
\nomenclature[H]{HD}{Half-duplex}
\nomenclature[I]{IRIS}{Illegal RIS}
\nomenclature[I]{ISAC}{Integrated sensing and communications}
\nomenclature[I]{IUI}{Inter-user interference}
\nomenclature[M]{MINLP}{Mixed-integer non-linear program}
\nomenclature[M]{MU-MISO}{Multi-user MISO}
\nomenclature[P]{PKG}{Physical layer key generation}
\nomenclature[R]{RPT}{Reverse pilot transmission}
\nomenclature[R]{RSS}{Received signal strength}
\nomenclature[S]{SDP}{Semi-definite program}
\nomenclature[S]{SNR}{Signal-to-noise ratio}
\nomenclature[S]{SPSC}{Strictly positive secrecy capacity}
\nomenclature[S]{SVD}{Singular value decomposition}
\nomenclature[A]{AML}{Adversarial ML}
\nomenclature[A]{AWGN}{Additive white Gaussian noise}
\nomenclature[B]{BIM}{Basic iterative method}
\nomenclature[C]{CNN}{Convolutional neural network}
\nomenclature[D]{DRL}{Deep reinforcement learning}
\nomenclature[F]{FGM}{Fast gradient method}
\nomenclature[F]{FGSM}{Fast gradient sign method}
\nomenclature[F]{FL}{Federated learning}
\nomenclature[I]{IoT}{Internet of things}
\nomenclature[L]{LID}{Local intrinsic dimension}
\nomenclature[M]{ML}{Machine learning}
\nomenclature[M]{MIM}{Momentum iterative method}
\nomenclature[M]{MLP}{Multi-layer perceptron}
\nomenclature[M]{MSE}{Mean squared error}
\nomenclature[N]{NID}{Network intrusion detection systems}
\nomenclature[O]{OFDM}{Orthogonal frequency division multiplexing}
\nomenclature[P]{PGD}{Projected gradient descent}
\nomenclature[Q]{QPS}{Quantized phase shift}
\nomenclature[R]{RL}{Reinforcement learning}
\nomenclature[C]{CD}{Constellation diagram}
\nomenclature[G]{GAN}{Generative adversarial network}
\nomenclature[S]{SNN}{Spiking neural network}
\nomenclature[A]{Adv.}{Advantage}
\nomenclature[A]{Approxn.}{Approximation}
\nomenclature[A]{Atk.}{Attack}
\nomenclature[D]{Def.}{Defense}
\nomenclature[I]{Inf.}{Interference}
\nomenclature[M]{Max.}{Maximum}
\nomenclature[M]{Min.}{Minimum}
\nomenclature[O]{Obj.}{Objective}
\nomenclature[O]{Opt.}{Optimization}
\nomenclature[R]{Ref.}{Reference}
\nomenclature[D]{DQN}{Deep Q-Networks}
\nomenclature[A]{AO}{Alternating optimization}
\nomenclature[W]{WPA3}{Wi-Fi Protected Access 3}
\nomenclature[T]{TLS}{Transport Layer Security}
\nomenclature[R]{Rach}{random access channel}
\nomenclature[B]{BD}{block diagonal}
\nomenclature[U]{UV}{Ultraviolet}
\nomenclature[A]{AoA}{Angle-of-arrival}
\nomenclature[H]{HST}{High-speed train}

% ================ Sec. I Introduction
\section{Introduction}\label{Sec.I}
Reconfigurable intelligent surfaces (RISs) have recently garnered significant attention in academia and industry due to their ability to reconfigure wireless propagation environments and create smart radio environments intelligently~\cite{9927314,9896889}. A RIS is a two-dimensional artificial electromagnetic (EM) metasurface that can dynamically adjust the phase shifts of reflected EM waves~\cite{8936989,9122596}, directing them towards desired directions without the need for decoding, amplifying, or introducing time delays~\cite{9927314,9569556,9707727}. This unique capability, achieved through low-cost passive reflecting meta-atoms controlled by microcontrollers, offers numerous advantages, including cost-effectiveness, high spectral efficiency (SE), energy efficiency, and flexible deployment~\cite{9569556,9852091,9837942,9295369}.

\begin{figure*}
  \begin{center}
  \includegraphics[width=0.9\textwidth]{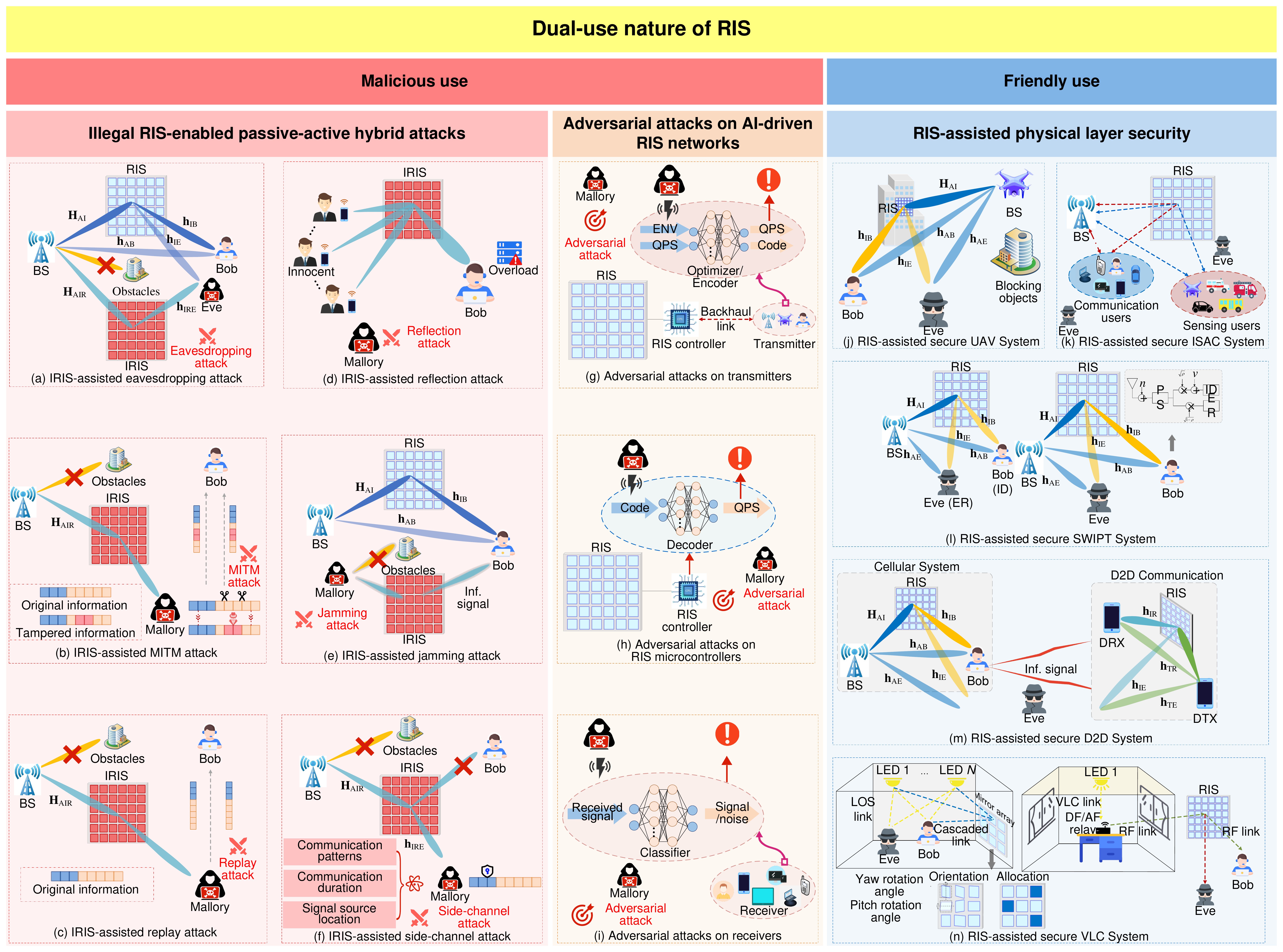}\\
  \vspace{-0.1in}
  \caption{Dual-use nature of RIS: The left part of the figure illustrates illegal RIS (IRIS)-assisted passive-active hybrid attacks including (a) eavesdropping attack; (b) MITM attack; (c) replay attack; (d) reflection attack; (e) jamming attack; and (f) side-channel attack. 
  The middle part of the figure demonstrates adversarial attacks in AI-driven RIS-assisted networks in which attackers can exploit the openness of wireless channel to insert adversarial perturbations and fool AI-based models into (g) predicting incorrect RIS's radiation pattern, or erroneously compressing the QPSs at the transmitter; (h) erroneously reconstructing the QPSs at the RIS microcontroller; and (i) miscalssifying the useful signal into the ``noise'' category at the receiver.
  The right part of the figure shows RIS-based defense mechanisms in which RISs can be integrated into diverse appealing wireless communication scenarios, including (j) RIS-assisted secure UAV systems; (k) RIS-assisted secure ISAC systems; (l) RIS-assisted secure SWIPT systems; (m) RIS-assisted secure D2D systems; and (n) RIS-assisted secure VLC systems to achieve vital security requirements.}\label{fig:I.0}
  \end{center}
  \vspace{-0.25in}
\end{figure*}

\begin{figure*}
  \begin{center}
  \includegraphics[width=0.8\textwidth]{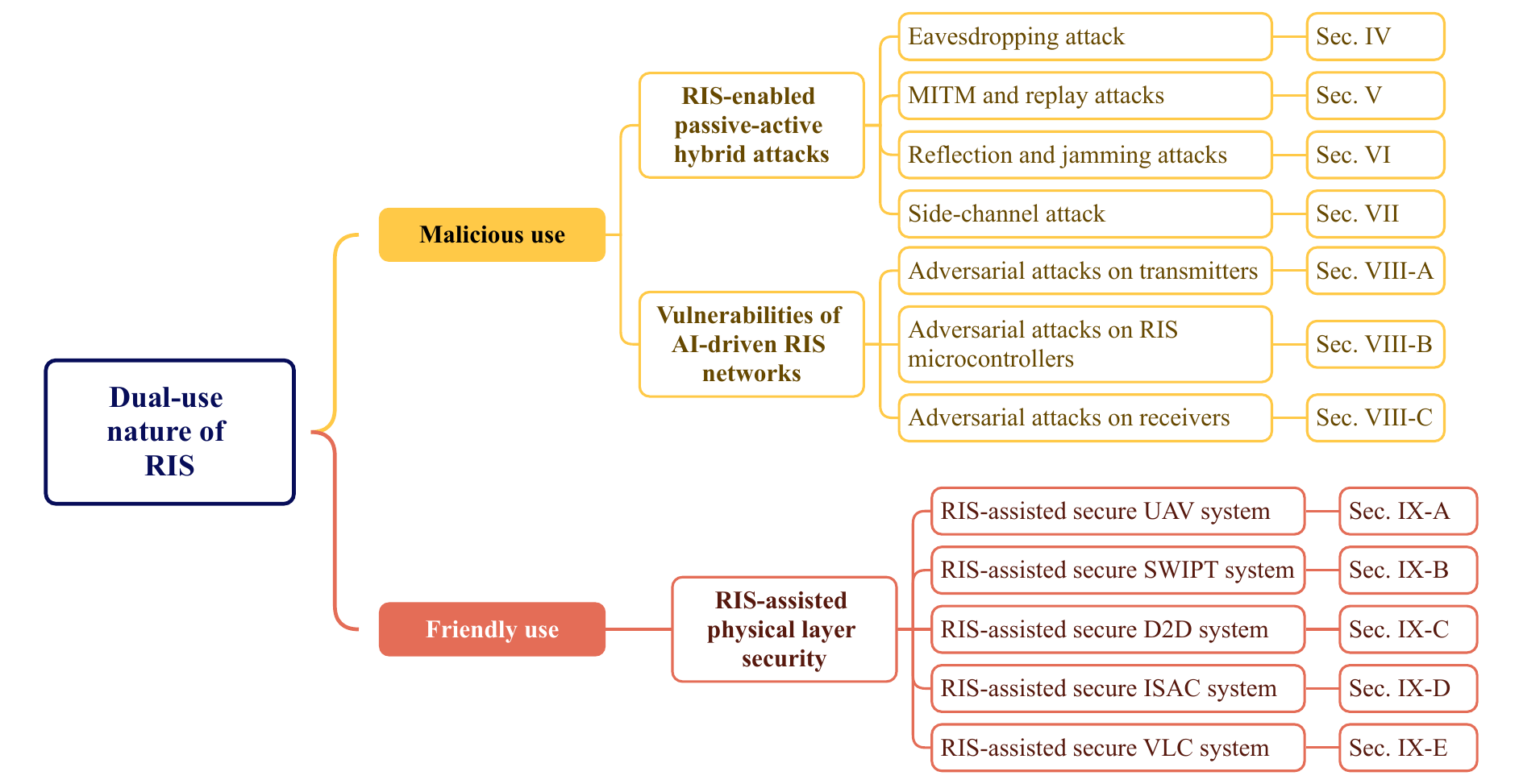}\\
  \vspace{-0.1in}
  \caption{The dual-use framework of RISs. (1) Malicious use: attackers can implement passive-active hybrid attacks by exploiting RIS's passivity to actively participate in malicious behavior due to reconfiguration, and achieve adversarial attacks on AI-driven RIS networks by exploiting the openness of wireless channels and the vulnerability of AI-driven models to adversarial.
  (2) Friendly use: RIS improves secrecy capacity by boosting the main channel and suppressing the eavesdropping channel in various RF and VLC scenarios.}\label{fig:I.0.A}
  \end{center}
  \vspace{-0.25in}
\end{figure*}

While RIS technology holds promise for enhancing wireless communication systems by boosting data rates, reducing power consumption, and ensuring secure transmission \cite{9896889, 9122596,9427098,9985456} as shown in Figs.~\ref{fig:I.0}(j)-\ref{fig:I.0}(n), it also introduces a new class of security vulnerabilities that were previously unexplored.
This reconfigurability endows the RISs with a dual-use nature that manifests in both constructive and destructive applications, as conceptualized in Fig.~\ref{fig:I.0.A}. On the friendly side, RIS capabilities can enhance legitimate communication through secrecy rate maximization, e.g., Section~\ref{Sec.IX}, and other security improvements. Conversely, these capabilities can potentially enable malicious applications when controlled by adversaries: (1) passive-active hybrid attacks through malicious reconfiguration, e.g., Sections~\ref{Sec.IV}-\ref{Sec.VII}, Figs.~\ref{fig:I.0}(a)-\ref{fig:I.0}(f); and (2) adversarial attacks on AI-driven RIS networks by exploiting wireless channel vulnerabilities, e.g., Section~\ref{Sec.VIII}, Figs.~\ref{fig:I.0}(g)-\ref{fig:I.0}(i). \par

Unlike traditional active attacks such as jamming and spoofing \cite{8334236,7840056}, or passive attacks \cite{NAYFEH2023103085} like eavesdropping, RIS can facilitate what can be described as \textit{passive implementations of active attacks}, or ``\textit{passive-active hybrid attacks}'', as demonstrated in Fig.~\ref{fig:I.0}.
For instance, an attacker named Mallory could exploit RIS to redirect legitimate signals towards unintended legitimate users \cite{10504558}, leading to unauthorized access or spoofing attacks \cite{9789438}. 
A RIS could also be configured to introduce destructive interference patterns that degrade communication quality for specific users \cite{9605003,10516473}, or enhance eavesdropping capabilities by directing more signal energy towards an eavesdropper's device \cite{10118920}. 
Furthermore, the attacker can exploit the openness of wireless channels and the susceptibility of AI models to adversarial perturbations, and can mislead AI-driven RIS networks into predicting incorrect RIS radiation patterns based on the environment descriptors, erroneously compressing or reconstructing the quantized phase shifts (QPSs) at the base stations (BSs) or RIS microcontrollers, respectively, and misclassifying useful signals into the ``noise'' category at the receivers. \par

\subsection{Passive-Active Hybrid Attacks: A New Security Challenge}\label{Sec.I.1}
RIS technology can modify the radio environment by altering propagation paths. This capability allows RISs to engage with signals in a manner that can potentially undermine communication security. 
This situation represents a hybrid attack scenario where the device, i.e., a RIS, is passive in terms of signal handling but actively participates in malicious behavior due to reconfiguration. The attack itself is active because it disrupts communication. The device used for the attack (the RIS) remains passive in terms of signal generation, relying instead on manipulating the environment around it.

Using RISs in this manner is an example of how passive technologies can be exploited for active malicious purposes. It underscores the dual-use nature of RIS technology, where its intended use for improving communication can be turned against legitimate users when controlled by an adversary. 
This type of attack blurs the lines between itself and traditional security threats, as it leverages the RIS's capability in controlling the wireless propagation environment to carry out ``\textit{passive-active hybrid attacks}''.  

These RIS-enabled passive-active hybrid attacks introduce novel challenges that fundamentally differ from conventional wireless threats. For example, unlike active jamming or spoofing attacks~\cite{8334236,7840056}, which require transmitters and leave detectable energy footprints, RIS-enabled attacks manipulate legitimate signals through augmentation of the wireless signal propagation environments and paths.
Compounding this stealth advantage is the RIS's inherent scalability: A single compromised surface with hundreds of elements can manipulate legitimate signals stealthily by controlling their reflection paths~\cite{10555049} to implement eavesdropping, man-in-the-middle (MITM), replay, reflection jamming, and side-channel attacks across spatial sectors, all while maintaining plausible deniability.

Specifically, eavesdropping attacks can potentially leverage RIS's passive beamforming to constructively focus signals toward eavesdroppers while avoiding active transmissions~\cite{10118920,10143983}; MITM attacks can stealthily intercept and manipulate communications by tuning RIS elements to redirect signals between unsuspecting parties; replay attacks can exploit a RIS's memoryless reflections to reflect and forward intercepted messages without leaving digital fingerprints; reflection and jamming attacks leverage a RIS's impedance matching to reflect and amplify ambient signals~\cite{9789438} into targeted denial-of-service (DoS) attacks; side-channel attacks utilize RIS-induced multipath distortions to expose subtle physical-layer leaks—all enabled by the RISs' ability to transform passive signal reflections into active threats without energy signatures~\cite{10081025,10424421}.
For security practitioners, these new forms of attacks demand new detection paradigms that correlate EM field anomalies with network traffic patterns, as neither physical nor network-layer monitoring alone can reliably expose these covert manipulations.

\subsection{Susceptibility to Single-Point Failure}\label{Sec.I.2}
A critical aspect of RIS security is the vulnerability of its microcontrollers, which serve as the central control points for configuring the reflection matrix. These microcontrollers dictate how signals are directed within the radio environment by adjusting the phase shifts of the reflecting elements~\cite{10328191,10516677}. Given their pivotal role, these microcontrollers represent a potential single-point failure within the RIS system. If compromised, a single microcontroller could allow Mallory to manipulate the entire RIS, leading to significant security breaches such as signal leakage, unauthorized redirection of signals, or complete disruption of communication channels~\cite{10504558,9789438}.

This vulnerability is further exacerbated by the typical deployment of multiple microcontrollers near the RIS. The physical proximity of these controllers makes them more accessible to adversaries, increasing the likelihood of unauthorized access or tampering. If Mallory gains control over one or more of these controllers, the attack success rate could significantly increase, as the compromised controllers could be used to coordinate a more sophisticated and widespread attack on the communication system.

The lack of adequate protection mechanisms for these microcontrollers compounds the risk. Without robust security measures such as encryption, secure access protocols, or physical protection, these controllers are vulnerable to being accessed or reprogrammed by attackers~\cite{9941040}. Additionally, the absence of a malfunction detection mechanism for microcontrollers is a significant concern. Without such a system, it may be challenging to identify when a controller has been compromised or behaved anomalously, allowing Mallory to carry out their activities undetected for extended periods.

\subsection{Need for New Control Mechanisms vs. Machine Learning Vulnerabilities}\label{Sec.I.3}
The sheer scale and complexity of RISs necessitate the development of new control mechanisms to manage the vast number of reflecting elements and optimize their configurations in real time. Traditional control methods may not suffice, given the dynamic nature of wireless environments and the intricate requirements of RIS. Consequently, machine learning (ML) and reinforcement learning (RL) have emerged as promising approaches to efficiently control RIS operations by automating decision-making processes and adapting to changing environmental conditions~\cite{10463689,10107766,10345491,9999559}.

However, the integration of ML and RL into RIS control systems exposes these networks to a range of sophisticated ML attacks. Backdoor attacks~\cite{9541185}, for instance, could allow adversaries to inject malicious triggers into the ML models, causing the RIS to behave unexpectedly when specific conditions are met, as shown in the middle part of Fig.~\ref{fig:I.0}. Similarly, poisoning attacks involve corrupting the training data used by ML models, leading to flawed decision-making and compromised RIS performance~\cite{10209197,10103793}. Attacks on RL could manipulate the learning process, causing the RIS to adopt sub-optimal or even harmful policies that degrade the security and efficiency of the communication system~\cite{rosenberg2021adversarial,du2024parl}.
These vulnerabilities highlight the paradox of relying on advanced ML and RL techniques to manage RIS: 

\textit{While these technologies are essential for handling the complexity of RIS, they also introduce new attack surfaces that adversaries could exploit.}

\subsection{Contributions of Our Survey}\label{Sec.I.4}

The emergence of \textit{passive-active hybrid attacks}, coupled with the vulnerabilities introduced by ML and RL control mechanisms, highlights the need for enhanced security measures in RIS-assisted next-generation wireless networks. To fully leverage the benefits of RIS technology while mitigating these new risks, it is essential to implement robust access controls, secure artificial intelligence (AI) algorithms resistant to adversarial manipulation, and continuously monitor RIS configurations. By addressing these challenges, the potential of RISs to revolutionize wireless communication can be realized without compromising security.

The contributions of this survey are summarized as follows. 
\begin{itemize}
\item 
We unveil the \textit{dual-use potential of RIS technology}, illustrating how adversaries can exploit its communication-enhancing capabilities to compromise legitimate users, posing significant security risks.

\item 
We identify a new class of security vulnerabilities termed ``\textit{passive-active hybrid attacks}'', where RIS, despite passively handling signals, can be reconfigured to actively facilitate malicious activities. This enables various RIS-assisted attacks, including eavesdropping, MITM, replay, reflection jamming, and side-channel attacks, emphasizing the need for stronger security frameworks.

\item 
We reveal that adversaries can exploit wireless channel openness to introduce adversarial perturbations in AI-driven RIS networks. These perturbations disrupt transmitters, RIS microcontrollers, and receivers, causing errors in RIS reflection predictions, disrupting QPS compression or reconstruction processes, and inducing signal misclassification, which underscores the urgency of developing resilient AI-RIS defenses.

\item 
We provide a comprehensive analysis of how RIS technology enhances security and privacy across radio frequency (RF) and visible light communication (VLC) systems, detailing its role in fortifying wireless communication networks.
\end{itemize}
A list of key findings and insights is provided, as follows:
\begin{itemize}
\item Cross-layer collaboration between the network and physical layers is the key to preventing ``passive-active hybrid attacks''. By integrating anomaly detection at the network layer with EM signal detection at the physical layer, a more comprehensive security situational awareness is desirable.

\item Effective detection and tracking of attackers and their controlled RISs are crucial. The passive nature of threats, coupled with the absence of active connections, can lead legitimate users to inadvertently engage in ``passive-active hybrid attacks''. This increases the difficulty in identifying attack types and tracking attack sources, highlighting the urgent need for advanced detection methods.

\item Enhancing adversarial defense techniques for AI-powered RIS-assisted communication networks is essential to counter evolving attacks. Recent breakthroughs in various AI domains can be leveraged to strengthen the security, robustness, and resilience of AI-RIS communications.

\item Trade-off between security and cost warrants thorough exploration. While recent research has focused on increasing RIS elements and optimizing joint objectives to enhance security and improve degrees of freedom (DoFs), this has led to substantial increases in computational complexity, power consumption, hardware overhead, and other associated costs.
\end{itemize}

\subsection{Paper Organization}\label{Sec.I.5}
\begin{figure*}
  \begin{center}
  \includegraphics[width=0.9\textwidth]{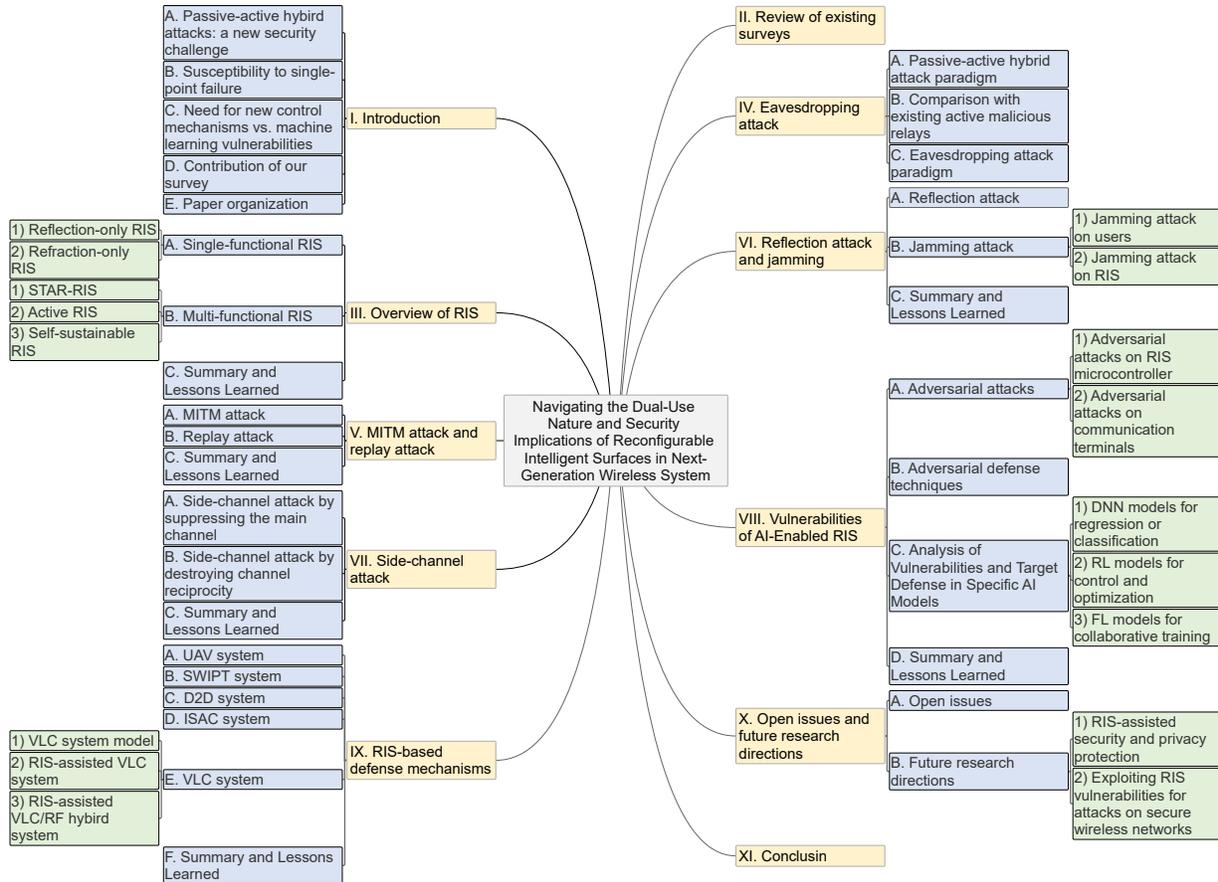}\\
  \vspace{-0.2in}
  \caption{Mind map of our survey: Section~\ref{Sec.I} is introduction. Section~\ref{Sec.II} is the review of existing surveys.
  Section~\ref{Sec.III} delineates the structures, principles and functions of different RIS types.
  The RIS-assisted ``passive-active hybrid attacks'' including eavesdropping, MITM, replay, reflection, jamming and side-channel attacks, are introduced in Sections~\ref{Sec.IV},~\ref{Sec.V},~\ref{Sec.VI} and~\ref{Sec.VII} in detail.
 Adversarial attacks against AI-powered RIS-assisted wireless networks are scrutinized in Section~\ref{Sec.VIII}. Section~\ref{Sec.IX} delves into the application of RIS in security enhancement and privacy protection across various scenarios. Subsequently, Section~\ref{Sec.X} consolidates the open issues and discusses future research directions. The survey concludes with Section~\ref{Sec.XI}, summarizing the key findings.}
 \label{fig:I.1}
  \end{center}
  \vspace{-0.25in}
\end{figure*}

The structure of this survey is outlined as follows and illustrated in Fig. \ref{fig:I.1}. Section \ref{Sec.II} reviews recent research on the applications of RIS in future wireless systems and secure communication networks. Section \ref{Sec.III} provides an overview of RIS covering the structure, principles, and functions of different RIS types. Section~\ref{Sec.IV},~\ref{Sec.V},~\ref{Sec.VI} and~\ref{Sec.VII} introduce RIS-assisted ``passive-active hybrid attacks'' including eavesdropping attack, MITM and replay attacks, reflection attack and jamming, and side-channel attack, respectively.
 In Section~\ref{Sec.VIII}, adversarial attacks on AI-driven RIS-assisted networks are examined. Section~\ref{Sec.IX} provides a comprehensive analysis of how RIS enhances security and privacy across various RF and VLC scenarios. Section~\ref{Sec.X} discusses open challenges and future research directions, focusing on RIS-assisted security and privacy measures, as well as adversarial exploitation of RIS vulnerabilities. Finally, Section~\ref{Sec.XI} concludes the survey by summarizing key findings. Table~\ref{Symbol comparison table.1} provides definitions for the notation used throughout the paper.

\begin{table}[t]
    \centering
    \caption{Notation and Definition}
    \vspace{-0.1in}
    \begin{tabular}{@{}m{1.8cm}l@{}}\hline
        \textbf{Notation} & \textbf{Definition}\\ \hline
        $N_\mathrm{t}$ & Transmission antenna \\
        $m$ & The $m$-th element in the RIS \\
        $\beta_m^t$ & Transmission coefficient of the $m$-th RIS element \\
        $\beta_m^r$ & Reflection coefficient of the $m$-th RIS element \\
        $\beta_{\mathrm{max}}$ & Amplification factor \\
        $\theta_m^t$ & Transmission phase shift of the $m$-th RIS element \\
        $\theta_m^r$ & Reflection phase shift of the $m$-th RIS element \\
        $\mathbf{H}_\mathrm{AI}$ & Equivalent channel link of BS-RIS \\
        $\mathbf{h}_\mathrm{IB}$ & Equivalent channel link of RIS-Bob \\
        $\mathbf{h}_\mathrm{IE}$ & Equivalent channel link of RIS-eavesdropper \\
        $L$ & The number of sectors in a multi-sector RIS \\
        $T_\mathrm{r}$ & RPT period \\
        $T_\mathrm{d}$ & DT period \\
        $\rho$ & PS ratio \\
        $\rho^*$ & Optimal PS ratio \\
        $K$ & Number of primary users \\
        $\mathcal{RV}$ & Gaussian distribution \\
        $C_{\mathrm{VLC},k}$ & Channel capacity of $k\in\{\mathrm{Bob},\mathrm{Eve}\}$ \\
        $\gamma_k$ & SINR of $k\in\{\mathrm{Bob},\mathrm{Eve}\}$ \\
        $B$ & Modulation bandwidth \\
        $e$ & Base of natural logarithms \\
        $\mathcal{O}(\cdot)$ & Computational complexity \\
        $d_{\mathrm{F}}$ & Rayleigh distance \\
        $D$ & RIS aperture \\
        $\lambda$ & Length of carrier wavelength \\ \hline
        \end{tabular}
    \nonumber
    \label{Symbol comparison table.1}
    \vspace{-0.1in}
\end{table}

% ==========Sec. II Review of Existing Surveys
\section{Review of Existing Surveys}\label{Sec.II}

As shown in~Table \ref{Table:II.1}, many studies have extensively examined the development and integration of RISs, emphasizing their potential and challenges. In~\cite{9122596}, the development and design considerations for RISs in upcoming wireless networks were provided, offering insights into the future of wireless communication. An overview of how to address the challenges faced by RIS-assisted hybrid wireless networks with passive and active components was presented in~\cite{9326394}. The design and applications of RISs to assist the fifth-generation (5G) and beyond 5G (B5G) wireless communication networks were thoroughly investigated in~\cite{okogbaa2022design}. A comprehensive tutorial of optical RISs and RIS-assisted indoor VLC systems was investigated in~\cite{9968053}, which also discussed the integration of optical RISs with other emerging technologies. A systematic overview and understanding of efficient optimization approaches for RIS-assisted envisioned six-generation (6G) networks was provided in~\cite{10361836} and included model-based, heuristic, and ML algorithms with secrecy rate (SR) maximization as one of the diverse objectives and constraints. A state-of-the-art survey in joint optimization designs and performance evaluation for RIS-assisted 6G wireless scenarios was afforded in~\cite{10356072}, focusing on optimizing system effectiveness and resource efficiency. \par

\begin{figure}
  \begin{center}
  \includegraphics[width=3.0in]{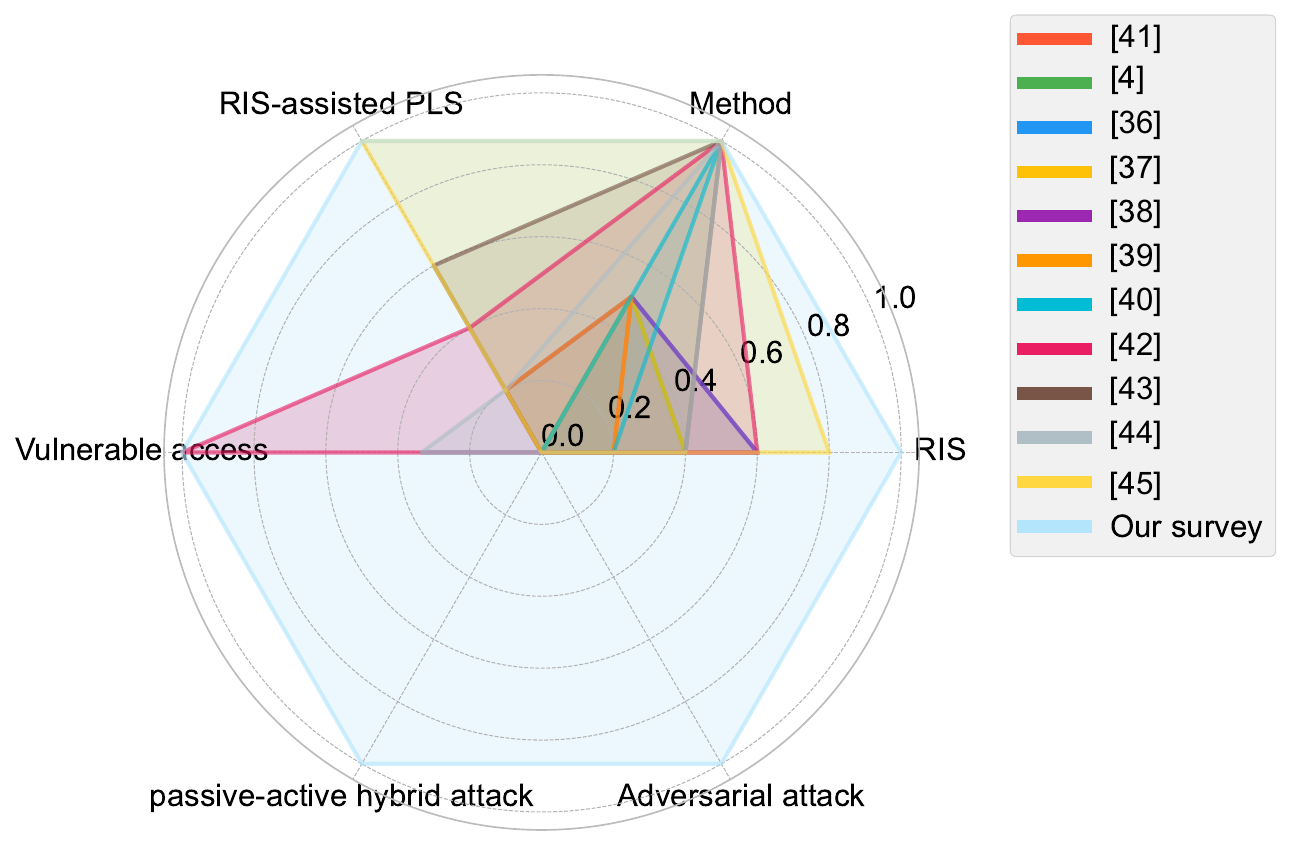}
  \vspace{-0.1in}
  \caption{Radar graph of existing surveys: the comparison of diverse research dimensions including RIS, method, RIS-assisted security enhancement, vulnerable access, passive-active hybrid attack, and adversarial attack between related works and our survey.}\label{fig:II.1}
  \end{center}
  \vspace{-0.25in}
\end{figure}

A wide range of literature has been reviewed on secure wireless communication networks. However, the potential security challenges have not been systematically summarized. 
The exploration of security enhancement in existing and emerging wireless networks, including employing RIS, was comprehensively explored in~\cite{sanenga2020overview}. A general framework for RIS-assisted security enhancement was proposed in~\cite{10188924} against eavesdropping and jamming attacks in 6G-internet of things (IoT) networks and focuses on secure resource allocation, beamforming, artificial noise (AN), and cooperative communications. A detailed overview of RIS-assisted security enhancement for 6G wireless communication was provided in \cite{10409564} and focuses on applying RIS in various 6G scenarios and wireless system topologies. A comprehensive survey for ML-empowered security enhancement techniques toward 6G was served in~\cite{zhang2024artificial} and focuses on ML-enabled intelligent privacy protection for 6G. A detailed literature review about the information-theoretic security of RIS-assisted emerging RF and optical scenarios was offered in~\cite{khoshafa2024ris}, and also discusses ML techniques for RIS-assisted security enhancement. The comparison of dimensions and depth of research between related works and our survey is illustrated as a radar chart shown in Fig.~\ref{fig:II.1}. \par

Our survey is motivated by the rapid development of emerging RIS technology in wireless communication systems. The union of appealing scenarios with the assistance of RIS can fully leverage the advantages of diverse systems and significantly enhance security performance. Except for the significant effects of RIS in boosting security enhancement, its potential threats cannot be ignored, mainly including vulnerable access, passive-active hybrid attacks, and adversarial attacks.
To the best of the authors' knowledge, this is the first article to discuss
the potential security risks of integrating RIS into wireless communication systems, as well as the capability of RIS in security enhancement.

\begin{table*}[!ht]
    \centering
    \caption{Comparative Review of Existing Surveys: Key Discussion Areas Cover RIS Functions, Optimization Methods, RIS-Assisted Security Enhancement Scenarios, Potential Threats for RIS-Assisted Wireless Networks, and Main Focus}
    \vspace{-0.1in}
    \begin{adjustbox}{width=0.9\textwidth}
    \begin{tabular}{|m{1cm}<{\centering}|m{1cm}<{\centering}|m{1cm}<{\centering}|m{1cm}<{\centering}|m{1cm}<{\centering}|m{1cm}<{\centering}|m{1cm}<{\centering}|m{1cm}<{\centering}|m{1cm}<{\centering}|m{1cm}<{\centering}|m{1cm}<{\centering}|m{1.3cm}<{\centering}|m{1.3cm}<{\centering}|m{1.3cm}<{\centering}|m{3.3cm}<{\centering}|}
    \hline
    \multirow{2}{*}[-1.5em]{\textbf{Ref.}} & \multirow{2}{*}[-1.5em]{\textbf{Year}} & \multicolumn{2}{c|}{\textbf{RIS}} & \multicolumn{2}{c|}{\textbf{Method}} & \multicolumn{5}{c|}{\textbf{RIS-assisted security enhancement}} & \multicolumn{3}{c|}{\textbf{\begin{minipage}{2.4cm}
\centering
Potential threat for RIS-assisted wireless networks
\end{minipage}
    }} & \multirow{2}{*}[-2.5em]{\textbf{Main focus}} \\[20pt] \cline{3-14}
    ~ & ~ & \textbf{Single-mode RIS} & \textbf{Enhanced RIS designs} & \textbf{Convex optimization} & \textbf{AI} & \textbf{UAV scenarios} & \textbf{SWIPT scenarios} & \textbf{D2D scenarios} & \textbf{ISAC scenarios} & \textbf{VLC scenarios} & \textbf{Vulnerable access} & \textbf{Passive-active hybrid attack} & \textbf{Adversarial attack} & ~ \\ \hline
    \cite{sanenga2020overview} & 2020 & \Checkmark & \XSolidBrush & \Checkmark & \XSolidBrush & \XSolidBrush & \XSolidBrush &  \XSolidBrush &  \XSolidBrush & \XSolidBrush & \XSolidBrush & \XSolidBrush & \XSolidBrush & An overview of security enhancement in existing and emerging wireless networks including employing RIS \\ \hline    
    \cite{9122596} & 2020 & \Checkmark & \XSolidBrush & \Checkmark & \XSolidBrush & \XSolidBrush & \XSolidBrush &  \XSolidBrush &  \XSolidBrush & \XSolidBrush & \XSolidBrush & \XSolidBrush & \XSolidBrush & An overview of development and design consideration for RIS in upcoming wireless networks \\ \hline
    \cite{9326394} & 2021 & \Checkmark & \Checkmark & \Checkmark & \XSolidBrush & \XSolidBrush & \XSolidBrush & \XSolidBrush &  \XSolidBrush & \XSolidBrush & \XSolidBrush & \XSolidBrush & \XSolidBrush & An overview of how to tackle challenges faced by hybrid RIS-assisted wireless network, including security enhancement as a future direction \\ \hline
    \cite{okogbaa2022design} & 2022 & \Checkmark & \XSolidBrush & \Checkmark & \XSolidBrush & \XSolidBrush & \XSolidBrush & \XSolidBrush & \XSolidBrush & \XSolidBrush & \XSolidBrush & \XSolidBrush & \XSolidBrush & An overview of design and application of RISs to assist the 5G and B5G communication networks \\ \hline
    \cite{9968053} & 2023 & \Checkmark & \Checkmark & \Checkmark & \XSolidBrush & \XSolidBrush & \XSolidBrush & \XSolidBrush & \XSolidBrush & \Checkmark & \XSolidBrush & \XSolidBrush & \XSolidBrush & An overview of optical RISs-assisted indoor VLC system \\ \hline
    \cite{10361836} & 2023 & \Checkmark & \XSolidBrush & \Checkmark & \XSolidBrush & \XSolidBrush & \Checkmark &  \XSolidBrush &  \XSolidBrush & \XSolidBrush & \XSolidBrush & \XSolidBrush & \XSolidBrush & An overview of optimization techniques for RIS-assisted envisioned 6G networks with security enhancement as one of the scenarios \\ \hline
    \cite{10356072} & 2023 & \Checkmark & \XSolidBrush & \Checkmark & \Checkmark & \XSolidBrush & \XSolidBrush & \XSolidBrush & \XSolidBrush & \XSolidBrush & \Checkmark & \XSolidBrush & \XSolidBrush & An overview of joint optimization designs and performance evaluation for RIS-assisted 6G wireless scenarios \\ \hline
    \cite{10188924} & 2024 & \Checkmark & \Checkmark & \Checkmark & \Checkmark & \Checkmark & \Checkmark & \XSolidBrush & \XSolidBrush & \XSolidBrush & \Checkmark & \XSolidBrush & \XSolidBrush & An overview of RIS-assisted security enhancement for 6G-IoT networks against eavesdropping and jamming attacks \\ \hline
    \cite{10409564} & 2024 & \Checkmark & \XSolidBrush & \Checkmark & \Checkmark & \Checkmark & \XSolidBrush & \Checkmark & \XSolidBrush & \Checkmark & \XSolidBrush & \XSolidBrush & \XSolidBrush & An overview of RIS-assisted security enhancement for 6G wireless communication including diverse application scenarios and wireless topologies \\ \hline
    \cite{zhang2024artificial} & 2024 & \Checkmark & \XSolidBrush & \Checkmark & \Checkmark & \Checkmark & \XSolidBrush & \XSolidBrush & \XSolidBrush & \XSolidBrush & \Checkmark & \XSolidBrush & \XSolidBrush & An overview of ML-empowered security enhancement toward 6G including key 6G radio techniques and ML-enabled privacy protection \\ \hline
    \cite{khoshafa2024ris} & 2024 & \Checkmark & \Checkmark & \Checkmark & \Checkmark & \Checkmark & \Checkmark & \Checkmark & \Checkmark & \Checkmark & \XSolidBrush & \XSolidBrush & \XSolidBrush & An overview of RIS-assisted security enhancement in emerging RF and VLC systems, and discusses ML techniques for RIS-assisted privacy protection \\ \hline
    \multicolumn{2}{|c|}{Our survey} & \Checkmark & \Checkmark & \Checkmark & \Checkmark & \Checkmark & \Checkmark & \Checkmark & \Checkmark & \Checkmark & \Checkmark & \Checkmark & \Checkmark & An overview of RIS technology including its various structures and functionalities, RIS-assisted security enhancement in diverse RF and VLC scenarios, potential security threats including vulnerable access, passive-active hybrid attacks, and adversarial attacks \\ 
    \hline
    \end{tabular}
    \end{adjustbox}
    \label{Table:II.1}
    \vspace{-0.1in}
\end{table*}

% ================= Sec. Overview of Reconfiguration Intelligent Surfaces
\section{Overview of Reconfigurable Intelligent Surfaces}\label{Sec.III}

A RIS is a two-dimensional (2D) artificial EM metasurface, including a RIS panel, copper backplane, and circuit board controlled by a microcontroller~\cite{9941040}. Several passive meta-atoms are printed on the RIS panel to directly and independently interact with incident signals~\cite{9982493}. The middle copper backplane is utilized to avoid signal leakage, and the circuit board is adopted to independently adjust the phase shift of each meta-atom and even amplify the power of the incident signal. Then, the RIS can manipulate the incident propagation towards the desired direction to enhance communication coverage, boost security performance, and improve spectrum and energy efficiency. \par

Multiple types of RIS can be adapted to satisfy the requirements of different communication systems.
Recent comprehensive studies have systematically characterized RIS hardware architectures and their corresponding evolving roles in emerging wireless networks. For instance, the study in~\cite{8741198} delves into the influence of RIS resolution on its power consumption, and devises energy-efficient schemes for both RIS phase shifts and BS power allocation to meet the green and sustainable criteria of wireless networks.
The work in~\cite{arXiv11136} summarizes the utilization of fabricated passive RISs in two representative indoor wireless trials at the WiFi frequency band for achieving spatiotemporal focusing and nulling, and a multipath scattering environment. The study~\cite{10596064} delves into diverse hardware architectures and resulting versatile operating modes of RISs, such as signal reflection and transmission, reception and amplification, sensing and computation, and corresponding potential applications, including integrating sensing and communications (ISAC), next-generation multiple access (NGMA). The study~\cite{s13638} formalizes RIS control interfaces and signaling protocols for heterogeneous architectures, including reflection, refraction, transmission, etc, while quantifying operational metrics like bandwidth/area of influence that are critical for deployment planning. \par

Recent advances in RIS technology continue to expand its capabilities and applicability toward the vision of integrated communication, sensing, and computing in next-generation wireless systems.
In line of this objective, the work in~\cite{10352433} proposes a hybrid RIS architecture that simultaneously performs signal reflection and sensing, and greatly enhances the self-configuring and adaptability of each meta-atom.
The work in~\cite{Alexandropoulos2025ReceivingRE} introduces a semi-passive RIS operating in a time-division duplex manner, with one phase estimating channel matrices through the tunable absorption phase profiles and the other facilitating inter-user communication via capacity-achieving reflection coefficients.
Regarding intelligent integration of communications and computing, reconfigurable intelligent computational surfaces are investigated in~\cite{10183797}. These surfaces leverage task-oriented computational metamaterials to enable intelligent spectrum sensing and secure wireless transmission concurrently with communication functions.
The study in~\cite{10922857} investigates a novel stacked intelligent surface design composed of multiple transmissive metasurfaces to improve signal processing capability in the EM domain. \par

To analyze the dual-use nature of RISs, we next focus on single-mode RISs, including reflection-only and refraction-only RISs, and enhanced RIS designs, including STAR-RIS, active-RIS, and self-sustainable RIS, which exhibit distinct trade-offs between functionality and security implications. \par

\begin{table*}[ht]
    \centering
    \caption{Comparison of Multiple RIS Structures: Compares Single-Mode RISs and Enhanced RIS Designs Based on Operation Mode, Amplitude Coefficient, Phase Shift, Hardware Cost, Complexity, Coverage Area, Path Loss, and Energy Dependency}
    \vspace{-0.1in}
    \begin{adjustbox}{width=0.9\textwidth}
    \begin{tabular}{|m{1.2cm}<{\centering}|m{1.2cm}<{\centering}|m{1.2cm}<{\centering}|m{2cm}<{\centering}|m{2cm}<{\centering}|m{1.2cm}<{\centering}|m{1.2cm}<{\centering}|m{1.2cm}<{\centering}|m{1.2cm}<{\centering}|m{1.2cm}<{\centering}|m{1.2cm}<{\centering}|}
    \hline
        \multicolumn{2}{|c|}{\textbf{RIS Structure}} & \textbf{Operation Mode} & \textbf{Amplitude Coefficient} & \textbf{Phase shift} & \textbf{Hardware Cost} & \textbf{Complexity} & \textbf{Coverage Area} & \textbf{Path Loss} & \textbf{Energy Dependency} & \textbf{Ref.} \\ \hline
        \multirow{2}{*}[-0.5em]{\begin{minipage}{1.2cm}
\centering
Single-mode RIS
\end{minipage}} & Reflection-only & S mode & $\beta_m^r=1$ & $\theta_m^r\in\left[0,2\pi\right)$ & Low & Low & Half-space & Double-fading & Dependent & \cite{10144102,9661068} \\ \cline{2-11}
         ~ & Refraction-only & S mode & $\beta_m^t=1$ & $\theta_m^t\in\left[0,2\pi\right)$ & Low & Low & Half-space & Double-fading & Dependent & \cite{10330576} \\ \hline
        ~ & STAR & S mode & $\beta_m^r+\beta_m^t=1$ & $\theta_m^r,\theta_m^t\in\left[0,2\pi\right)$ & Middle & Middle & Full-space & Double-fading & Dependent & \cite{9690478,10288376,10436358,10133841,10133841} \\ \cline{2-11}
        Enhanced RIS designs & Active & S mode & $\beta_m^r/\beta_m^t\leq\beta_{\mathrm{max}}$ or $\beta_m^r+\beta_m^t\leq\beta_{\mathrm{max}}$  & $\theta_m^r/\theta_m^t\in\left[0,2\pi\right)$ & Middle & Middle & Full-space or Half-space & Additive & Dependent & \cite{10153967,10198355,10188853,9568854} \\ \cline{2-11}
         ~ & Self-sustainable & S mode \& H mode & $\beta_m^r+\beta_m^t\leq\beta_{\mathrm{max}}$  & $\theta_m^r/\theta_m^t\in\left[0,2\pi\right)$ & High & High & Full-space & Additive & Independent & \cite{10370741,10198324,10225701} \\ \hline
    \end{tabular}
    \end{adjustbox}
    \label{Table:III.1}
    \vspace{-0.2in}
\end{table*}

 \begin{figure}[t!]
    \centering
    \begin{minipage}[t]{0.8\linewidth}
    \centering
            \includegraphics[width=1.0\linewidth]{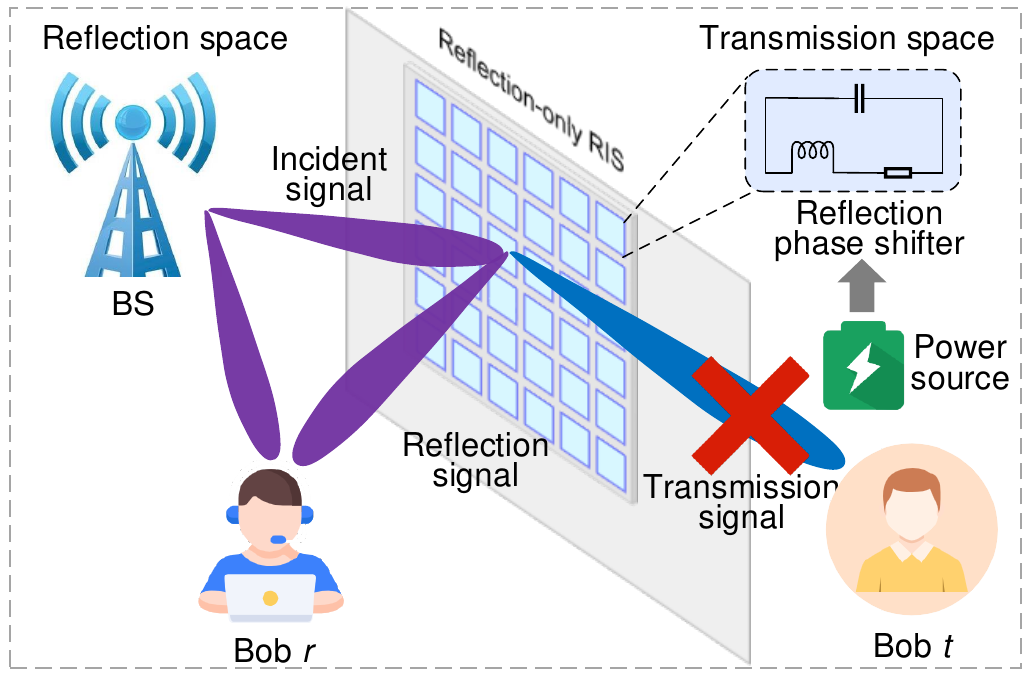}\label{fig:III.1.1}
            \vspace{-0.02in}
            \footnotesize{(a) Reflection-only RIS.}\\
    \end{minipage}\\
    \begin{minipage}[t]{0.8\linewidth}
    \centering
            \includegraphics[width=1.0\linewidth]{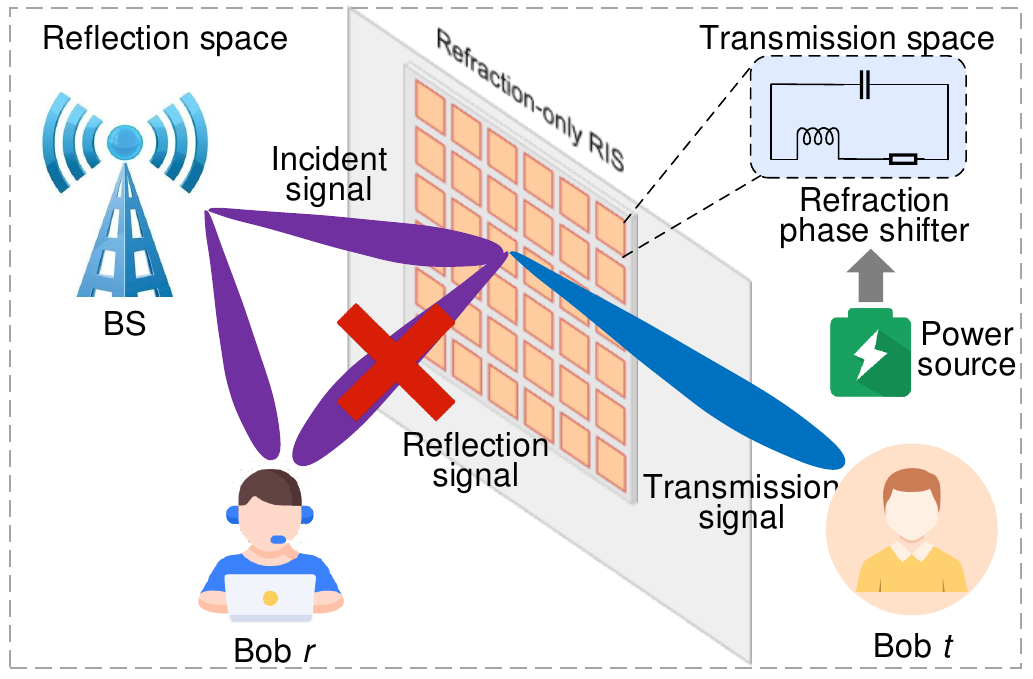}\label{fig:III.1.2}
            \vspace{-0.02in}
            \footnotesize{(b) Refraction-only RIS.}\\
    \end{minipage}
    \caption{Single-mode RIS: (a) Reflection-only RIS: fully reflect incident signals into reflection space without any refraction; (b) Refraction-only RIS: fully refract incident signals into transmission space without any reflection.}
    \label{fig:III.1}
    \vspace{-0.25in}
 \end{figure}

 \begin{figure*}[t!]
    \centering
    \begin{minipage}[t]{0.7\linewidth}
    \centering
            \includegraphics[width=1.0\linewidth]{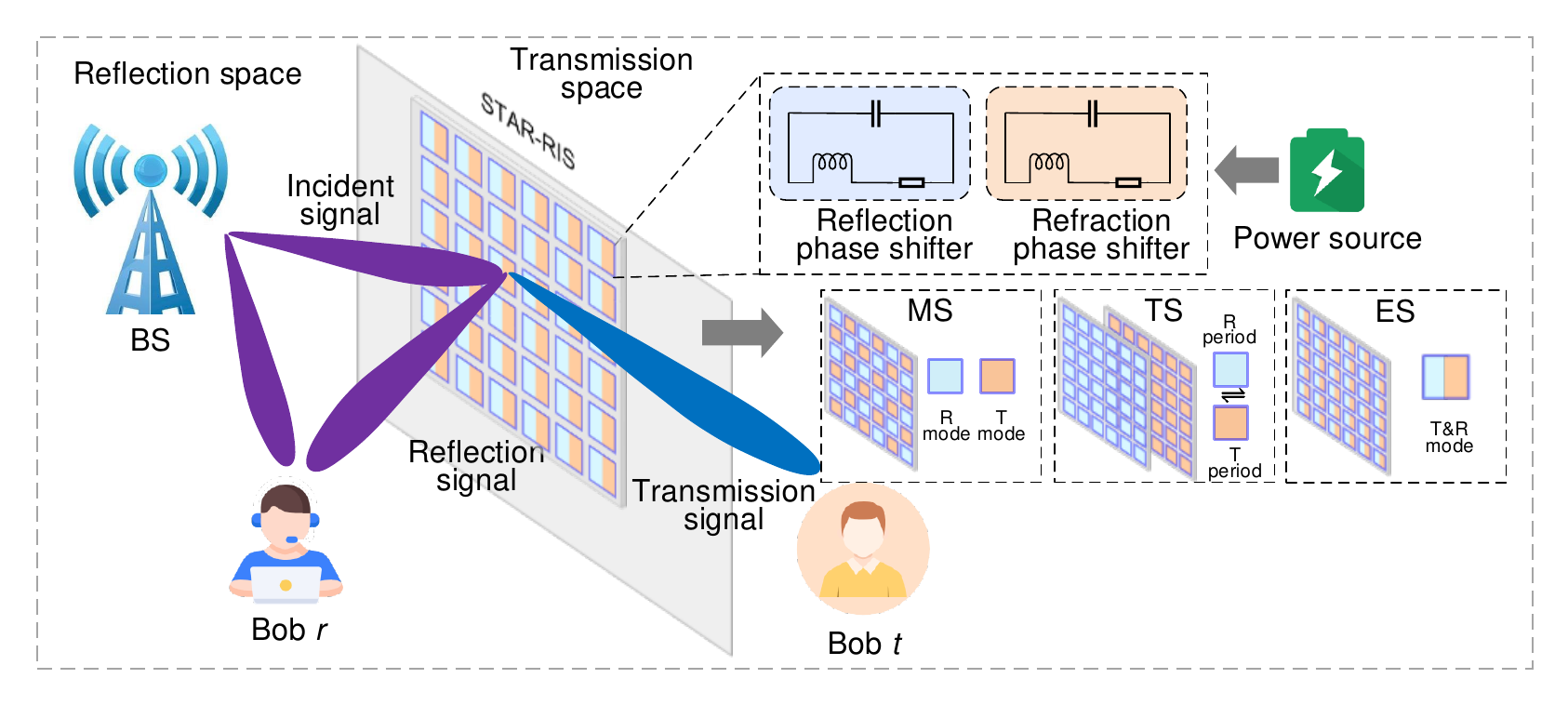}\label{fig:III.2.1}
            \vspace{-0.02in}
            \footnotesize{(a) STAR-RIS.}\\
    \end{minipage}
    \begin{minipage}[t]{0.7\linewidth}
    \centering
            \includegraphics[width=1.0\linewidth]{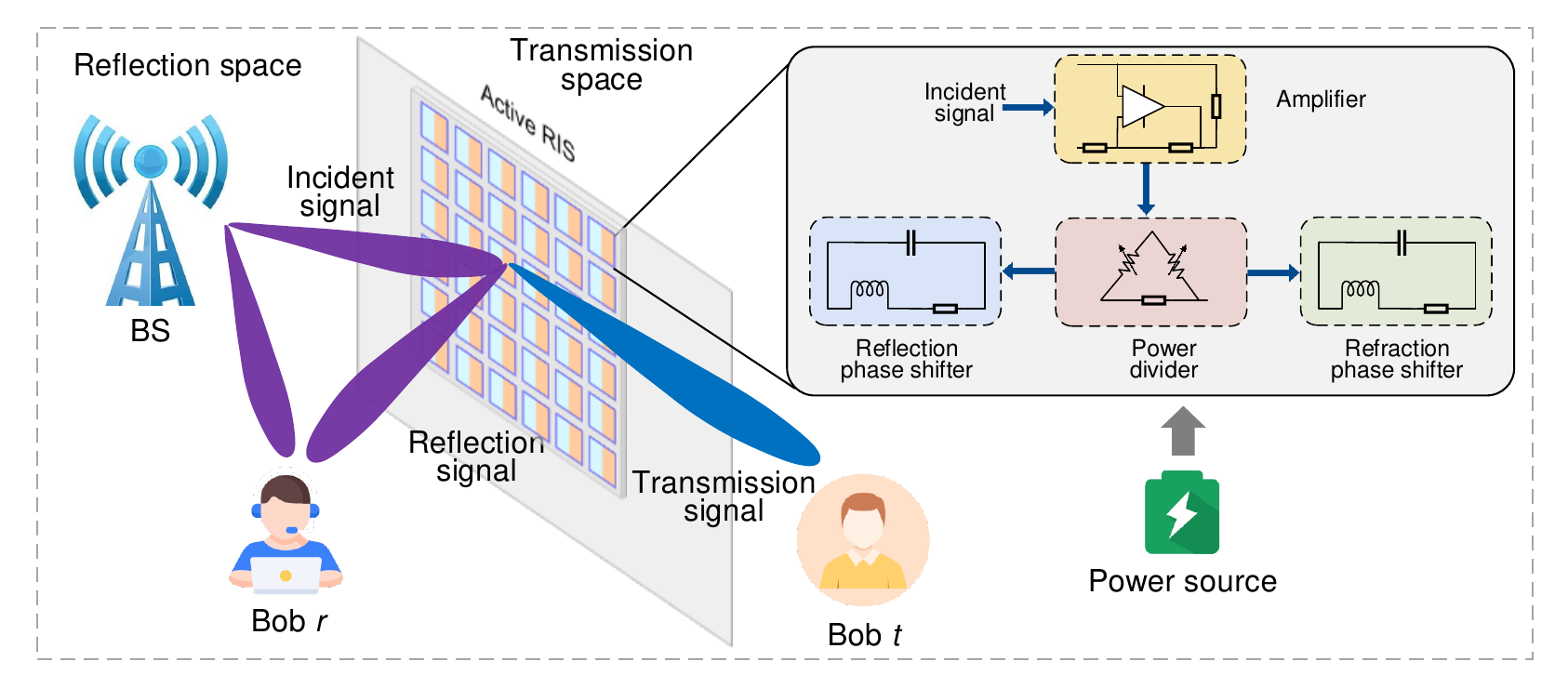}\label{fig:III.2.2}
            \vspace{-0.02in}
            \footnotesize{(b) Active RIS.}\\
    \end{minipage}
    \begin{minipage}[t]{0.7\linewidth}
    \centering
            \includegraphics[width=1.0\linewidth]{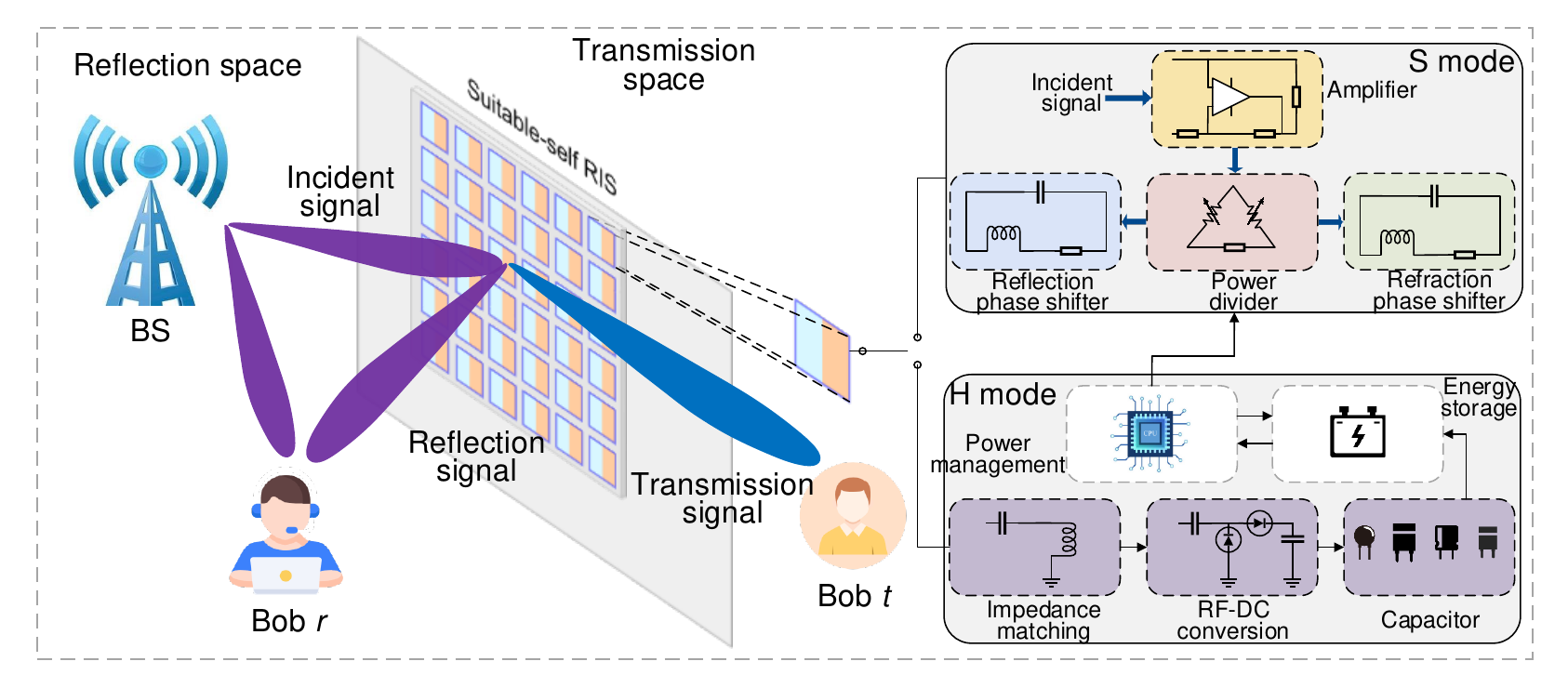}\label{fig:III.2.3}
            \vspace{-0.02in}
            \footnotesize{(c) Self-sustainable RIS.}\\
    \end{minipage}
    \caption{Enhanced RIS designs: (a) STAR-RIS: includes three practical operating protocols called MS, TS, and ES to achieve signal refraction and reflection dependent on power source simultaneously; (b) Active RIS: amplifies the transmission and reflection signals with low-cost hardware dependent on power source; (c) Self-sustainable RIS: supplies signal amplification, reflection, refraction in S mode by harvesting energy in H mode without power source.}
    \label{fig:III.2}
    \vspace{-0.25in}
 \end{figure*}

\subsection{Single-Mode RIS}\label{Sec.III.1}
A single-mode RIS can only perform reflection or refraction functions on the incident signal and can be divided into reflection-only and refraction-only RIS.

\subsubsection{Reflection-only RIS}\label{Sec.III.1.1}
A reflection-only RIS is the most classic type adopted in the communication system to manipulate the incident propagation~\cite{10144102}, as shown in Fig.~\ref{fig:III.1}(a). Each passive RIS element is uniquely connected to an individual and reconfigurable impedance network~\cite{9514409}, enabling it to fine-tune the phase shift of incoming signals independently and then totally reflect them in the intended direction within the unit-modulus reflection coefficient~\cite{10964081}. The BS and its associated users are situated on the same side as the RIS, and the eavesdropper may be located at the reflection space to eavesdrop on the confidential signals for the legitimate users. RISs can not only facilitate the formation of a virtual line-of-sight (LoS) connection between the BS and the users to enhance communication coverage and improve the quality of service (QoS) for users, but also can manipulate the incident propagation towards the desired directions to focus only on authorized users and superimpose destructively at unauthorized users.

\subsubsection{Refraction-only RIS}\label{Sec.III.1.2}
The metasurface of a refraction-only RIS is encapsulated with a transparent layer composed of glass material~\cite{9690478} and can make all incident signals from the reflection space pass through it and access the transmission space without any reflection. The BS and users are divided into independent reflection and transmission spaces by the RIS, as demonstrated in Fig.~\ref{fig:III.1}(b), and the eavesdropper may position themselves within the transmission area to intercept confidential information intended for authorized users. The transmission signal is manipulated by the massive independent and passive refracting elements~\cite{9661068} with tunable phase shifts and unit-modulus and can propagate in the desired direction with optimal refraction matrix of RIS to enhance the security capacity, reliability, throughput, transmission rate and other metrics of the wireless communication system~\cite{10330576}. The refraction-only RISs enable efficient wireless communication between distinct environments, such as outdoor BSs serving indoor users in buildings or vehicles, ensuring seamless signal transmission across spatial boundaries.

\subsection{Enhanced RIS Designs}\label{Sec.III.2}
Though the reflection mentioned above or refraction RIS has played dramatically significant roles in covering communication blind areas, tuning wireless propagation environment, and improving spectrum and energy efficiency, there are still some limitations that should be further investigated to improve the performance of wireless communication systems.

As for a reflection/refraction-only RIS, the BS and users can only be distributed on the same or opposite side of RIS. The half-space coverage~\cite{10153967} seriously affects the flexibility and effectiveness of the RIS, limiting its application scenarios. Due to the passive characteristic of RIS, the path loss of RIS-assisted cascaded link is inversely proportional to the product-distance instead of sum-distance~\cite{10243567}, and then leads to the double-fading attenuation~\cite{10198355} which influences the strength of reflection/refraction signal~\cite{10198324}, and limits the coverage area of RIS~\cite{hameed2024enhancing}. Though the ``passive'' meta-atoms directly reflect/refract the incident signal immediately without any signal processing and time delay~\cite{cao2024intelligent, 9373634}, the operational power of each mate-atom with 5-bit resolution is 1.5 mW, and the operational control of the RIS with 200 elements is up to meet 1.2 W, which is on par with its energy supply and demands attention~\cite{10198324}. The embedded batteries and the external grid are commonly adopted to supply energy for RIS but limit RIS's operation life and deployment flexibility in real-world applications~\cite{9424177}.

Consequently, due to the limitations mentioned above of RIS, including half-space coverage, double-fading attenuation, and energy dependency, there are various enhanced RIS designs, such as simultaneous transmitting and reflecting RIS (STAR-RIS), active RIS, and self-sustainable RIS, have been investigated to improve the performance of wireless communication systems further and satisfy the requirements of emerging wireless communication scenarios.

\subsubsection{STAR-RIS}\label{Sec.III.2.1}
A STAR-RIS is designed to address the limitation of half-space coverage, further improve the DoF for the RIS, and then manipulate signals into intelligently propagating in the full-space~\cite{9690478,9437234}, as illustrated in Fig.~\ref{fig:III.2}(a). The transparent substrate is adopted to integrate massive tunable elements and simultaneously divide the incident signal into reflection and transmission space, which can be imagined as ``ice cubes in a glass of water''~\cite{9690478}. The transmission and reflection coefficients of the $m$-th element $\sqrt{\beta_m^t}e^{j\theta_m^t}$ and $\sqrt{\beta_m^r}e^{j\theta_m^r}$ are imposed on the incident signal to intelligently tune the amplitude and phase shift of the transmission and reflection part, where $\beta_m^t$, $\beta_m^r$$\in\left[0,1\right]$ and $\theta_m^t$, $\theta_m^r$$\in\left[0,2\pi\right)$ represent the transmission or reflection coefficient and phase shift of the $m$-th element, respectively. According to the law of energy conservation, the sum of $\beta_m^t$ and $\beta_m^r$ should be equal to one~\cite{9690478}.

STAR-RIS-assisted wireless communication systems employ three operational protocols: mode switching (MS), time switching (TS), and energy splitting (ES), respectively~\cite{10133841}. MS divides STAR-RIS elements into transmission mode (T mode) $(\beta_m^t=1)$ and reflection mode (R mode) 
$(\beta_m^r=1)$, akin to combined traditional RISs. TS periodically switches all elements between T and R modes. ES operates all elements concurrently in transmission and reflection mode (T\&R mode), splitting signals based on coefficients and phase shifts. This expands coverage from half to full space, boosting QoS, bridging outdoor-indoor gaps, mitigating wall penetration loss, and enhancing signal strength by reducing propagation distance. STAR-RIS's optical transparency suits aesthetic building integration. However, the STAR-RIS introduces new security risks. The confidential signals can also leak to the eavesdropper distributed in both the reflection and transmission spaces.

\subsubsection{Active RIS}\label{Sec.III.2.2}
An active RIS is designed to overcome the challenges of double-fading attenuation by amplifying the transmission/reflection signals with low-cost hardware~\cite{10188853,9896755}, as depicted in Fig.~\ref{fig:III.2}(b). As for each active RIS element, an integrated amplifier first magnifies the incident signals, and then the phase shift circuit imposes the phase shift in the transmission or/and reflection signals, similar to a conventional passive RIS element~\cite{10153967}. Due to signal amplification, the multiplicative channel fading is converted to the additive form~\cite{9568854}, then the received signal strength is enhanced. 
However, the active RIS introduced non-negligible power consumption compared with traditional passive RIS.
In order to balance the cost and efficiency, active RIS elements can be integrated with passive elements, for example, the semi-passive RIS, and the study in~\cite{9758764} designs a single and variable gain amplifier for reflection amplification to reduce number of active RIS elements.
Meanwhile, active RISs can amplify the interference signals to jam legitimate users.

\subsubsection{Self-Sustainable RIS}\label{Sec.III.2.3}
A self-sustaining RIS is designed to support RIS operations independently, eliminating reliance on energy from embedded batteries or the external power grid~\cite{10225701}, as depicted in Fig.~\ref{fig:III.2}(c). Specifically, each element of self-sustainable RIS possesses both the energy harvesting mode (H mode) and the signal relay mode (S mode) and can be freely switched between the two operation modes by flexibly adjusting the circuit connection. As for the H mode, the element rectifies the incident RF signals into direct-current (DC) signals and ultimately converts them into energy by the energy harvesting (EH) circuit~\cite{10225701}, which includes an impedance matching network, RF-DC conversion circuit, capacitor, power management module, and energy storage module. In terms of the S mode, the incident signals are amplified, transmitted, and/or reflected by the element with the transmission or reflection coefficient $\beta_m^t$ or $\beta_m^r$ through the harvested energy during the H mode.
Furthermore, the study in~\cite{10693440} presents fabricated prototypes of self-sustaining RIS with experimental measurements, providing proof-of-concept validation for these enhanced RIS designs and establishing a foundation for their practical implementation. \par

Those above single-mode RISs and enhanced RIS designs can be regarded as exceptional cases of self-sustainable RIS. For example, the reflection-only or refraction-only RIS can be treated as all elements of RISs in the S mode with $\beta_m^r$ or $\beta_m^t$ equal to one, and the STAR-RIS can be recognized as all elements of RISs in the S mode with the sum of $\beta_m^r$ and $\beta_m^t$ equal to one. However, the eavesdropper can be distributed in both the transmission and reflection spaces, and the energy harvested in the H mode can also be utilized to amplify interference signals to suppress secrecy capacity.

\subsection{Summary and Lessons Learned}\label{Sec.III.3}

Various types of RISs have been developed to meet the diverse needs of modern networks. Single RISs use low-cost passive/refraction elements to redirect signals but limit coverage by splitting the area into zones~\cite{10153967}. In cascaded RIS-assisted links, multiplicative path loss weakens signals~\cite{10243567,10198355}, and although passive elements simplify signals, they still require power for phase shifting. As passive elements increase, there is an increasing need for more energy-efficient RIS designs~\cite{10198324}.

Enhanced RIS designs have been developed to address these limitations. STAR-RISs enable full-space coverage by independently tuning reflected and refracted signals~\cite{9690478,9437234}. Active RISs use integrated amplifiers to convert multiplicative path loss to additive, reducing double-fading effects~\cite{10188853}. Self-sustainable RISs can harvest energy from incident signals to power themselves and manage signals without external power sources~\cite{10225701}. 
These enhanced RIS designs introduce new security concerns. Confidential signals could leak to the eavesdropper distributed across both the reflection and transmission zones, while active RISs may amplify interference signals, thus compromising secrecy capacity.
Table~\ref{Table:III.1} compares single-mode RISs and enhanced RIS designs with their structures, benefits, and limitations outlined.

\begin{table*}[!ht]
    \centering
    \caption{Description of IRIS-Assisted Attacks Including Eavesdropping Attack, MITM Attack and Replay Attack, Reflection Attack and Jamming Attack, Side-Channel Attack}
    \vspace{-0.1in}
    \begin{adjustbox}{width=0.9\textwidth}
    \begin{tabular}{|m{2cm}<{\centering}|m{2cm}<{\centering}|m{4cm}<{\centering}|m{4cm}<{\centering}|m{3cm}<{\centering}|m{3cm}<{\centering}|m{1cm}<{\centering}|}
    \hline
\textbf{Atk. Type} & \textbf{Atk. mode} & \textbf{Definition} & \textbf{The role of IRIS} & \textbf{Potential risk} & \textbf{Countermeasures} & \textbf{Ref.} \\ \hline

    Eavesdropping attack & Enhance eavesdropping signal & Expand wiretap signal coverage area and enhance its strength. & Eavesdropper can adjust wireless signal paths to favorable positions. & Except for eavesdropping confidential information, attackers can cause more serious attacks, such as MITM and replay attacks. & AN technology, advanced signal encryption masking. & \cite{10118920} \\ \hline

    \multirow{2}{*}[-3em]{\begin{minipage}{2cm}
    \centering
    MITM attack and replay attack
    \end{minipage}} & MITM attack & Attackers insert between two legitimate parties to intercept, modify, or manipulate communication content without their awareness. & Attackers manipulate IRIS to amplify or suppress signals for themselves or legitimate users. After intercepting signals, they can attach malicious messages to mislead victims further. & \multirow{2}{*}[2em]{\begin{minipage}{3cm}
    \centering
    The interception of information from legitimate terminals via IRISs complicates tracing and neutralizing the threat, and passive IRISs remain undetectable to legitimate terminals without active connections.
    \end{minipage}} & \multirow{2}{*}[-1em]{\begin{minipage}{3cm}
    \centering
    Signal encryption, integrity verification, and strengthening the security protocols of wireless communication systems.
    \end{minipage}}
    & \cite{10693994} \\ \cline{2-4,7}
    ~ & Replay attack & Attackers capture legitimate traffic and reuse it at a specific time & Attackers exploit IRIS to capture legitimate communication signals within the network and then replay the captured data to victims at the opportune time. & ~ & ~ & \cite{9827890} \\ \hline
    
    \multirow{2}{*}[-3em]{\begin{minipage}{2cm}
    \centering
    Reflection attack and jamming attack
    \end{minipage}} & Reflection attack & Reflect and amplify substantial request traffic toward a target device or network node, overwhelming it and depleting its resources. & Attackers exploit IRIS to reflect signals transmitted from legitimate users onto the attack target, causing a sharp increase in traffic reaching the target. & The reflected signals originate from legitimate terminals, inadvertently involving them in the attack and increasing their complexity and stealth. & \multirow{2}{*}[0em]{\begin{minipage}{3cm}
    \centering
    Cross-layer collaboration between the network and physical layers involves integrating anomaly detection at the network layer with EM signal detection at the physical layer.
    \end{minipage}}
    & \cite{10316535} \\ \cline{2-5,7}
    ~ & Jamming attack & Attackers deliberately disrupt or obstruct signal transmission and reception by sending interference signals to the victim. & Attackers exploit IRIS to forge virtual illegitimate links, transmitting interference to legitimate users or jamming signals to undermine the RIS's reflective capabilities. & These jamming signals can reduce the main channel capacity for legitimate users and potentially disrupt communications. & ~ & \cite{9789438,9693184,wang2022intelligent} \\ \hline

    \multirow{2}{*}[-3em]{\begin{minipage}{2cm}
    \centering
    Side-channel attack
    \end{minipage}} & Suppress main channel & Attacker can force legitimate communication parties to make adjustments at the physical layer. & 
    The attacker suppresses the capacity of the main channel by manipulating the RIS. It can't decode information directly, but can infer intelligence from signal characteristics. & \multirow{2}{*}[-1em]{\begin{minipage}{3cm}
    \centering
    According to the passive nature of the RIS, Mallory can implement the attack without increasing the radio footprint.
    \end{minipage}} & \multirow{2}{*}[0em]{\begin{minipage}{3cm}
    \centering
    Advanced signal encryption masking, frequency hopping, a combination of ISAC technology, and environmental monitoring and response.
    \end{minipage}}
    & \cite{9605003,10516473,9112252} \\ \cline{2-4,7}
    ~ & Destroy channel reciprocity & Attacker can force the target communication system to react in specific ways or make configuration changes to cope with channel state degradation. & Attacker can adopt the RIS to disrupt channel reciprocity, degrade channel state, and access sensitive information. & ~ & ~ & \cite{10081025,10424421} \\ \hline
    \end{tabular}
    \end{adjustbox}
    \label{Table:VII.1}
    \vspace{-0.2in}
\end{table*}

% ============ Sec. IV Eavesdropping Attack
\section{Eavesdropping Attack} \label{Sec.IV}
\subsection{Passive-Active Hybrid Attack Paradigm}\label{Sec.IV.1}
The RIS's ability to dynamically reconfigure signal propagation paths can be exploited by the attacker called Mallory to cause a new attack paradigm defined as the ``passive-active'' hybrid attack. Mallory can leverage the RIS's passive signal manipulation to actively disrupt communication, and blurs the lines between benign and malicious behaviors~\cite{10081025,10424421}. For instance, Mallory controlling a RIS can passively reflect confidential signals to eavesdrop~\cite{10118920,10143983}, as shown in Fig.~\ref{fig:VI.1}, or actively redirect them to launch MITM or replay attacks. Meanwhile, Mallory can also exploit third-party systems to reflect and amplify a large amount of request traffic towards a specific victim device or network node to conduct the ``reflection attack'' and jamming attack~\cite{9789438}. Furthermore, even if the signal is encrypted, it is difficult for Mallory to decipher the encrypted information, and can still infer confidential information, such as EM emissions, power consumption, and time delays by monitoring and analyzing the unauthorized redirection signals, which can be classified as the side-channel attack.

In addition, the enhanced RIS designs introduced in Section~\ref{Sec.III.2} have also brought new security risks while satisfying the requirements of emerging wireless communication scenarios. Specifically, though the STAR-RIS can provide full-space communication coverage~\cite{9690478,9437234}, Mallory can also manipulate it to achieve passive-active hybrid attacks in both the reflection and transmission spaces. As for the reflection-only or the refraction-only RISs, Mallory must be located on the same side of the RIS as the transmitting terminals, due to the limitation of half-space coverage. In contrast, the STAR-RIS can expand the attack range to $360^{\circ}$ all-round attack. Mallory can exploit the active RIS to further enhance the eavesdropped signals by amplifying the unintended signals and reducing their double-fading effects~\cite{9896755,9758764}, and the active RIS can also reveal more side-channel information according to the strengthened signals. The self-sustainable RIS can eliminate power-dependent detection footprints, enabling persistent attacks without energy supply requirements~\cite{10225701,10693440}. According to the characteristic of energy-autonomous, the self-sustainable RIS is difficult to be detected malicious behavior through energy consumption monitoring, which significantly improves the concealment of attacks. \par

\begin{table}[!ht]
    \centering
    \caption{\centering{Characteristics and Corresponding Security Issues of Enhanced RIS Designs}}
    \vspace{-0.1in}
    \begin{adjustbox}{width=0.45\textwidth}
    \begin{tabular}{|m{1.75cm}<{\centering}|m{3.5cm}<{\centering}|m{3.5cm}<{\centering}|m{1.75cm}<{\centering}|}
    \hline
        \textbf{Enhanced RIS designs} & \textbf{Characteristics} & \textbf{Security issue} & \textbf{Attack Enhancement Properties}\\ \hline 
        STAR RIS & Provide full-space communication coverage & Mallory can expand the attack range to $360^{\circ}$ all-round attack & Complexity \& diversity \\ \hline
        Active RIS & Amplify incident signals & Mallory can further enhance the eavesdropped signals and reveal more side-channel information & Complexity \& effectiveness \\ \hline
        Self-sustainable RIS & Support RIS operation independently & Mallory can bypass power-dependent detection, enabling persistent attacks without energy constraints & Complexity \& stealthiness \& persistence \\ \hline
    \end{tabular}
    \end{adjustbox}
    \label{Table:VI.A.1}
    \vspace{-0.2in}
\end{table}

Tables~\ref{Table:VII.1} and~\ref{Table:VII.2} introduce RIS-assisted passive-active hybrid attacks, including eavesdropping attacks discussed in this section, as well as MITM and reply attacks, reflection and jamming attacks, and side-channel attacks that will be introduced in subsequent sections, to raise awareness of the significant security risks posed by the dual-use of RIS technology. \par

\subsection{Comparison with Existing Active Malicious Relays}\label{Sec.IV.2}
As for active malicious relays, like~\cite{8334236}, a malicious amplify-and-forward (AF) relay exhibits fundamentally different characteristics compared to passive RIS attacks: It can actively connect with the source to send incorrect pilot signals during the reverse pilot transmission (RPT) period, and then during the data transmission (DT) phase it can actively inject malicious data to spoof the destination. This active nature makes relays inherently more detectable but also more capable of direct signal manipulation. The active relay includes many RF chains to achieve signal processing, and can actively decode, amplify, and retransmit malicious signals. The active communication with terminals and signal processing makes the traditional malicious relay relatively easily traceable via transmission signatures, RF fingerprinting, or energy monitoring. In contrast, the passive nature of a RIS makes it less capable of active signal alteration but significantly more stealthy. \par

Mallory and his manipulated IRIS are not actively engaged in information transmission within the wireless communication networks, and remain undetectable by legitimate terminals without active connections. As for the MITM and replay attacks, the intercepted signals originate from legitimate terminals and are reflected by the malicious RIS, making tracing and neutralizing the threat more complex. With regard to reflection and jamming attacks, legitimate signals are reflected onto the target victim. Not only can Mallory hide its real internet protocol (IP) address, but the legitimate terminals can also unknowingly participate in the attack, dramatically improving the attack's complexity and stealth. A comparison of active malicious relay-assisted attacks and IRIS-assisted passive-active hybrid attacks is summarized in Table~\ref{Table:VI.B.1}. \par

\begin{table}[!ht]
    \centering
    \caption{\centering{Comparison of Active Malicious Relay-Assisted Attacks and IRIS-Assisted Passive-Active Hybrid Attacks}}
    \vspace{-0.1in}
    \begin{adjustbox}{width=0.45\textwidth}
    \begin{tabular}{|m{1.75cm}<{\centering}|m{3.5cm}<{\centering}|m{3.5cm}<{\centering}|}
    \hline
        ~ & \textbf{Active malicious relay-assisted attacks} & \textbf{IRIS-assisted passive-active hybrid attacks} \\ \hline 
        Operation mechanism & Active signal processing & Passive signal redirection \\ \hline
        Communication mechanism & Actively communicate with terminals & Not actively engage in communication \\ \hline
        Signal origin & Actively retransmit self-generated malicious content & Reflect and redirect signals from legitimate terminals \\ \hline
        Real IP address & Relatively easily be traced & Not exposed \\ \hline
    \end{tabular}
    \end{adjustbox}
    \label{Table:VI.B.1}
    \vspace{-0.2in}
\end{table}

\subsection{Eavesdropping Attack Paradigm}\label{Sec.IV.3}
\begin{figure}
  \begin{center}
  \includegraphics[width=3.0in]{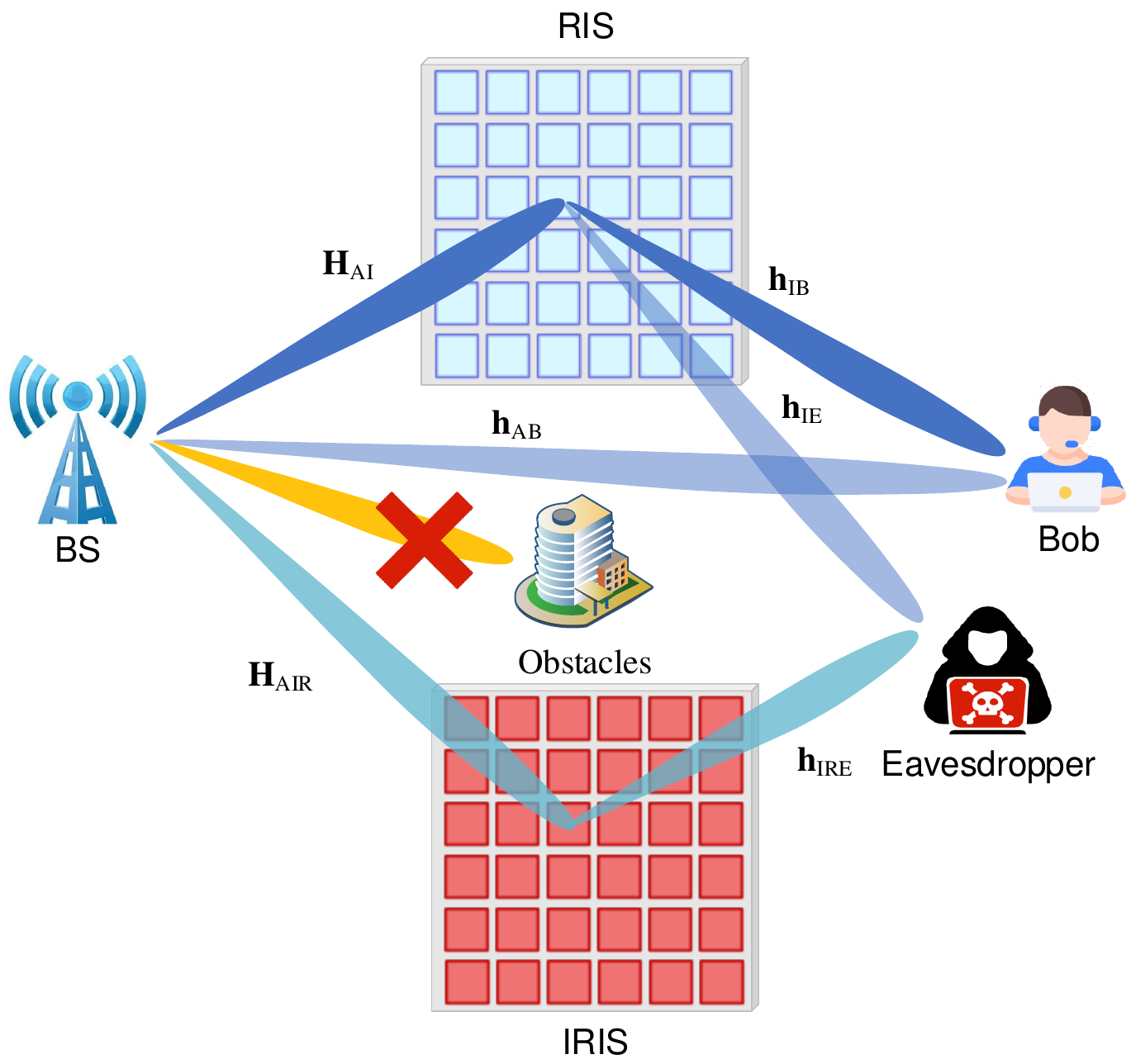}\\
  \vspace{-0.1in}
  \caption{IRIS-assisted eavesdropping attack paradigm: eavesdropper can achieve an eavesdropping attack by manipulating the IRIS to obtain or enhance the signals that cannot be detected or are previously weak.}
	\label{fig:VI.1}
  \end{center}
  \vspace{-0.25in}
\end{figure}

A passive eavesdropper can leverage RIS functionalities to expand the wiretap signals coverage area and cause the eavesdropping attack, as shown in Fig.~\ref{fig:I.0}(a). Specifically, the eavesdropper can detect and amplify weak or previously undetectable signals by manipulating the reflection matrix, e.g., through unauthorized access to legitimate RISs or through own RISs; see Fig.~\ref{fig:VI.1}. 

In~\cite{10118920}, the eavesdropper has illegally accessed a RIS microcontroller to enhance its eavesdropping capability in the typical RIS-assisted millimeter-wave (mmWave) multiple input multiple output (MIMO) wiretap system.
The legitimate user named Bob adopts a legitimate RIS to boost its confidential information transmission rate in the presence of the IRIS, which is pretty sequestered and intensifies the difficulty in obtaining channel state information (CSI) for BS.
In this event, both the legitimate system and the eavesdropper fully maximize their potential to boost the received signal strength (RSS) themselves, which can be formulated as a strategic game.
The water-filling strategy and Lagrangian multiplier are introduced to optimize the discrete reflection matrix of IRIS according to the singular value decomposition (SVD) of the composite channel and the intrinsic sparse characteristic of mmWave propagation~\cite{9217160}, assuming that the CSI of the wiretap channel can be acquired by the eavesdropper.
Simulations indicate that signal leakage rises with the IRIS element increasing, and the SR cannot be dramatically boosted just by optimizing the legitimate RIS and maximizing the confidential information transmission rate. \par

After intercepting the confidential information, Mallory can cause more serious attacks, such as the man-in-the-middle (MITM) attack and the replay attack. Specifically, Mallory can concatenate malicious messages to confuse the victim, or repeatedly send the intercepted messages in the specific slots to mislead the victim into executing unauthorized actions. Moreover, the eavesdropper is passive and the RIS does not actively engage in transmitting information within the wireless communication network. They can nearly conceal themselves perfectly, posing a significant security risk. \par

The AN technology and advanced signal encryption masking serve as effective approaches to ensuring the confidentiality of legitimate communication links when the eavesdropper's and IRIS's locations are unknown. The generation of AN signals is independent of the instantaneous CSI of the eavesdropping channel or the cascaded channel assisted by the IRIS, thereby effectively countering the high concealment capability of eavesdroppers and IRIS deployments. In~\cite{10143983}, the AN covariance matrix generated in the BS is jointly optimized with the legitimate active and passive precoding matrices at the BS and RIS, respectively, the number of data streams, and the linear combiner at the receiver to maximize the secrecy rate in the presence of an eavesdropper, Eve, and its controlled IRIS. 
The study in~\cite{9501003} demonstrates that even without the BS's awareness of the IRIS presence, a joint design of the legitimate precoding/combining matrices, AN covariance matrix, and the phase shifts of an $L$-element RIS can effectively secure the MIMO system against an eavesdropper employing an IRIS with more than $5L$ elements.
Meanwhile, the advanced signal encryption masking can prevent Eve from decoding intercepted signals to some extent in situations where signal leakage is difficult to detect and prevent, thereby improving the security performance of wireless communication systems. \par

% =========== Sec. V MITM attack and replay attack
\section{MITM attack and Replay attack} \label{Sec.V}
RISs can be exploited to facilitate MITM attacks. In such attacks, attackers insert between two legitimate parties to intercept, modify, or manipulate communication content without their awareness~\cite{7442758}. 
RISs can also be used to launch replay attacks. A replay attack captures legitimate traffic and reuses it later without modification~\cite{10701455}. 

\subsection{MITM Attack}\label{Sec.V.1}
In wireless communication systems, attackers can leverage the signal reflection capabilities of RISs to conduct MITM or relay attacks.
Precisely, the attackers can manipulate the reflection matrix of the RISs to illegally enhance the signals received by themselves while suppressing those received by legitimate users at the intended devices~\cite{10402016}. 
Once the signals are intercepted, it is possible for attackers to concatenate malicious messages to confuse victims~\cite{tyagi2024securing}, as shown in Fig.~\ref{fig:I.0}(b).
The attackers can alter the phase and amplitude of the incident signal to tamper with or forge information. They can intentionally introduce signal transmission delays by controlling the reflection path of the RISs, causing time differences between communication parties. \par

In~\cite{10693994}, a benign RIS and a malicious RIS are controlled by Bob and Mallory, respectively; both aim to enhance the received signals at their owner and suppress the received signals at their opponent under a multiple input single output (MISO) communication system.
Mallory optimizes the reflection matrix of the malicious RIS to decrease the secrecy rate by enhancing eavesdropping signals and diminishing communication signals. Concurrently, Bob adjusts the benign RIS to maximize the worst-case secrecy rate by jointly optimizing transmitter beamforming and the benign RIS reflection matrix, countering Mallory's influence.
A max-min secrecy rate problem is formulated to maximize the worst-case secrecy rate, and this non-convex optimization problem is divided into two sub-problems using an alternating optimization (AO) approach. The semidefinite relaxation (SDR) and Charnes-Coopers (CCP) techniques are adopted to transform each non-convex max-min sub-problem into the min-max convex-concave sub-problem, and then solved by the gradient-descent-ascent (GDA) algorithm.

After Mallory eavesdrops and intercepts confidential information with the assistance of malicious RISs, it can manipulate communication content and even concatenate malicious messages to confuse victims~\cite{10700793}. In this scenario, Mallory fully monitors the communication between the BS and the legitimate user through its controlled malicious RIS. This allows Mallory to intercept sensitive information, such as address, details of the grid account, and local network parameters~\cite{10577212}. Similarly to altering hyperlinks on public Wi-Fi networks~\cite{10406178}, this manipulation can confuse and mislead victims and even lead to substantial economic loss for users.

\subsection{Replay Attack}\label{Sec.V.2}
In the MITM attack, Mallory intercepts communication between two parties, eavesdropping on sensitive information and potentially altering the content of the messages. 
In contrast, a replay attack involves the attacker capturing legitimate traffic and repeatedly sending intercepted messages to deceive victims, even without any modification, as shown in Fig.~\ref{fig:I.0}(c). Replay attacks are used to mislead the receiver or execute unauthorized actions by replaying valid data packets. 

In~\cite{9827890}, the interaction between multiple communication parties, the legitimate user and RIS, Mallory, and the illegitimate RIS is investigated in a MISO wiretapped channel. Mallory optimizes the reflection matrix of the illegitimate RIS to try their best to maximize the wiretap rate. In contrast, the legitimate user jointly optimizes the transmitter beamforming and the reflection matrix of the legitimate RIS to maximize the worst-case secrecy rate caused by the malicious RIS.
Under the assumption of all information available, a max-min secrecy rate problem is formulated, and three algorithms are utilized to tackle the max-min problem called GDA, AO algorithm, and the mixed Nash equilibrium (NE) in zero-sum games in strategic form, respectively. Simulations show that AO fails to converge with continuous phase shifting, while GDA can; discrete phase shifts improve convergence for both. 

After Mallory captures legitimate communication signals within the network with the help of the malicious RIS, it can replay the captured data to the target system at the opportune time. It deceives the receiver into believing it is a new, valid request or response. Since replayed data are legitimate, many defense systems would fail to detect the existence of an attack. This may allow the attackers to carry out more destructive attacks undetected. Meanwhile, the attackers can replay sensor data to fool the system into thinking that everything is normal while malicious operations are being carried out~\cite{10516611}.

\subsection{Summary and Lessons Learned}\label{Sec.V.3}
Attackers can execute MITM or replay attacks using RISs while remaining undetected by legitimate terminals without active connections, posing significant security risks~\cite{10402016}. The intercepted information's origin from legitimate terminals and reflection by RISs complicates tracing and neutralizing the threat~\cite{9827890}.
Measures, such as signal encryption, integrity verification, and strengthening wireless communication security protocols, can be employed to counter RIS-assisted MITM and replay attacks. 

Advanced encryption ensures that data transmitted over the network layer remains protected from interception and tampering. While Mallory might exploit the RIS to intercept confidential signals~\cite{10693994}, he can by no means decode or alter the information due to strong encryption. Integrity verification mechanisms are crucial in helping the victim detect and prevent data tampering launched by Mallory during an MITM attack. Enhanced security protocols, such as Wi-Fi Protected Access 3 (WPA3)~\cite{10274082} and Transport Layer Security (TLS)~\cite{10621017}, can be integrated into clients, servers, and RIS microcontrollers. This integration strengthens access authentication and prevents unauthorized access by attackers. Additionally, regular updates and patches to address protocol vulnerabilities are essential for maintaining resilience against evolving attack technologies.

% =============== Sec. VI Reflection attack and jamming
\section{Reflection Attack and Jamming Attack} \label{Sec.VI}
In this attack, attackers adopt third-party systems to conduct the ``reflection attack''~\cite{10092781} and jamming attack~\cite{9733393, 10699421}, as shown in Figs.~\ref{fig:I.0}(d) and \ref{fig:I.0}(e), respectively, by reflecting and amplifying a large amount of request traffic towards a specific target device or network node, overwhelming and jamming the target device and exhausting computing and communication resources.

\subsection{Reflection Attack}\label{Sec.VI.1}
Attackers can exploit RISs to launch ``reflection attacks'' in wireless communication systems, increasing attack effectiveness and concealment.
Attackers can change the reflection matrix of a RIS to reflect signals transmitted from legitimate users onto the attack target, causing a sharp increase in traffic reaching the target. 
Specifically, there are multiple BSs and users, and the RIS can re-route the traffic of the users to the same BS for their network access requirements. Thus, the target BS may be overwhelmed by processing excessive traffic, and the RIS has changed the usual random access channel (RACH) access process and judgment criteria.
The attackers can transmit interference signals to victims with the help of their controlled RISs, increasing network traffic at the target and interrupting network services for the victims. 

In~\cite{10316535}, an $L$-sector mode RIS with block diagonal (BD) reflection matrix is designed. The $L$-sector RIS includes several cells, and in each cell, there are $L$ antennas deployed at each vertex of an $L$-side polygon that is fully connected. Based on the architecture, the entire space can be divided into $L$ sectors, where $L\geq2$. 
The incident signal can be partially reflected toward its original sector and scattered into the other $L-1$ sectors. Once Mallory controls the $L$-sector mode RIS, the incident signals can be scattered in all directions of the entire space, and even include the coverage hole of traditional RISs, such as the reflection-only RIS shown in Section~\ref{Sec.III.1.1} and refraction-only RIS shown in Section~\ref{Sec.III.1.2}. The scattered signals can transmit serious interference signals to victims distributed throughout the space. Furthermore, the original incident signals originate from the legitimate user, and the passive $L$-sector mode RIS lacks interaction with communication parties. This absence of interaction makes it challenging to identify and trace the source of interference and Mallory, thereby exacerbating the concealment and potential destructiveness of the attack. 

Since the RISs reflect incident signals, instead of actively communicating with terminals and victims. Mallory's IP address is not revealed at any point, making it nearly impossible to detect its presence. The reflected signals may originate from multiple legitimate terminals and make legitimate terminals participate in the attack, making it difficult to trace the sources of the attacks and increasing the complexity and concealment of the attacks.

\subsection{Jamming Attack}\label{Sec.VI.2}
\begin{figure}
  \begin{center}
  \includegraphics[width=3.0in]{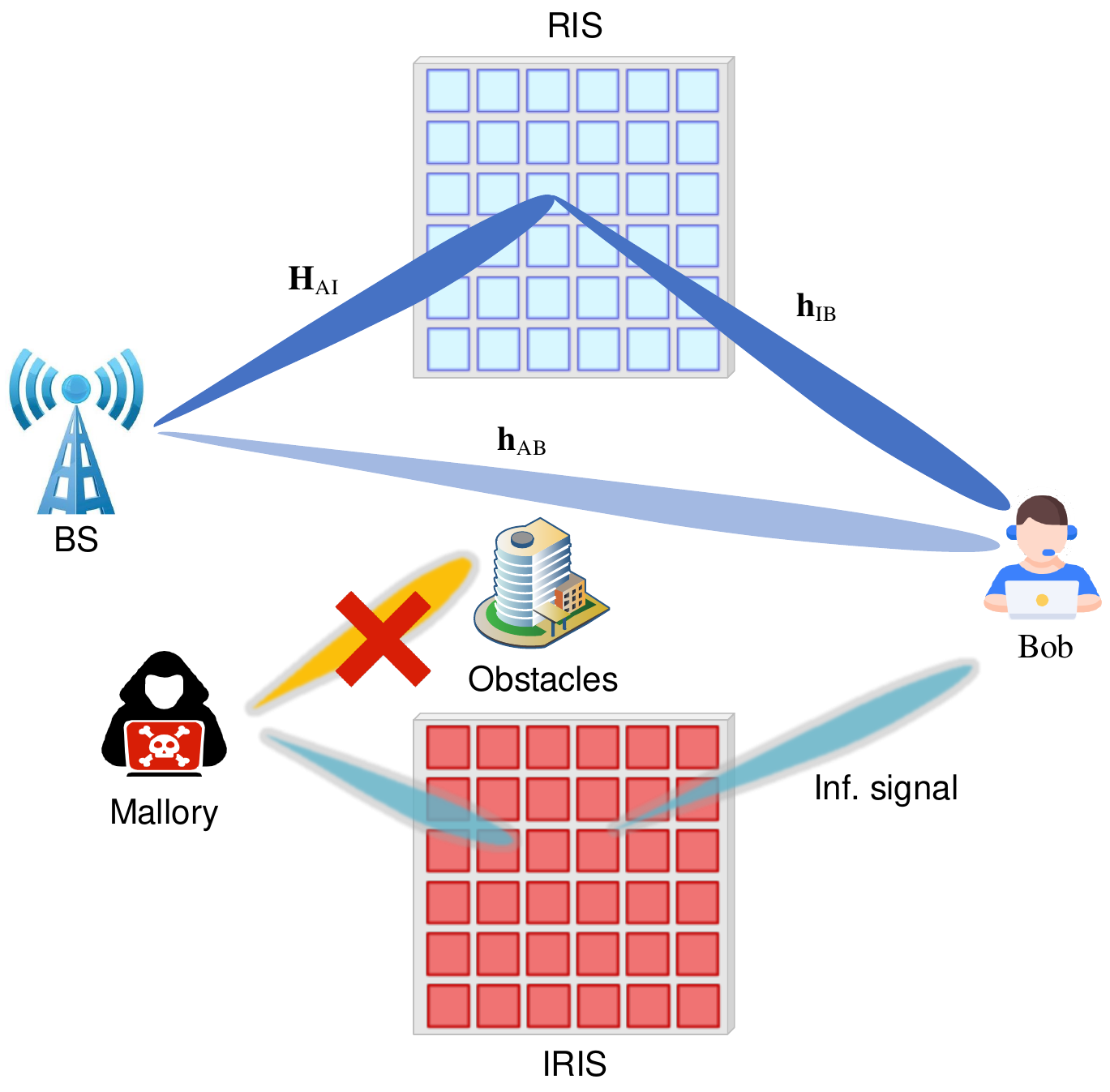}\\
  \vspace{-0.1in}
  \caption{IRIS-assisted jamming attack: the attack can illegally manipulate the IRIS to establish the virtual illegitimate link and transmit interference signals to disturb Bob.}
	\label{fig:VI.2}
  \end{center}
  \vspace{-0.25in}
\end{figure}

Active attackers can illegally access a RIS to establish a virtual illegitimate link and transmit interference signals to disturb Bob, as shown in Fig.~\ref{fig:VI.2}. The RIS can also lose its reflective capabilities when interfered with by active hackers~\cite{wang2022intelligent}.

\subsubsection{Jamming Attacks on Users}\label{Sec.VI.2.1}
In~\cite{9789438}, Mallory can manipulate a RIS to minimize the secrecy performance in a MISO wireless communication system.
Both Bob and Mallory can receive signals transmitted by LoS and cascaded links from BS, and a RIS is deployed beside Mallory to jam Bob by extending the interference signal coverage.
The reflection matrix of the RIS is optimized to reflect more interference signals and minimize the data rate for Bob.
The proposed scheme is suitable for RIS-assisted systems with known CSI of illegitimate parts and does not support systems without CSI.
Simulations illustrate that the system SR decreases with an increase in the number of IRIS elements and the improvement of the interference signal. Meanwhile, the SR can only be boosted slightly by deployed distributed legitimate RIS because the channel capacity of Mallory is accordingly improved with the increasing data rate of Bob.

In~\cite{9693184}, a jamming-assisted proactive attack is introduced to maximize the sum eavesdropping rate in a frequency division multiple access (C) based wireless surveillance scenario under Rayleigh-Rician fading. A suspicious BS and several suspicious users, all with single antennas, are monitored by a RIS-assisted half-duplex (HD) monitor. The monitor jams certain users to force the BS to reallocate more transmit power to other users, optimizing the jamming sets, power allocation, and RIS reflection pattern. This strategy aims to increase the sum ergodic eavesdropping rate from the desired users.
The AO algorithm iteratively optimizes the objectives. With a given jamming set and power allocation, the successive convex approximation (SCA) method searches for the optimal reflection pattern by iteratively updating slack variables based on statistical CSI. The jamming power allocation sub-problem is convex and solved via convex optimization. Optimal jamming sets are selected using an exhaustive search and a designed heuristic scheme.

\subsubsection{Jamming Attacks on RIS}\label{Sec.VI.2.2}
In~\cite{wang2022intelligent}, a min-max two-layer optimal linear programming is developed to cut off the transmission link between the BS and user group assisted by RIS in the free space loss channel.
The BS has multiple antenna groups, serving multiple user groups via one-to-one correspondence with the assisted RIS. 
Meanwhile, a hacker transmits interference to jam the victim RISs selected by its attacking decision vector.
The hacker uses greedy and robust min-max linear programming to minimize the BS gain. This optimal programming involves a confrontation between the hacker and the BS, where the hacker attempts to find an attacking decision vector to reduce the BS gain. In response, the BS adjusts its resource allocation according to the hacker’s attack strategy to maximize communication efficiency.
The complex min-max optimal problem can be converted into a single-level mixed-integer linear programming that can be solved using linear programming methods by introducing equivalent substitution, dual transformation, and linear transformation.

\subsection{Summary and Lessons Learned}\label{Sec.VI.3}
Attackers can exploit the ability of the RISs to manipulate signal reflection paths~\cite{10316535}, leading to reflection attacks by directing legitimate signals towards victims. 
The victim receives a substantial number of reflected signals from legitimate terminals, which exhaust resources such as bandwidth and processing power. This results in decreased performance or an inability to function properly~\cite{10092781}.
Furthermore, Mallory does not directly communicate with the victim, and its real IP address is not exposed throughout the entire process, while legitimate terminals unknowingly participate, dramatically increasing the attack's complexity and stealth~\cite{10092781}. 

To effectively counter reflection attacks using a RIS, enhancing information sharing and collaboration between the network layer and the physical layer is crucial. By integrating EM signal detection with network traffic monitoring, a more comprehensive security situational awareness can be achieved. Specifically, at the network layer, ML algorithms can be employed to enhance traffic monitoring and anomaly detection, particularly in situations of sudden traffic surges. Strengthening intrusion detection, such as using network intrusion detection systems (NIDs)~\cite{s24175516}, can reduce the chances of attackers masquerading as legitimate users. At the physical layer, integrating communication and sensing technologies can enhance the monitoring of EM signals, providing real-time feedback and environmental analysis. By analyzing abnormal reflection paths, potential RIS-controlled environments can be identified. 

% ============== Sec. VII Side-channel attack
\section{Side-channel Attack} \label{Sec.VII}
A side-channel attack refers to a method in which Mallory does not directly target the encryption algorithm but instead exploits vulnerabilities in the physical implementation process to obtain sensitive data~\cite{10.1145/3645109}, as shown in Fig.~\ref{fig:I.0}(f).
Mallory infers confidential information by monitoring and analyzing unintended information leaked by a device during the communication process, such as EM emissions, power consumption, and timing delays. These analyses can reveal communication patterns, changes in the location of the signal source, or even the duration of communications~\cite{s22218564,7835151}.  

By controlling RISs, Mallory can carry out ``side-channel'' attacks and adjust the reflection paths of wireless signals to positions more favorable. Mallory can exploit the RISs to degrade the secrecy capacity of legitimate users by introducing interference signals or destroying the channel reciprocity. Although Mallory cannot directly decode the information, he can still capture signal characteristics and infer valuable intelligence by analyzing the signals' timing characteristics, strength variations, and spectral features. 

\subsection{Side-Channel Attack by Suppressing the Main Channel}\label{Sec.VII.1}
Mallory can exploit a RIS to degrade the secrecy capacity of legitimate users by suppressing the capacity of legitimate channels, and Mallory can force legitimate communication parties to make adjustments at the physical level. Examples of such adjustments include altering transmission power, modifying frequency, or adjusting signal transmission parameters or path selection. By observing and analyzing these adjustments, Mallory can gain additional insights into the communication system's operational status, infer the system’s internal state, or even access sensitive information. 

In~\cite{9112252}, an IRIS is adopted as a green attacker to minimize the signal-to-interference-plus-noise ratio (SINR) at Bob by jointly optimizing the amplitude and phase shifts of its elements in a MIMO wireless communication system under the Rayleigh fading channel.
The IRIS is deployed between the BS and Bob to destructively superpose the direct and cascaded links at Bob and degrade its QoS without any energy footprint.
The block coordinate descent (BCD) method decouples the mixed-integer non-linear program (MINLP) into two sub-problems, assuming known CSI. SDR converts the phase shift optimization sub-problem into a convex semi-definite program (SDP), solved using the Gaussian randomization method for a rank-one approximate solution. The reflection coefficient amplitude optimization sub-problem is a convex problem directly solvable by the CVX tool.
The proposed scheme is suitable for scenarios in which IRIS knows the CSI and does not support most cases with CSI unknown.
Simulations showcase that the SINR at Bob degrades with the IRIS element increasing and the distance between the IRIS and BS decreasing. 
The attack performance of the proposed IRIS-based green attacker can even outperform that of the active jamming attack scheme in some situations. 

In~\cite{9605003}, the RIS illegally accessed by Mallory is adopted to degrade the signal-to-noise ratio (SNR) at Bob, and it is located between the legitimate BS and Bob.
The IRIS is deployed into two different positions with different element numbers, and the BS and Bob are fixed to assess its attack efficiency.
The phase shifts are set the same, and the SNR at Bob is evaluated versus different transmit powers.

The authors of~\cite{10516473} further investigate the performance degradation at the legitimate user caused by IRIS under channel estimation errors (CEEs). 
The hacker controls an IRIS to degrade the SNR at the specific user without being aware of the single/multi-user secure wireless communication system under the Rician fading channel.
The CEE is introduced into the static path to relax the assumption of perfect CSI and formulate the optimization problem as a min-max problem to minimize the SNR at the specific user with the worst channel uncertainties under the constraint of minimum SNR for others.
Nemirovski's lemma is adopted to express the non-convex constraints equivalently.
The proposed scheme is suitable for situations where the hacker knows the upper limit of CEE.
Simulations demonstrate that the CEEs will not change the inverse relationship between the SNR at the victim and the number of RIS elements. 
Nonetheless, the impact of CEEs cannot be ignored, and their effects on single- and multi-user secure wireless communication networks are also tricky.

\subsection{Side-Channel Attack by Destroying Channel Reciprocity}\label{Sec.VII.2}
The RIS can be manipulated by Mallory to destroy the channel reciprocity without leaving any energy footprint. Concretely, the RIS can be designed to destroy the channel reciprocity~\cite{10081025,10424421,10145059,staat2022mirror} or disrupt the physical layer key generation (PKG) between the legitimate BS and users~\cite{9771319}, e.g., by producing different phase shifts during the bi-direction channel probing. This strategy can force the target communication system to react in specific ways or make configuration changes. For instance, they might change transmission power and transmission parameters, select different paths, or adjust frequency, thereby exposing more side-channel information~\cite{9069285}. This strategy allows Mallory to infer the system's internal state further or acquire sensitive information. \par

In~\cite{10081025}, serious inter-user interference (IUI) without CSI and extra power in a multi-user MISO (MU-MISO) system via the RIS-assisted fully passive attacker is proposed to raise awareness for potential security threats.
A one-bit, controllable RIS-based fully passive attacker is located between the BS and Bob to produce active channel aging (ACA) and destroy the orthogonality between MU's precoder matrix and co-user channels. Specifically, during the communication period, there are two communication phases, including the RPT period and the DT phase.
The reflection matrix is randomly formed following the uniform distribution during the period of RPT and DT, called $T_\mathrm{r}$ and $T_\mathrm{d}$, respectively, under the assumption of $T_\mathrm{r}\leq T_\mathrm{d}$ and the change cycle of reflection matrix not exceeding $T_\mathrm{r}$. \par

Since the reflection matrix of the RIS is different between the two phases, the MU's precoder matrix is not orthogonal to the subspace of co-user channels. Then the ACA is caused, which brings serious IUI and results in a dramatic decrease in communication performance at legitimate MUs.
Simulations demonstrate that the proposed RIS-based fully passive attacker can be independent of CSI and extra power and have much lower complexity and more efficiency compared to CSI-based passive attacker in~\cite{9112252}. Meanwhile, the attack produced by the proposed RIS-based fully passive attacker increases with the RIS element and cannot be mitigated by increasing transmit power. Furthermore, the proposed scheme is robust to the quantization level of the reflection matrix. \par

Compared with~\cite{10081025}, whose RIS reflection matrix changes only once during the DT phase, a persistent RIS-based fully passive attacker is described in~\cite{10424421} to continuously cause ACA and be unrealistic for Bob to acquire the CSI of ACA in MU-MISO systems under the Rayleigh-Rician fading channel.
Mallory adjusts the RIS's phase shift and amplitude multiple times during the DT phase without synchronization requirements.
Due to various delays in RF propagation and computation, the CSI can age rapidly~\cite{9806300}, and legitimate users cannot acquire the CSI of ACA, making it impossible to mitigate the IUI.

\subsection{Summary and Lessons Learned}\label{Sec.VII.3}
Mallory can exploit the RIS to suppress the capacity or destroy the channel reciprocity of legitimate channels~\cite{10516473,10424421}, reducing the communication capacity of legitimate users. This forces them to make changes at the physical level, revealing more side-channel information by analyzing the signal timing, strength variations, and spectral properties, which can be used to infer the system's operational status and acquire sensitive information~\cite{10.1145/3645109}. Although Mallory cannot have direct access to the data, it can still pose serious security threats by obtaining sensitive information, including communication patterns, communication duration, or even the location of the signal source~\cite{s22218564,7835151}. 

Due to the passive nature of the RIS, both passive attackers and RISs do not actively engage in information transmission within the wireless communication network, allowing them to remain nearly undetectable~\cite{10118920,9206122}, unless the BS employs active access methods~\cite{10424421}.
Communication systems can adopt advanced techniques, including signal encryption masking, frequency hopping, and environmental monitoring with response strategies to counter side-channel attacks and ensure information security. Encryption techniques obscure signals, complicating the interpretation of side-channel information~\cite{9686688}, while continuous variation of transmission signal frequency hinders interception of consistent side-channel data~\cite{10545402}. Furthermore, ISAC~\cite{9737357} enhances environmental change detection, enabling immediate adjustments in communication parameters, such as frequency and power, or switching communication paths when detecting interference or potential side-channel activities to mitigate attack risks.

\begin{table*}[!ht]
    \centering
    \caption{IRIS-Assisted Attacks Including Eavesdropping Attack, MITM Attack and Replay Attack, Reflection Attack and Jamming Attack, Side-Channel Attack: Include the Illegitimate and Legitimate Parts. As for the illegitimate part, the IRIS is controlled by attackers, and it summarizes the attack means, attack target, the role of IRIS, CSI, optimization objectives, and optimization algorithms. In terms of the legitimate part, there are legitimate users and/or legitimate RIS, and concludes the system scenario, role of RIS, and channel model.}
    \vspace{-0.1in}
    \begin{adjustbox}{width=0.9\textwidth}
    \begin{tabular}{|m{1.2cm}<{\centering}|m{1.2cm}<{\centering}|m{1.2cm}<{\centering}|m{1.2cm}<{\centering}|m{1.2cm}<{\centering}|m{1.2cm}<{\centering}|m{1.2cm}<{\centering}|m{1.2cm}<{\centering}|m{1.2cm}<{\centering}|m{1.2cm}<{\centering}|m{1.2cm}<{\centering}|m{1.2cm}<{\centering}|m{1.2cm}<{\centering}|m{1.2cm}<{\centering}|}
    \hline
        \multicolumn{9}{|c|}{\textbf{Illegitimate part}} & \multicolumn{4}{c|}{\textbf{Legitimate part}} & \multirow{2}{*}[-0.5em]{\textbf{Ref.}} \\[5pt] \cline{1-13}
        \textbf{Atk. type} & \textbf{Power} & \textbf{Atk. mode} & \textbf{Atk. Target} & \textbf{IRIS} & \textbf{CSI} & \textbf{Metrics} & \textbf{Opt. Obj.} & \textbf{Method} & \textbf{Scenario} & \textbf{RIS} & \textbf{LoS} & \textbf{Cascaded} & ~ \\ \hline
     
    Eavesdrop-ping attack & Not required & Enhance eavesdropping & Legitimate users & Signal leakage & Required & Max. wiretap rate & IRIS reflection matrix & Water filling strategy; Lagrangian & Eve-MIMO & Enhance coverage & Rayleigh & Rician & \cite{10118920}  \\ \hline
    
    \multirow{2}{*}[-1em]{\begin{minipage}{1.2cm}
    \centering
    MITM attack and replay attack
    \end{minipage}} & \multirow{2}{*}[-1em]{\begin{minipage}{1.2cm}
    \centering
    Not required
    \end{minipage}} & MITM attack & Legitimate users & Eavesdrop and intercept confidential signals & Required & Max. wiretap rate & IRIS reflection matrix & AO; GDA; SDR & Eve-MISO & Max. worst-case secrecy rate & Rayleigh & Rayleigh & \cite{10693994} \\ \cline{3-14}
    ~ & ~ & Replay attack & Legitimate users & Intercept communication context & Required & Max. wiretap rate & IRIS reflection matrix & GDA; AO; NE & Eve-MISO & Max. worst-case secrecy rate & Rayleigh & Rayleigh & \cite{9827890} \\ \hline

    \multirow{4}{*}[-6em]{\begin{minipage}{1.2cm}
    \centering
    Reflection attack and jamming attack
    \end{minipage}} & Not required & Reflection attack & Legitimate users & Reflect numerous signals from legitimate users & Required & Max. interference signal rate & IRIS reflection matrix & BCD; FP & MU-MISO & / & Rician & / & \cite{10316535} \\ \cline{2-14}
    ~ & \multirow{3}{*}[-6em]{Required} & \multirow{3}{*}[-6em]{\begin{minipage}{1.2cm}
    \centering
    Jamming attack
    \end{minipage}} & Legitimate users & Reflect more Inf. signal & Required & Min. SR & IRIS reflection matrix & Extend Inf. signal coverage & Eve, Mallory-MISO & Enhance coverage & Rayleigh & Rician & \cite{9789438} \\ \cline{4-14}
    ~ & ~ & ~ & Legitimate users & Strengthen jamming effect & Required statistical CSI & Max. sum ergodic Eve rate & Jamming sets; power allocation; IRIS reflection matrix & AO; SCA; exhaustive search, heuristic scheme & Monitor-SISO & / & Rayleigh & / & \cite{9693184} \\ \cline{4-14}
    ~ & ~ & ~ & RIS & / & Required distances & Min. BS gain & Hacker's attacking decision & Equivalent substitution; dual transformation; linear transformation & Multiple user groups-MISO & Enhance coverage & / & Free space loss & \cite{wang2022intelligent} \\ \hline

    \multirow{5}{*}[-7em]{\begin{minipage}{1.2cm}
    \centering
    Side-channel attack
    \end{minipage}} & \multirow{5}{*}[-7em]{\begin{minipage}{1.2cm}
    \centering
    Not required
    \end{minipage}} & \multirow{3}{*}[-2.5em]{\begin{minipage}{1.2cm}
    \centering
    Suppress the main channel
    \end{minipage}} & \multirow{3}{*}[-2.5em]{\begin{minipage}{1.2cm}
    \centering
    Legitimate users
    \end{minipage}} & \multirow{3}{*}[-2.5em]{\begin{minipage}{1.2cm}
    \centering
    Decrease SINR at Bob
    \end{minipage}} & Required & Min. SINR at Bob & IRIS reflection matrix & BCD; SDR; Gaussian randomization & MISO & / & Rayleigh & / & \cite{9112252} \\ \cline{6-14}
    ~ & ~ & ~ & ~ & ~ & Required statistical CSI & Min. SNR at Bob & IRIS reflection matrix and position & Set the same phase shifts & Eve-SISO & / & Rayleigh & / & \cite{9605003} \\ \cline{6-14}
    ~ & ~ & ~ & ~ & ~ & Required imperfect CSI & Min. SNR at specific Bob & IRIS reflection matrix & Nemiro-vski's lemma & Single / multi-user-MISO & / & Rician & / & \cite{10516473} \\ \cline{3-14}
    ~ & ~ & \multirow{3}{*}[-2em]{\begin{minipage}{1.2cm}
    \centering
    Destroy channel reciprocity
    \end{minipage}} & \multirow{2}{*}[-2em]{CSI} & \multirow{2}{*}[-2em]{\begin{minipage}{1.2cm}
    \centering
    Persistent ACA
    \end{minipage}} & Not required & Cause ACA & Random IRIS reflection matrix & Generate random RIS reflection matrix & MU-MISO & / & Rayleigh & / & \cite{10081025} \\ \cline{6-14}
    ~ & ~ & ~ & ~ & ~ & Not required & Cause unavailable ACA & Random IRIS reflection matrix & Change random IRIS reflection matrix multi-times & MU-MISO & / & Rayleigh & / & \cite{10424421} \\ \hline

    \end{tabular}
    \end{adjustbox}
    \label{Table:VII.2}
    \vspace{-0.2in}
\end{table*}

%=====================Sec. VIII Vulnerabulities of AI-enabled RIS
\section{Vulnerabilities of AI-Enabled RIS}\label{Sec.VIII}

\begin{table*}[!ht]
    \centering
    \caption{AML Attacks on AI-Powered RIS-Assisted Wireless Networks: Include Attack Means, System Scenario, Victim AI Model, Defense Mode Adopted, and Performance Metrics}
    \vspace{-0.1in}
    \begin{adjustbox}{width=0.9\textwidth}
    \begin{tabular}{|m{1.2cm}<{\centering}|m{1.5cm}<{\centering}|m{1.2cm}<{\centering}|m{1.2cm}<{\centering}|m{1.2cm}<{\centering}|m{1.2cm}<{\centering}|m{1.2cm}<{\centering}|m{1.2cm}<{\centering}|m{1.2cm}<{\centering}|m{1.2cm}<{\centering}|m{1.2cm}<{\centering}|m{1.2cm}<{\centering}|m{1.2cm}<{\centering}|m{1.2cm}<{\centering}|m{1.2cm}<{\centering}|m{1.2cm}<{\centering}|}
    \hline
    \multicolumn{5}{|c|}{\textbf{Attack}} & \multicolumn{1}{c|}{\textbf{System}} & \multicolumn{2}{c|}{\textbf{AI Data}} & \multicolumn{5}{c|}{\textbf{AI Model}} & \textbf{Defense} & \multirow{2}{*}[-0.5em]{\textbf{Metrics}} & \multirow{2}{*}[-0.5em]{\textbf{Ref.}} \\[5pt] \cline{1-14}
       \textbf{Atk. Target} & \textbf{Atk. Type} & \textbf{Atk. Mode} & \textbf{Atk. Obj.} & \textbf{Purpose} & \textbf{Scenario} & \textbf{Dataset} & \textbf{Training Ratio} & \textbf{Training Model} & \textbf{Model Role} & \textbf{Input} & \textbf{Output} & \textbf{Loss Function} & \textbf{Def. Mode} & ~ & ~ \\ \hline
        Transmitters & White-box attack & FGSM, BIM, PGD, MIM & Input samples & Error predict reflection matrix & RIS-assisted mmWave SISO & DeepMIMO & 85\% & MLP & Opt. reflection matrix & Environment descriptors & Achievable rate & Cross-entropy & Defensive distillation mitigation & MSE & \cite{9889707} \\ \hline
        \multirow{2}{*}[-3em]{\begin{minipage}{1.2cm}
        \centering
        RIS microcontrollers
        \end{minipage}} & White-box attack & FGSM & Input samples & Misclassify the decoded QPSs & RIS-assisted SISO system with band-limited feedback channel & Synthetic & 50\% & MLP & Reconstruct the compressed QPSs & Received  code & QPSs & MSE & / & BLER & \cite{son2024adversarial} \\ \cline{2-16}
     ~ & Black-box attack & FGSM trained in substitute network & Input samples & Misclassify the decoded QPSs & RIS-assisted SISO system with band-limited feedback channel & Synthetic & 50\% & CNN & Reconstruct the compressed QPSs & Received  code & QPSs & MSE & / & BLER & Proposed scheme 1  \\ \hline
        Receivers & Grey-box attack & FGM & Input samples & Misclassify the ``signal'' class into ``noise'' class & An Eve-SISO & Synthetic & 50\% & CNN & Distinguish ``signal'' or ``noise'' & Received signal & ``signal'' or ``noise'' class & Cross-entropy & / & Min. loss function & \cite{9771899} \\ \hline
    \end{tabular}
    \end{adjustbox}
    \label{Table:VIII.1}
    \vspace{-0.1in}
\end{table*}

Numerous studies have successfully integrated AI to control and configure RISs in wireless communication systems~\cite{9734045,9500409,10279487,9784887}. This integration enhances the efficiency of wireless communications by adapting to the dynamic environment, detecting and predicting potential security threats, and optimizing signal propagation and resource allocation~\cite{10463689,10463696}. 
For example, the study in~\cite{8815412} demonstrates a deep neural network (DNN)-based coordinate mapping method for real-time RIS beam focusing in dynamic indoor environments, optimizing signal propagation through position-aware phase configurations. Furthermore, the study in~\cite{9864655} provides a comprehensive overview of deep reinforcement learning (DRL) methods for optimizing wireless networks with RISs, and highlights its significant roles in achieving high sum-rate and energy efficiency in promising 6G era. \par

The integration also overcomes challenges that conventional communication systems often grapple with or perceive as insurmountable barriers, such as handling complex optimization problems involving multiple objectives and constraints, which are common in RIS-assisted wireless communication systems~\cite{10107766,10345491}. 
For example, the study presented in~\cite{9410457} addresses the NP-hard beamforming challenge in multi-hop RIS-assisted terahertz (THz) networks via DRL-based hybrid beamforming, while the work~\cite{10930892} advances the field by solving the non-differentiable discrete-phase optimization in the multi-RIS-assisted MISO network through a neuroevolution-optimized multi-branch attention convolutional neural network (CNN) architecture. \par

On the other hand, the RISs controlled and configured by AI are susceptible to the contamination of malicious data, commonly referred to as adversarial attacks~\cite{10263803}. These attacks utilize the gradient information of the input data to create small and elaborate perturbations~\cite{10263803,10092277}. The elaborate perturbations can fool AI-based models into predicting the incorrect reflection matrices of the RISs based on the environment descriptors, erroneously compressing or reconstructing the QPSs at the transmitters or RIS microcontrollers, and misclassifying the useful signal into the ``noise'' category at the receivers. Such adversarial attacks markedly escalate the vulnerability of the AI-powered RIS-assisted wireless communication model~\cite{10460991,10416752}.

\begin{figure*}
  \begin{center}
  \includegraphics[width=0.9\textwidth]{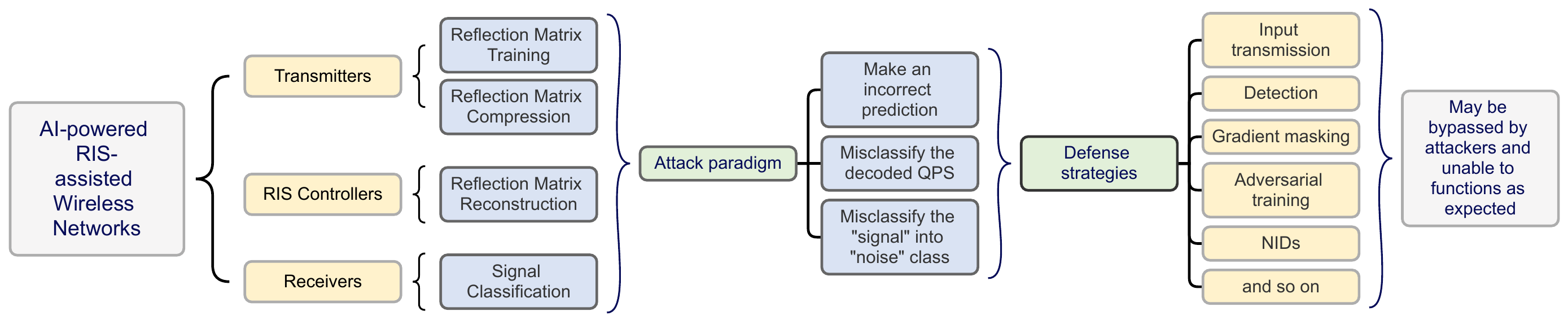}\\
  \vspace{-0.1in}
  \caption{AI-powered RIS-assisted wireless networks: AI models can be trained to predict the optimal RIS reflection matrices, compress and reconstruct QPSs at BSs and RIS controllers, detect and classify transmission signals at receivers. However, these models are susceptible to adversarial attacks, where attackers exploit wireless channel openness to inject perturbations that confuse the models. Though a plethora of adversarial defense strategies exist to neutralize adversarial attack methodologies, attackers can often bypass these defenses, rendering them ineffective.}
	\label{fig:VIII.1}
  \end{center}
  \vspace{-0.25in}
\end{figure*}

\subsection{Adversarial Attacks}\label{Sec.VIII.1}

Adversarial attacks on AI-powered RIS-assisted wireless networks can exploit wireless channel openness to inject perturbations that mainly confuse the transmitters, RIS microcontrollers and receivers to make incorrect predictions or misclassification~\cite{schwinn2023exploring,tsai2023adversarial}, and can take various forms, including white-box, black-box, and grey-box attacks, each with its level of knowledge about the system's internals. \par

The gradient-based attack paradigm exploits the AI model's differential response to generate adversarial examples, and four principal methods dominate this approach, called fast gradient sign method (FGSM), basic iterative method (BIM), projected gradient descent (PGD), and momentum iterative method (MIM)~\cite{9889707}. Specifically, FGSM stands as one of the most widely used methods, and it adds one-step gradient with a certain step size to the input samples as the adversarial perturbations to deceive the model~\cite{9984916}. 
As an iterative variant of FGSM, BIM applies the single-step gradient perturbation iteratively, and gradually refines adversarial samples through multiple small-step updates~\cite{11012053}.
PGD extends BIM by introducing stochastic noise during gradient computation to enhance attack robustness. Meanwhile, its projection step enforces strict constraints to balance attack potency and visual stealth~\cite{9984916}.
The MIM attack enhances BIM by integrating momentum into gradient updates, stabilizing perturbation directions to escape local optima, while optionally adding noise for robustness~\cite{10278178}. \par

The perturbation direction depends on attack objectives. Non-targeted attacks follow the positive gradient to induce arbitrary misclassification, while targeted attacks employ negative gradients to drive specific erroneous outputs~\cite{10839056}. These gradient manipulations can utilize minor perturbations to confuse AI models into predicting the incorrect RIS's radiation patterns, erroneously compressing or reconstructing the QPSs at the BSs or RIS microcontrollers, and misclassifying the useful signal into the ``noise'' category at the receivers. \par

\subsubsection{Adversarial Attacks on Transmitters}\label{Sec.VIII.1.0}
To overcome the real-time bottleneck in RIS deployment, which mainly includes the computation complexity of RIS optimization and transmission delays, AI models can be employed to predict RIS reflection matrices based on the environmental descriptors at the BSs~\cite{9889707}. Subsequently, BSs transmit the compressed QPSs to the RIS microcontrollers via the backhaul links~\cite{9592779}. However, attackers can exploit the susceptibility of AI models to adversarial perturbations, and can mislead BSs into incorrect reflection matrix prediction and erroneous QPS compression, as shown in Fig.~\ref{fig:I.0}(g).

In~\cite{9889707}, an AI-power RIS-assisted wireless communication system undergoing white-box adversarial ML (AML) attacks is proposed to investigate its vulnerability against AML attacks using the ray-tracing-based DeepMIMO dataset~\cite{alkhateeb2019deepmimo}.
There is a fixed BS, and several candidate users are equipped with a single antenna. The RIS is deployed to reflect the orthogonal frequency division multiplexing (OFDM) signals. Meanwhile, the DNN is trained to build a function mapping from the environment descriptors to the reflection matrix and then predict the system's achievable rates.
The aforementioned adversarial attack methods, including FGSM, BIM, PGD, and MIM are adopted to attack the DNN-based wireless system, and the defensive distillation mitigation method is adopted to mitigate the adversarial attacks. Meanwhile, the mean squared error (MSE) is utilized to assess the vulnerability and robustness of the system under defended and undefended scenarios.

Simulations illustrate that attacks of MIM, FGSM, and PGD perform similar attack effects under different attack powers, and the attack effect of BIM will strengthen with the increase of power, which indicates the considerable vulnerability of AI-powered RIS-assisted wireless systems against various AML attacks. Though the defensive distillation mitigation method can improve the system's robustness, its impact is different for all adversarial attack types. BIM and MIM attacks perform the most efficient attack effects for undefended and defended systems, respectively.

\subsubsection{Adversarial Attacks on RIS Microcontrollers}\label{Sec.VIII.1.1}
Attackers can exploit wireless channels' openness to inject perturbations that confuse RIS microcontrollers into erroneously reconstructing QPSs which are trained and compressed at BSs and then transmitted to RIS microcontrollers through the backhaul links, as shown in Fig.~\ref{fig:I.0}(h).

In~\cite{son2024adversarial}, the multi-layer perceptron (MLP) auto-encoder developed in~\cite{8651357} is adopted to break the bottleneck of RIS-assisted single input single output (SISO) band-limited channel in~\cite{9592779}.
Specifically, the QPSs are available at RIS through the feedback channel, which is band-limited. This channel cannot accommodate extensive feedback overhead, primarily because of the substantial quantity of RIS elements and the quantization level associated with each component. The MLP auto-encoder is introduced to compress and reconstruct the QPSs on the BS and RIS side.
Nevertheless, the MLP is vulnerable to adversarial attacks, prompting the adoption of the FGSM as a white-box adversarial attack to assess the impact of adversarial attacks compared to jamming attacks on the MLP-based auto-encoder within a RIS-assisted wireless network.

The adversarial perturbation generated by FGSM is applied to the decoder input to elevate the loss function, determined by the MSE of the QPSs, to cause misclassification of the decoded QPSs.
The proposed adversarial attack is suitable for scenarios where Mallory possesses comprehensive knowledge of the auto-encoder, including its weights, bias parameters, and the number of layers. However, the proposed attack scheme is not applicable when Mallory does not have access to the specifics of the victim models.
Simulations demonstrate that adversarial attacks' impact escalates as the noise and interference variance rise. Furthermore, under comparable conditions, this effect surpasses traditional additive white Gaussian noise (AWGN) jamming attacks.

Drawing inspiration from the aforementioned MLP-based auto-encoder and RIS-assisted SISO band-limited feedback channel, the black-box adversarial attacks, which are more common in real-world scenarios \cite{son2024adversarial} can be introduced into the decoder block of the CNN-based auto-encoder communication system.

In particular, the QPSs are under-completely mapped to a feature space code with a lower dimension at the BS side via the encoder block. Subsequently, the code is transmitted through the band-limited feedback channel and reconstructed to the estimated QPSs at the RIS side using the decoder block.
Although the CNN model is well-suited for auto-encoders, prioritizing low error rates and fast processing speeds, it is notably prone to adversarial attacks.
The black-box attack on the auto-encoder in~\cite{8651357} can be introduced in this system. According to the transferability of adversarial attacks, attacks designed for a specific model are likely also effective for other models.
Consequently, though Mallory has little knowledge of the CNN model in the decoder block, it can design a white-box attack based on its substitute auto-encoder, such as an MLP-based auto-encoder, to produce corresponding adversarial perturbations, which are subsequently added to the input of the unknown CNN-based auto-encoder to fool the network into producing the misclassified QPSs.

\subsubsection{Adversarial Attacks on Receivers}\label{Sec.VIII.1.2}
AI models can be employed at the receivers to detect and classify the received signals, but these models are susceptible to adversarial attacks due to the openness of wireless channels, which can mislead the receivers into misclassifying proper signals as noise, as shown in Fig.~\ref{fig:I.0}(i).

In~\cite{9771899}, AML is added to the transmitter signals to fool an eavesdropper in a RIS-assisted covert communication system under the Rician channel fading.
Both Bob and the eavesdropper try to classify their received signals into ``noise'' or ``signal'' classes via their DNN-trained classifier. 
The targeted attack is adopted against the eavesdropper. The adversarial perturbations are added to the transmission signals to enforce the specific misclassification from label ``signal'' to label ``noise'' under the assumption of knowing all channel information.
The fast gradient method (FGM) is introduced to linearize the loss function, and the opposite direction of the gradient is added to input samples of the eavesdropper DNN to decrease its loss function with the misclassified class.
Simulations illustrate that the adversarial perturbations have little influence on Bob, even with the increasing perturbations. However, it plays a dramatically significant role in reducing the accuracy of the classifier at the eavesdropper.

Analogously, in addition to the transfer-based attack mentioned earlier, there are score-based attacks where attackers observe the model’s output scores and behavior, generate adversarial samples, and continuously refine attack strategies to achieve greater effectiveness. There are also decision-based attacks where attackers traverse the decision boundary by generating perturbations to craft adversarial examples just outside the decision boundary~\cite{9984916}. All of the black-box attacks~\cite{9889707} can be adapted to perturb the AI-powered RIS-assisted wireless networks, such as the compression of QPSs~\cite{9592779} and classification of received signals~\cite{9771899}. Consequently, according to the characteristics of a black-box attack, it no longer relies on detailed model information and can be more suitable for real-world scenarios.

\subsection{Adversarial Defense Techniques}\label{Sec.VIII.2}
Analogous to the aforementioned defensive distillation mitigation approach in~\cite{9889707}, a plethora of adversarial defense strategies exist to neutralize adversarial attack methodologies and enhance the robustness of the system~\cite{10298624}, such as adversarial attack detection~\cite{10129254}, NIDs~\cite{debicha2023tad}, gradient masking, adversarial training~\cite{10263803}, and input transformation~\cite{9984916}. \par

In~\cite{9889707}, a defensive distillation mitigation method is proposed to enhance the robustness of the DNN-driven RIS network under gradient-based adversarial attacks. This method includes two training models, denoted as the teacher training model and the student training model. The teacher model generates softened output distributions via high-temperature Softmax, and the student model learns to mimic while simultaneously minimizing standard classification loss which is the weighted total loss of cross-entropy and Kullback-Leibler (KL) divergence losses. The method can soften the RIS prediction model's output probabilities using high-temperature Softmax, and this smoothing effect achieves the gradient masking to obscure the gradients that attacks rely on. The student model can be updated by adding the loss functions of the original sample and the adversarial sample, and the process of adversarial training can further force the model to learn stable decision boundaries, thereby improving resilience against both gradient-based and transfer-based attacks. \par

Furthermore, adversarial attack detection and input transformation can be cascaded with the aforementioned defensive distillation method to form a collaborative defense framework. By computing the gradient information of input environment descriptor samples, high-gradient regions, e.g., indicating potential adversarial perturbations, are localized and reconstructed or replaced using generative adversarial networks (GANs) to generate samples conforming to the clean data distribution. This cascaded approach not only purifies contaminated inputs but also synergizes with the decision-smoothing property of defensive distillation, significantly enhancing the overall robustness of the system. \par

However, advanced attackers can circumvent these defenses, rendering them ineffective. Take adversarial attack detection as an example. The monitor first trains the clean and disturbed samples and measures the differences between them to detect the adversarial samples caused by the subtle perturbations. If the attackers discover attack failure, they change the corresponding perturbations by changing the budget or attack modes to bypass the defenses of adversarial attack detection. Although the NIDs can initially defend against malicious network traffic, attackers may adapt by tweaking a small subset of the traffic characteristics based on prior feedback until circumvent the NIDs.

\begin{table}[!ht]
    \centering
    \caption{\centering{Analysis of Vulnerabilities and Defense Mechanisms in Different AI Models}}
    \vspace{-0.1in}
    \begin{adjustbox}{width=0.45\textwidth}
    \begin{tabular}{|m{1cm}<{\centering}|m{3cm}<{\centering}|m{3cm}<{\centering}|m{3cm}<{\centering}|}
    \hline
        \textbf{AI Model} & \textbf{Vulnerabilities} & \textbf{Implication} & \textbf{Defense Mechanisms}\\ \hline 
        DNN Model & Gradient-based attacks: FGSM, PGD, and so on & RIS reflection matrix erroneous prediction; QPSs erroneous compression or reconstruction; signal misclassification & Adversarial training; defense distillation \\ \hline
        RL Model & Policy manipulation attacks: environment poisoning, data poisoning & Erroneous knowledge formation; sub-optimal policy learning & Multi-stage defense framework: reward anomaly detection, adversarial model verification, and failure-independent model verification ensemble \\ \hline
        FL Model & Backdoor attacks: data poisoning, model poisoning & Undermine model integrity and availability; prevent global AI convergence and cause faulty RIS configuration & Integrated strategies combining anomaly detection and robust FL models \\ \hline
    \end{tabular}
    \end{adjustbox}
    \label{Table:VIII.C.1}
    \vspace{-0.2in}
\end{table}

\subsection{Analysis of Vulnerabilities and Target Defense in Specific AI Models}
The susceptibility of AI-powered RIS-assisted wireless networks to adversarial attacks is not uniform, and different AI models exhibit distinct attack surfaces due to their unique learning mechanisms and operational roles within the network. Consequently, a one-size-fits-all defense is ineffective. The following analysis delineates the primary vulnerabilities and correspondingly recommended targeted defense mechanisms for three predominant AI models, e.g., DNN, RL and federated learning (FL) models, in RIS-assisted wireless networks, and summarized in Table~\ref{Table:VIII.C.1}.

\subsubsection{DNN Models for Regression or Classification}
DNNs are predominantly vulnerable to gradient-based attacks, such as FGSM, PGD~\cite{9889707,9984916}. Attackers exploit DNN models' differentiability to craft minimal elaborately adversarial perturbations to the input data CSI or received signals, causing erroneous prediction of RIS phase shifts~\cite{9889707}, misclassification of received signals [116], erroneously compressing or reconstructing QPSs~\cite{son2024adversarial}.

Recommended defenses include adversarial training~\cite{10263803} to enhance model robustness by exposing it to adversarial examples during training, and defensive distillation~\cite{9889707} to smooth the output decision surface, making it harder for gradients to be exploited.

\subsubsection{RL Models for Control and Optimization}
RL-driven RIS controllers are highly vulnerable to policy manipulation attacks through malicious environmental feedback or poisoned training data~\cite{9536399}. For example, the attacker continuously monitors the RIS's action. When the RIS applies a configuration that improves the capacity of legitimate channels, the attacker transmits the jamming signal to drop the SNR at legitimate users, and then provides a negative reward to the RL agent. Over time, this misleads the agent into adopting sub-optimal policies, consistently impairing network performance. Alternatively, through data poisoning, the attacker can directly inject fabricated transition tuples, such as pairing a beneficial RIS configuration with an artificially low reward, introducing spurious correlations that hinder the learning of effective policies and lead to sub-optimal performance.

To mitigate policy manipulation and data poisoning in RL-enabled RIS-assisted networks, \textit{a multi-stage defense framework is crucial, spanning the training, pre-deployment, and runtime phases of the RL agent’s lifecycle}. During online learning, reward anomaly detection protects the RL agent by modeling the distribution of legitimate reward, such as functions of SNR or throughput~\cite{9956995}, and filtering real-time outliers to prevent poisoned rewards from corrupting policy updates. Pre-deployment adversarial model verification tests the trained RL model in a high-fidelity simulation environment against diverse adversarial scenarios to ensure robust beamforming under malicious conditions. At runtime, a failure-independent model verification ensemble~\cite{9006090} employs multiple RIS configuration predictors, each resilient to specific interference or deception types. Through weighted consensus, this framework ensures reliable reflective beamforming even under unseen attacks, thereby maintaining system security and performance through functional redundancy and enhanced generalization.

\subsubsection{FL Models for Collaborative Training}
In FL-driven RIS systems, multiple BS-RIS clients collaboratively train a global AI model by uploading local gradient updates to a central server for aggregation~\cite{9415623}. Adversarial attacks exploit this distributed learning framework, with attackers acting as ``Trojan horse'' to manipulate local updates through backdoor attacks, mainly including data poisoning attacks and model update poisoning attacks, thereby undermining the integrity and availability of the global model~\cite{10872904}. For example, malicious clients may embed a latent ``backdoor'' during the local training~\cite{9806416}, which remains undetectable until activated by a specific trigger, such as a unique signal feature. This trigger prompts the global AI model to reconfigure the RIS, redirecting confidential signals to an eavesdropper and enabling an eavesdropping channel. Alternatively, malicious clients can upload model updates that starkly contradict those of legitimate clients~\cite{10872904}, iteratively skewing the global AI model during aggregation rounds, preventing convergence and impairing RIS radiation pattern configuration, thus degrading the entire network's performance.

State-of-the-art defense approaches against adversarial attacks on FL-enabled RIS-assisted networks include integrated strategies combining anomaly update detection and robust FL models~\cite{9806416}. Concretely, the central serve can identify malicious updates by detecting significant deviations in model geometry or latent embeddings which can reveal attackers via high reconstruction errors in the encoder-decoder model, or by recognizing behavioral consistency among attackers which is absent in benign clients. 
Meanwhile, robust and secure FL models can be further enhanced by jointly injecting artificial and wireless differential privacy noise into the clipped gradients, suppressing anomalous magnitudes and mitigating potential backdoor patterns~\cite{11040056}, while feedback-based validation~\cite{10630602} leverages participant evaluations to reject globally aggregated models exhibiting sudden performance degradation on tasks like optimization of the RIS radiation pattern.

\subsection{Summary and Lessons Learned}\label{Sec.VIII.3}

AI models can be trained to optimize the RIS reflection matrix based on environmental data, classify transmission signals at the receiver, and manage the compression and reconstruction of QPSs at BSs and RIS controllers, as shown in Fig.~\ref{fig:VIII.1}. However, DNN-based models~\cite{9984916,9889707} are susceptible to adversarial attacks due to their dependence on gradient information~\cite{10263803}. Attackers may execute attacks by altering model inputs to create perturbations, leading to incorrect predictions~\cite{9889707}. By exploiting model access, adversaries can trick AI models into making erroneous decisions regarding RIS reflection matrices and signal classification~\cite{9771899}. Such attacks can be more damaging than traditional AWGN jamming~\cite{8651357}. 

Despite defenses like defensive distillation, input transformation, and adversarial training, these methods often fail to effectively protect AI-powered RIS-assisted communication systems. Attackers can bypass defenses by subtly adjusting traffic characteristics, rendering traditional strategies ineffective. Conventional defense techniques tend to rely on target models and are less effective against transfer attacks, showing weak generalization. Fig.~\ref{fig:VIII.1} and Table~\ref{Table:VIII.1} illustrate these adversarial attacks. Addressing these security threats is crucial for academia and industry to improve the robustness of AI-powered RIS-assisted secure wireless communication systems.

% ============ Sec. RIS-based defense mechanisms
\section{RIS-Based Defense Mechanisms}\label{Sec.IX}
Sections~\ref{Sec.IV}-\ref{Sec.VIII} delineate on the malicious implications of RIS dual-use nature, and include IRIS-enabled passive-active hybrid attacks where adversaries exploit passive RIS to actively launch malicious activities, and adversarial attacks on AI-driven RIS networks via the openness of the wireless channel and adversarial perturbations. This section highlights the friendly role of RIS in enhancing physical layer security across diverse wireless communication scenarios. By intelligently reconfiguring propagation environments, RISs can concentrate the reflective propagation on authorized users and cause destructive interference for unauthorized users~\cite{9807309,10604844}, resulting in the amplification of the main channel and the attenuation of the wiretap channel~\cite{10328191}. We systematically analyze RIS-assisted security enhancements in five typical scenarios, including unmanned aerial vehicle (UAV)~\cite{10289638,10146001}, simultaneous wireless information and power transfer (SWIPT)~\cite{10453453}, device-to-device (D2D)~\cite{10102557}, ISAC~\cite{10193812,10817325}, and VLC~\cite{10375270}. Each scenario demonstrates the RIS's ability to address unique security challenges~\cite{you2021towards,10251433,10054381} while leveraging its inherent advantages for next-generation wireless systems~\cite{10555049,9765815}. \par

\subsection{RIS-Assisted Secure UAV System}\label{Sec.IX.1}
UAV communication has been widely applied in civilian applications~\cite{10114467,info13080389,9975284}, such as transportation~\cite{10637488,10102437}, search and rescue, agriculture, forestry, environmental protection, and public safety~\cite{10102437}, due to its advantages of low cost, lightweight, high maneuverability, longer battery life, swift deployment, and convenience. \par

\subsubsection{Unique Security Challenges}\label{Sec.IX.1.A}
UAV communications are particularly vulnerable to eavesdropping due to their reliance on LoS-dominated air-to-ground channels, and Eve may exploit the inherent openness, broadcast nature, and signal superposition properties to intercept signals. Additionally, the constrained size and onboard power capacity of UAVs impose significant limitations on their ability to generate AN, necessitating careful optimization to maintain adequate security performance while preserving operational efficiency. \par

\begin{figure}
  \begin{center}
  \includegraphics[width=3.0in]{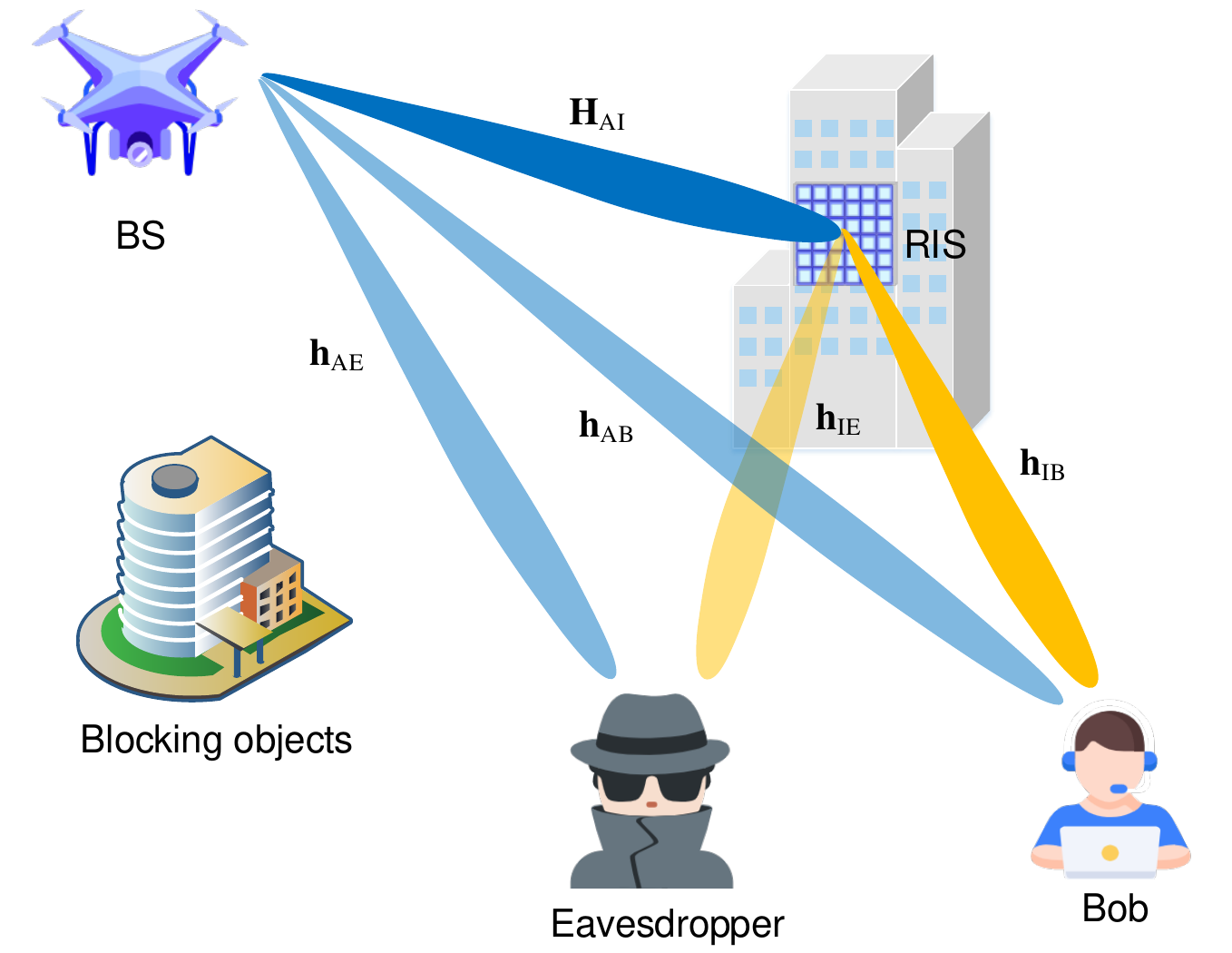}\\
  \vspace{-0.1in}
  \caption{RIS-assisted secure UAV-BS communication network: UAV-mounted BS assisted by RIS establishes cascaded links to overcome blockages and against the eavesdropper.}\label{fig:IX.1.1}
  \end{center}
  \vspace{-0.25in}
\end{figure}

\subsubsection{RIS's Defensive Roles}\label{Sec.IX.1.B}
Recent advances in RIS-assisted UAV communication networks have demonstrated significant improvements in security performance by leveraging two key advantages: The UAV's inherent mobility and RIS-enabled joint optimization capabilities. Research efforts have focused on optimizing critical system parameters, including UAV transmit power allocation, flight trajectory design, AN power distribution, RIS phase shift configurations, and dynamic user scheduling. These coordinated optimization approaches~\cite{HAILONG2024537} enable simultaneous enhancement of both communication reliability and physical-layer security in challenging aerial environments. \par

A secure RIS-assisted UAV SISO communication system is designed in~\cite{9209992} as shown in Fig.~\ref{fig:IX.1.1}, which adopts a UAV-mounted BS to replace the fixed BS, and transmits signals to the legitimate mobile user accompanied by a passive malicious eavesdropper. In this system, the UAV is constrained by flying at a fixed height to avoid collisions with buildings. It can effectively overcome blockages and information leakage according to its flexible mobility characteristics and the LoS-dominated air-to-ground channels. The SR is maximized by jointly optimizing the UAV transmit power, trajectory, and RIS passive beamforming (PBF). The sub-problems are alternatively optimally tackled by the Karush-Kuhn-Tucker (KKT) conditions, SCA, and phase alignment, respectively, to solve the non-convex and NP-hard global optimization formulation and obtain the approximation solution. To close the application, the authors of~\cite{9880605} ulteriorly consider the multiple mobile users scenario and optimize the user scheduling, whose secrecy performance is better than without scheduling, and the UAV trajectory is also different from the single mobile user scenario. \par

Meanwhile, mmWave technology and THz communication will occupy the leading position for the 5G and B5G wireless communication networks and have the potential to achieve gigabits-per-second (Gbps) transmission rate and ultra-low latency. The large-scale antenna array technology is adopted to compensate for the disadvantages of mmWave communication, such as atmospheric and rain attenuation and blockage effect~\cite{9528924}. Then, a MISO, RIS-assisted, secure UAV-BS mmWave communication network is designed in~\cite{9528924}, where the AN is exploited to enhance the secrecy performance. Without loss of generality, both the unblocked and blocked links between the UAV-BS and the mobile user are considered. To maximize the SR, the positions and beamforming of UAV-BS and RIS are alternatively optimized by SDR under the constraints of maximum transmit power, the UAV flight altitude range, and the legitimate mobile user minimum rate. \par

\subsubsection{Future Improvements}\label{Sec.IX.1.C}
Future enhancements for RIS-assisted UAV communication systems should address several critical challenges. While current implementations optimize security and performance through RIS reflection matrix adjustment~\cite{9528924}, intelligent user scheduling, and UAV trajectory planning~\cite{9209992,9880605} while capitalizing on UAV advantages like cost-effectiveness and mobility~\cite{10528789}, operational constraints remain. The requirement for fixed-altitude flight to prevent collisions significantly limits deployment flexibility. Advanced obstacle avoidance systems could expand operational space while mitigating blockage-induced information leakage. Furthermore, DRL techniques~\cite{10380323,10234427}, including Deep Q-Networks (DQN) and deep deterministic policy gradient (DDPG)~\cite{10021680}, show a strong potential to enable real-time adaptive security against mobile eavesdroppers. These approaches could dynamically optimize RIS configurations in response to environmental changes and threat patterns, significantly enhancing system resilience. \par

\subsection{RIS-Assisted Secure SWIPT System}\label{Sec.IX.2}
According to~\cite{han2020energy}, each 5G BS will consume approximately four times more energy than 4G BS, and the overall power consumption of 5G BSs will be 12 times than that of 4G BSs due to the dense deployment of 5G BSs. Furthermore, as reported in~\cite{20222112136564}, 990,404 tonnes of annual carbon emissions will be indirectly caused by 5G network operations under the medium-demand scenarios by 2030. It is important to develop green and cost-effective wireless communication technologies to dramatically reduce economic and industrial costs and achieve sustainable development. SWIPT can achieve receiving information signals and harvesting energy simultaneously and then can enhance the energy efficiency (EE) of wireless networks. \par

\subsubsection{Unique Security Challenges}\label{Sec.IX.2.A}
\begin{figure}[t]
	\centering{}\includegraphics[width=3.0in]{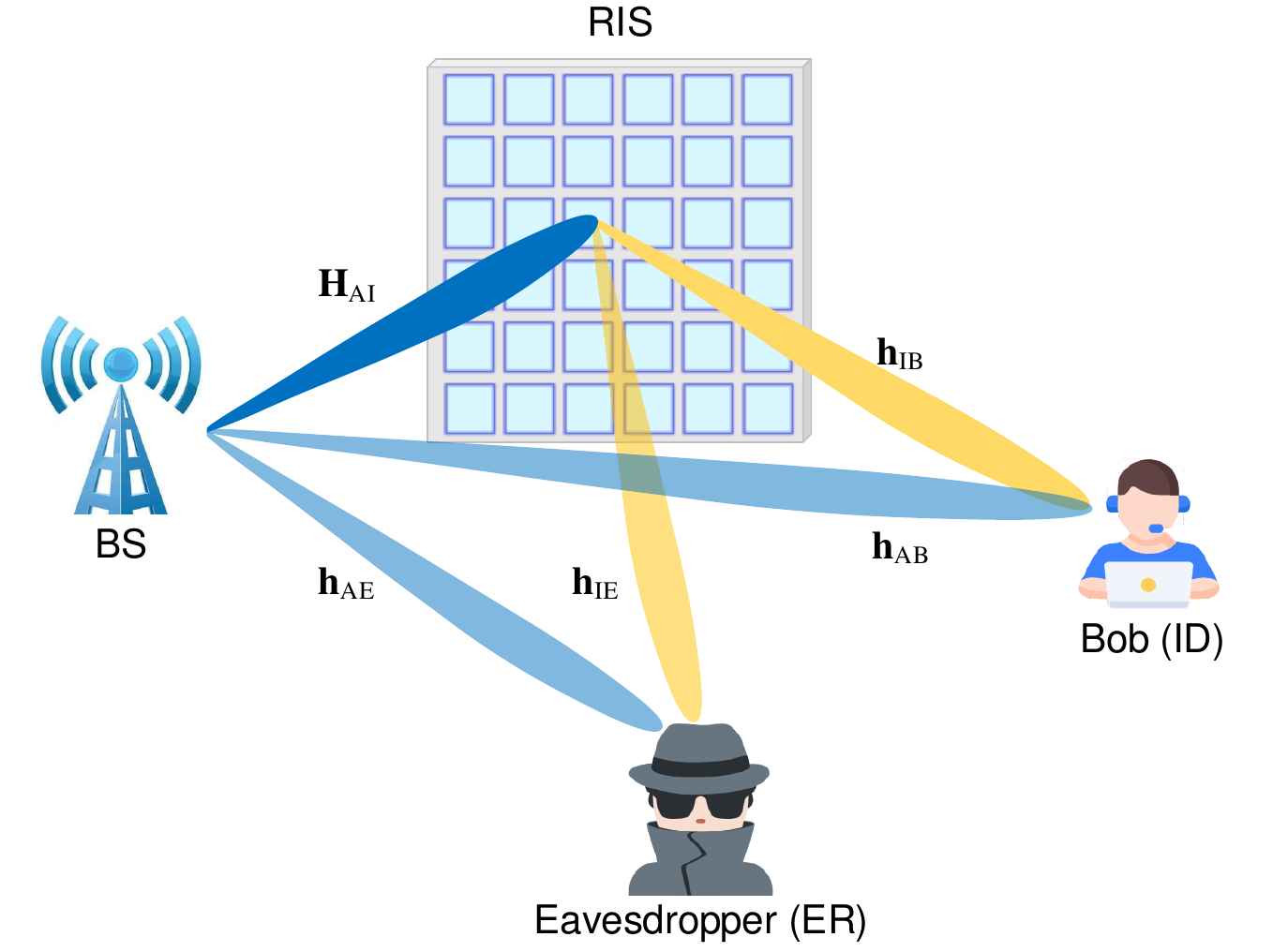}
    \vspace{-0.1in}
	\caption{RIS-assisted secure SWIPT system with separate ID and ER: BS assisted by the RIS concurrently dispatches data-bearing signals to an ID while delivering energy-laden signals to an ER, the latter of which could potentially transform into an eavesdropper.}
	\label{fig:IX.2.1}
    \vspace{-0.15in}
\end{figure}

\begin{figure}[ht]
	\centering{}\includegraphics[width=3.0in]{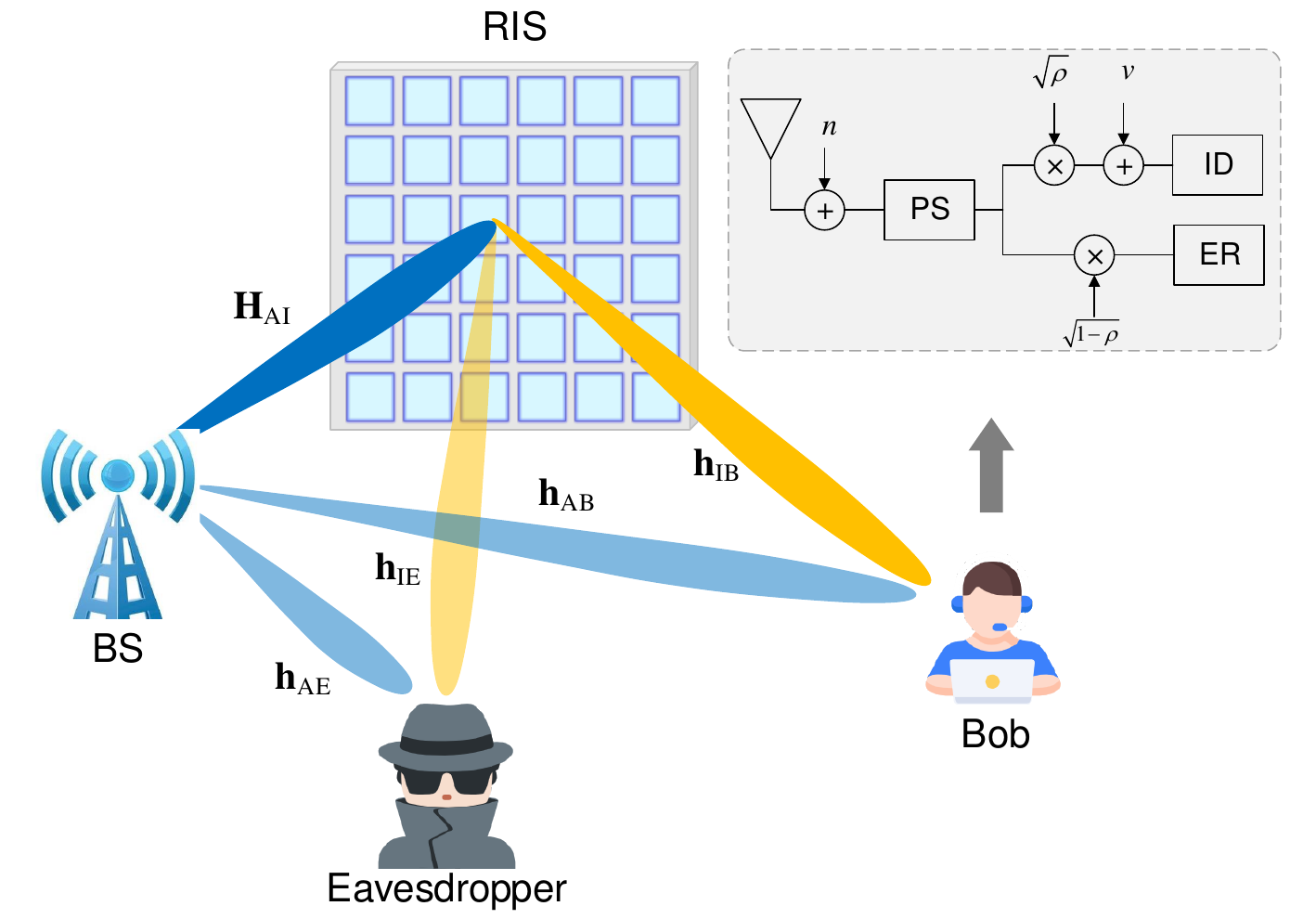}
    \vspace{-0.1in}
	\caption{RIS-assisted secure SWIPT system with the PS user: Bob is a unified user with the PS protocol, which can decode information and harvest energy.}
	\label{fig:V.2.2}
    \vspace{-0.25in}
\end{figure}

SWIPT systems face unique security challenges due to their dual functionality. As shown in Fig.~\ref{fig:IX.2.1}, conventional architectures employ separate information decoders (IDs) and energy receivers (ERs)~\cite{9852091}, where ERs may potentially eavesdrop on information-bearing signals. Alternatively, Fig.~\ref{fig:V.2.2} illustrates unified receivers employing power splitting (PS) protocols~\cite{9734045,9538921,9766101}, where the unified user can simultaneously decode information and harvest energy by itself with PS ratios $\rho$ and $1-\rho$, respectively. While the harvested energy can be repurposed to generate AN against eavesdroppers, RIS technology plays crucial roles in dynamically managing the information-energy signal mixture to enhance among legitimate uses' QoS, energy collection efficiency and secure performance. \par

\subsubsection{RIS's Defensive Roles}\label{Sec.IX.2.B}
\paragraph{RIS-assisted secure SWIPT system with separate ID and ER}\label{Sec.IX.2.B.1)}
In~\cite{9852091}, the precoder matrix, the AN covariance at the BS, and the phase shifts at the RIS are jointly optimized to maximize the achievable SR in a secure cognitive radio (CR) RIS-assisted SWIPT MIMO system with perfect CSI. Secondary users (SUs) include an ID and an ER, receiving information and harvesting energy, while $K$ primary users (PUs) share the spectrum and tolerate interference. The inexact block coordinate descent (IBCD) method alternately optimizes variable sets, using auxiliary variables and convexification via CVX for the precoder matrix and AN covariance, and the majorization-minimization (MM) algorithm for RIS phase shifts. Multiple random initial points ensure convergence to the global optimum. 

\paragraph{RIS-assisted secure SWIPT system with PS user}\label{Sec.IX.2.B.2)}
In~\cite{9538921}, an angle-aware user cooperation (AAUC) scheme maximizes average SR in a RIS-assisted SWIPT secure MISO system. The BS multicasts a common signal to all users, except the one monitored by the eavesdropper. Cooperative users forward the decoded signal to the monitored Bob using the RIS and harvested energy. The MM-based AO algorithm optimizes the precoder matrix and RIS phase shifts, reducing CPU time compared to the traditional second-order cone program (SOCP) algorithm. However, the AO algorithm faces challenges in practical applications due to complex transitions and numerous iterations. \par

In~\cite{9734045}, the PS factor at Bob, along with transmitter power and phase shift at the BS and RIS, are jointly optimized to enhance equipment EE and prolong service life in a secure SWIPT network. The system includes a BS and an eavesdropper with a single omnidirectional antenna. The feasible point pursuit-successive convex approximation (FPP-SCA)-based AO algorithm reformulates the non-convex objective function into an approximate convex problem using slack variables, solving it iteratively with the interior-point method. Despite its effectiveness, the AO algorithm is computationally intensive. A deep learning (DL)-based scheme is introduced to address this, significantly reducing computational time with five types of data and DNN structures, while maintaining security performance.
The proposed AO- and DL-based algorithms are suitable for SISO communication systems. However, these approaches should be further extended to the more common MIMO system to improve communication capacity and throughput.
Simulations show that increasing RIS elements can enhance security performance, and the DL-based approach matches AO algorithm security while significantly improving computational efficiency. \par

In~\cite{9766101}, a full-duplex cooperative jamming (FD-CJ) scheme using SWIPT technology is investigated to enhance a discrete RIS-assisted secure communication network over a Rician fading channel. Bob, acting as the FD-CJ, has a dual separate antenna, while the BS and the eavesdropper have an $N_\mathrm{t}$-antenna uniform linear array (ULA) and a single antenna, respectively. The BS beamformer, RIS phase shifts, and Bob’s AN transmitter power are jointly optimized to maximize network SR, constrained by the BS transmitter power, RIS reflection coefficient, and Bob’s AN power. The AO algorithm handles the mixed-integer non-convex objective function with perfect CSI at the BS. The beamformer and phase shift optimization sub-problems are convexified using SDR and CCP and solved via the interior point method. The continuous phase shift is quantized, and the optimal AN power is derived from its first-order derivative. The algorithm achieves the highest SR compared to benchmarks but converges only to local optima. Simulations indicate that continuous phase discretization causes performance loss, increasing with discrete steps. A digital system can design discrete phase shifts more quickly, making it suitable for practical scenarios. \par

\subsubsection{Future Improvements}\label{Sec.IX.2.C}
Future research directions for RIS-assisted secure SWIPT systems should further focus on overcoming current limitations in the security-energy trade-off. While existing approaches demonstrate improved SR and EE through either separate receiver architectures~\cite{9852091} or unified PS protocols~\cite{9734045,9538921,9766101}, critical challenges remain in practical implementation. Simulations in~\cite{9734045} show that the conventional AO algorithms face computational complexity issues that hinder real-time deployment. Emerging ML techniques offer promising solutions by enabling intelligent PS ratio adaptation to dynamically balance SR and EE under varying channel conditions, and efficiently solving multi-objective optimization problems with complex constraints. Furthermore, advanced protocol designs could incorporate dynamic switching between the PS and TS modes based on real-time security requirements and EH demands, potentially achieving superior performance compared to static schemes. \par

\subsection{RIS-Assisted Secure D2D System}\label{Sec.IX.3}
D2D communications is regarded as a critical technique to improve the spectral efficiency (SE) in cellular communications and relieve the problem of scarce spectrum resources. There are pairs of D2D transmitter (DTX) and D2D receiver (DRX) that reuse the same spectrum as the cellular users to directly deliver the content~\cite{9625248}, which can considerably enhance the SE, expand cellular convergence, and decrease the delay~\cite{9305710}. \par

\subsubsection{Unique Security Challenges}\label{Sec.IX.3.A}
D2D communication systems face distinctive security challenges stemming from their inherent spectrum sharing architecture and weak D2D link encryption. Particularly, the reuse of cellular spectrum resources introduces co-channel interference between D2D pairs and conventional cellular links, creating a complex trade-off between SE and interference management. Meanwhile, compared to cellular links with robust encryption protocols, D2D links often adopt lightweight or even omitted encryption schemes to prioritize low-latency and EE, as Eve may exploit the broadcast nature of wireless channels and weak signal protection to intercept D2D communications. \par

\subsubsection{RIS's Defensive Roles}\label{Sec.IX.3.B}
\begin{figure}
  \begin{center}
  \includegraphics[width=3.0in]{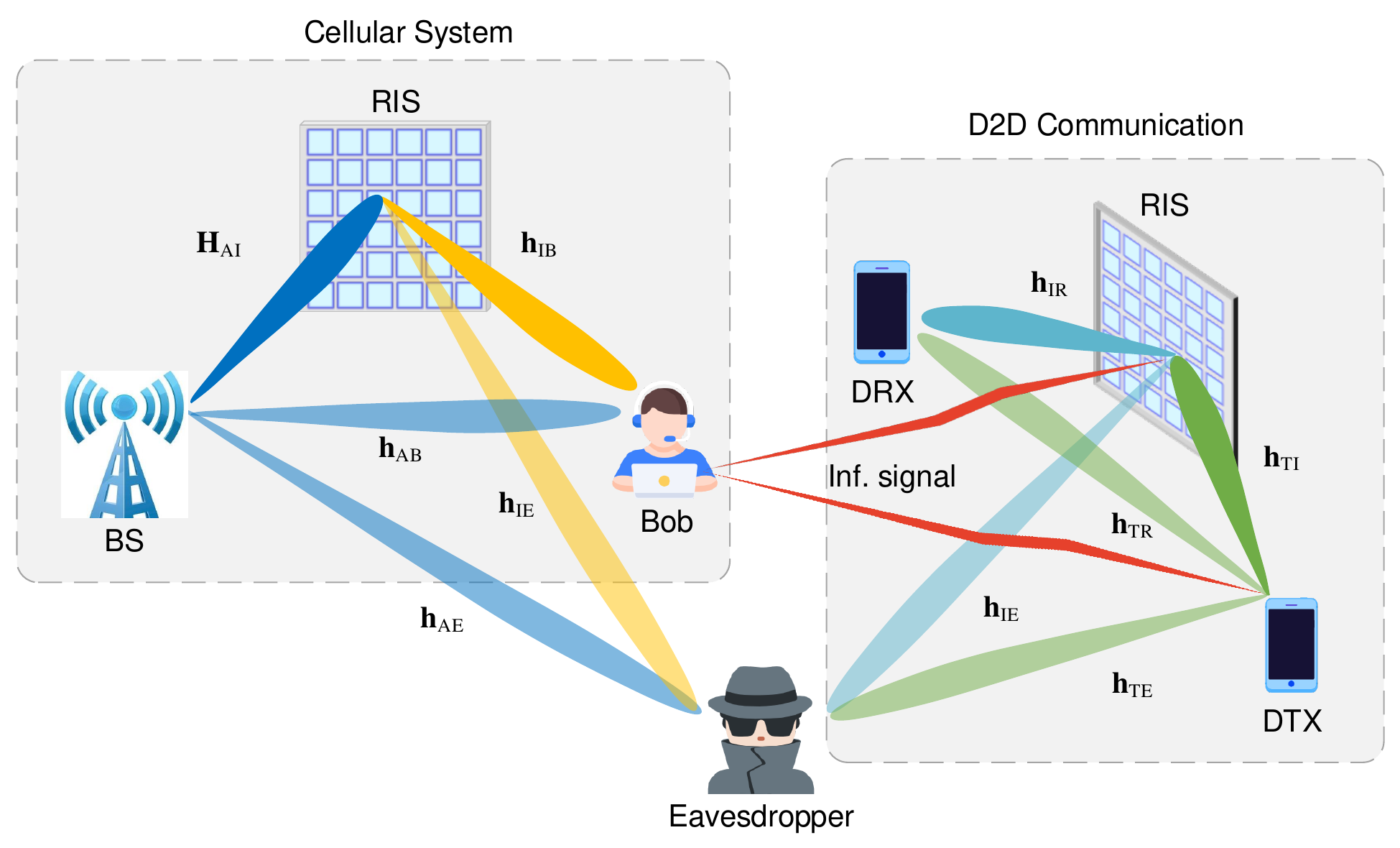}\\
  \vspace{-0.1in}
  \caption{D2D communication underlying cellular system: there are pairs of DTX and DRX that reuse the same spectrum as the cellular users to deliver the content assisted by the RIS directly, and the RIS can dramatically suppress the serious interference between the D2D and cellular links caused by spectrum reusing while concurrently countering potential eavesdropping threats posed by an eavesdropper.}
	\label{fig:IX.3.1}
  \end{center}
  \vspace{-0.25in}
\end{figure}
Signal manipulation capabilities of the RIS not only enable effective interference suppression between legitimate D2D and cellular links~\cite{10595399}, but also provide opportunities to strategically direct interference towards potential eavesdroppers, thereby enhancing physical layer security. This unique ability to simultaneously manage interference for legitimate users while generating targeted interference against eavesdroppers represents a paradigm shift in securing D2D communication, as demonstrated in Fig.~\ref{fig:IX.3.1}. \par

In~\cite{9305710}, analytical expressions for D2D outage probability, secrecy outage probability (SOP), and probability of non-zero secrecy capacity (PNSC) in a RIS-assisted downlink D2D underlay cellular system are derived. A single-antenna DTX and DRX communicate via a RIS-cascaded virtual link within a cellular network. A multi-antenna system selects the best antenna to transmit to the single-antenna Bob via a direct link, while the eavesdropper eavesdrops on the cellular network. The RIS phase shift is optimized through phase alignment, assuming RIS has complete CSI of the cascaded channels. Communication and security metrics are derived using the cumulative distribution function (CDF) of SINR for D2D and cellular links. These expressions suit D2D communication with a RIS-cascaded virtual link under a cellular network but do not support more complex scenarios. Simulation results show that increasing RIS elements improves EE, and system security performance is enhanced by increasing BS transmit power, with RIS outperforming relay-assisted scenarios. \par

In~\cite{9478912}, an analytical framework for spectrum sharing is proposed to enhance the robustness and security of underlay D2D networks with a RIS and a full-duplex (FD) jamming DRX. A BS and a DTX transmit signals to Bob and a DRX simultaneously via the same spectrum. The DRX, equipped with multiple antennas, selects the one with maximum reception to receive the DTX’s information, while others emit AN to confuse the eavesdropper using designed beamforming. The total power for D2D communications is fixed, and the power for the DTX and the DRX to emit private information and AN is assigned by PS factor $\rho$. Phase alignment adjusts the RIS phase under partial CSI. The framework derives the achievable ergodic SR based on statistical characterization of the D2D underlying cellular system with Rayleigh and Gaussian distributed $\mathcal{RV}$s. This framework suits RIS and FD jamming DRX combination in D2D underlay cellular networks but does not consider bidirectional communication and EH~\cite{9705482}. Simulations show a security-reliability trade-off with an eavesdropper’s attack, an optimal PS, $\rho^*$, with a fixed RIS element number, and a positive impact of RIS elements on system SR. \par

In~\cite{9625248}, a D2D underlying cellular system is introduced in the RIS-assisted uplink single input multiple output (SIMO) secure communication network to enhance the security performance and SE. Except for a BS with multiple antennas, Bob, and an eavesdropper with a single antenna, there is a pair of DTX and DRX, which reuse the same spectral resource with the cellular user Bob. The BCD algorithm is adopted to optimize the BS's beamforming vector, RIS's phase shift, and power allocation for Bob and the DTX to gain the maximum SR. Then, the non-convex objective function is decoupled into three sub-problems according to the corresponding optimization objectives. The optimal beamforming and power allocation solution can be obtained by the Rayleigh quotient maximization problem and linear program, respectively, and the optimization problem of phase shift is dealt with the auxiliary variables and SDR technique and then solved by the CVX tool. Simulation results show that the SR can be improved with increased RIS element number and maximum transmitter power. \par

\subsubsection{Future Improvements}\label{Sec.IX.3.C}
Future research in RIS-assisted secure D2D communications should address several critical directions to enhance both SE and SR. First, advanced RIS deployment strategies, such as multi-RIS coordination and optimized placement between D2D pairs and cellular infrastructures, can be investigated to achieve extensive communication coverage, suppress multi-user interference (MUI), and improve secure performance~\cite{9519722}. This includes modeling the interplay between LoS and cascaded links in hybrid cellular-D2D topologies. Afterwards, FD-enabled D2D architectures integrated with RIS could significantly improve SE and SR by enabling simultaneous bidirectional communication while suppressing MUI. Furthermore, dynamic RIS configuration protocols need to be developed to real-time channel conditions with mobility D2D patterns, ensuring robust, secure performance in practical deployment scenarios. \par

\subsection{RIS-Assisted Secure ISAC System}\label{Sec.IX.4}
ISAC has emerged as a transformative paradigm that enables simultaneous target sensing and user communication through shared spectrum and infrastructure utilization~\cite{10443321,9303435,9847217,10012421}. This joint design approach achieves synergistic performance gains by co-optimizing communication and sensing resources~\cite{11025997}. \par

\subsubsection{Unique Security Challenges}\label{Sec.IX.4.A}
\begin{figure}
  \begin{center}
  \includegraphics[width=3.0in]{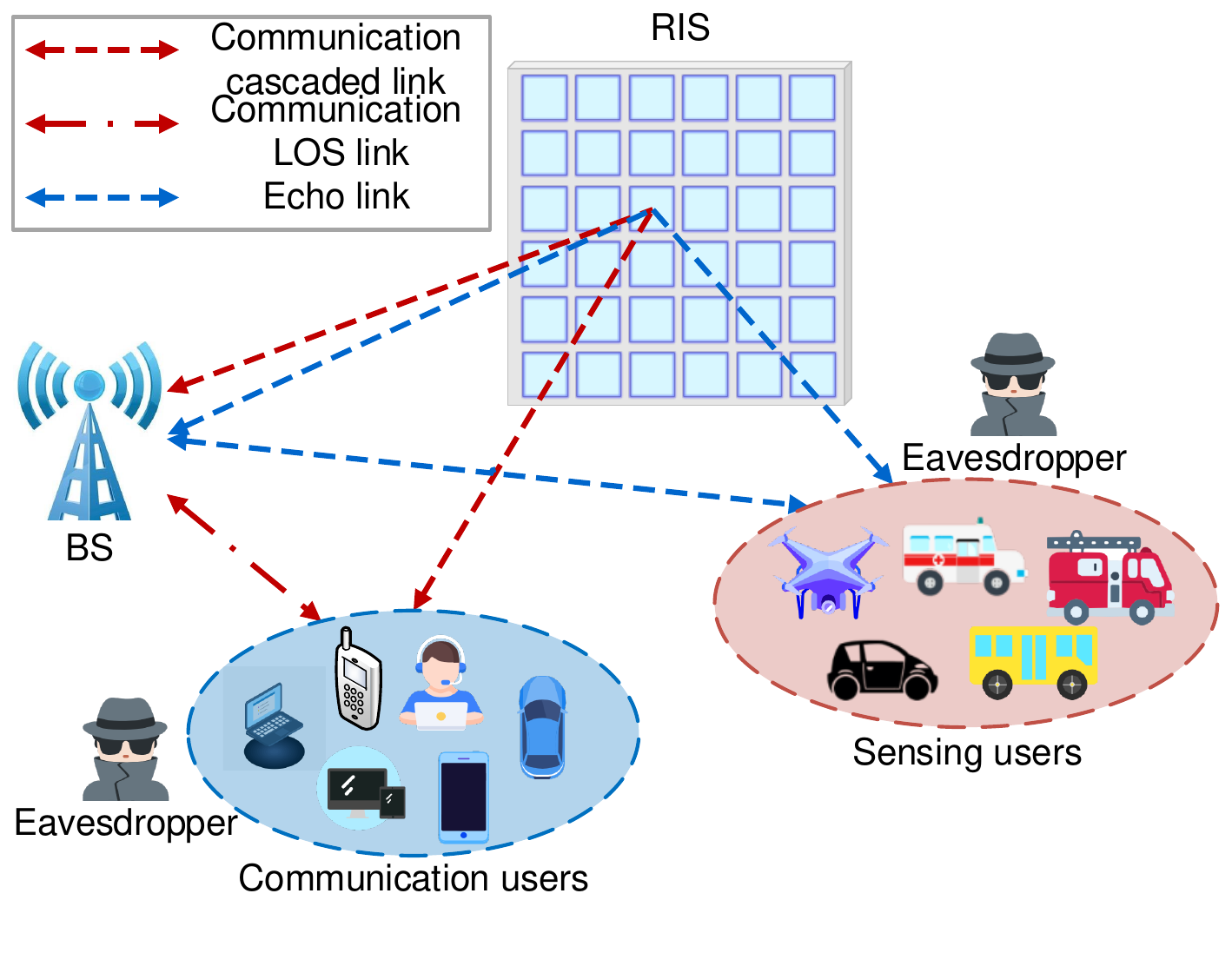}\\
  \vspace{-0.1in}
  \caption{RIS-assisted secure ISAC communication network: BS simultaneously propagates communication and sensing signals to communication and sensing users, respectively, in the presence of an eavesdropper beside them, and the sensing users may become potential eavesdroppers.}
	\label{fig:IX.4.1}
  \end{center}
  \vspace{-0.25in}
\end{figure}
The inherent openness of ISAC systems introduces unique security challenges. As illustrated in Fig.~\ref{fig:IX.4.1}, potential eavesdroppers may exist among legitimate communication and sensing users, and the sensing target may be a suspicious Eve potentially intercepting information-bearing signals transmitted for the communication users~\cite{10143420}. This dual-functional architecture creates novel attack surfaces that demand novel physical-layer protection mechanisms. \par

\subsubsection{RIS's Defensive Roles}\label{Sec.IX.4.B}
By dynamically reconfiguring the EM environment, RISs can strengthen legitimate communication links while strategically suppressing signal propagation toward suspicious sensing targets or unauthorized receivers, thereby achieving secure coexistence of sensing and communication functions. \par

In~\cite{9973319}, an optimization framework maximizes the SR of an MU-MISO ISAC system, utilizing an active RIS to counter eavesdropping by a malicious UAV. The model features a multi-antenna dual-function BS serving single-antenna users in a secure zone, with a UAV eavesdropper operating in Rayleigh fading. Fractional programming (FP) and MM techniques address the non-convex optimization problem under perfect CSI, subject to radar detection SNR thresholds and total power constraints. The SR maximization problem is reformulated into a tractable form, guaranteeing beamformers and RIS coefficients satisfy both communication and radar needs within power limits. This approach benefits active RIS-assisted ISAC security but assumes ideal CSI and neglects hardware-induced reconfiguration errors. Simulations confirm the active RIS-assisted system achieves superior SR over passive RIS and non-RIS benchmarks, validating the framework's anti-eavesdropping efficacy.

In~\cite{2025arXiv250415033R}, a RIS-assisted uplink privacy-preserving ISAC system is proposed to maximize the achievable sum rate while concealing users' spatial signatures from a wiretapper. The scenario involves multiple single-antenna users transmitting synchronously to a multi-antenna BS, with a wiretapper physically connected to the BS attempting to extract user location information. A trade-off parameter is introduced to incorporate the projection constraint into the objective function, and the Riemannian manifold optimization is employed to transform the non-convex constant modulus constraint into an unconstrained problem within Riemannian space. Simulations demonstrate that the wiretapper is unable to accurately detect user locations, confirming the scheme’s effectiveness in maintaining high communication performance while ensuring user occultation.

\subsubsection{Future Improvements}\label{Sec.IX.4.C}
In the ISAC system, the target sensing can be enabled with user communication simultaneously by sharing the same spectrum and infrastructures~\cite{10443321,10012421}, which can be adapted to detect the eavesdropper. Then, the RIS can further assist the system in enhancing security performance with the prior knowledge of the eavesdropper. \par

However, more stringent requirements have been put forward for hardware facilities. The BS should be equipped with transmitting and receiving antennas or dual-functional antennas that can achieve communication and sensing simultaneously~\cite{9973319}. Meanwhile, the RIS-assisted secure ISAC scenario can detect and track potential eavesdroppers by leveraging the RIS's beam steering and coverage extension capabilities to expand sensing ranges and achieve adaptive beam scanning~\cite{10284917,9913311}. According to the detected location information of potential eavesdroppers, the communication-sensing resources can be jointly dynamically optimized to further improve security performance. \par

\subsection{RIS-Assisted Secure Optical Communication System}\label{Sec.IX.5}
\subsubsection{Optical System Model}\label{Sec.IX.5.A}
\begin{figure}
  \begin{center}
  \includegraphics[width=3.5in]{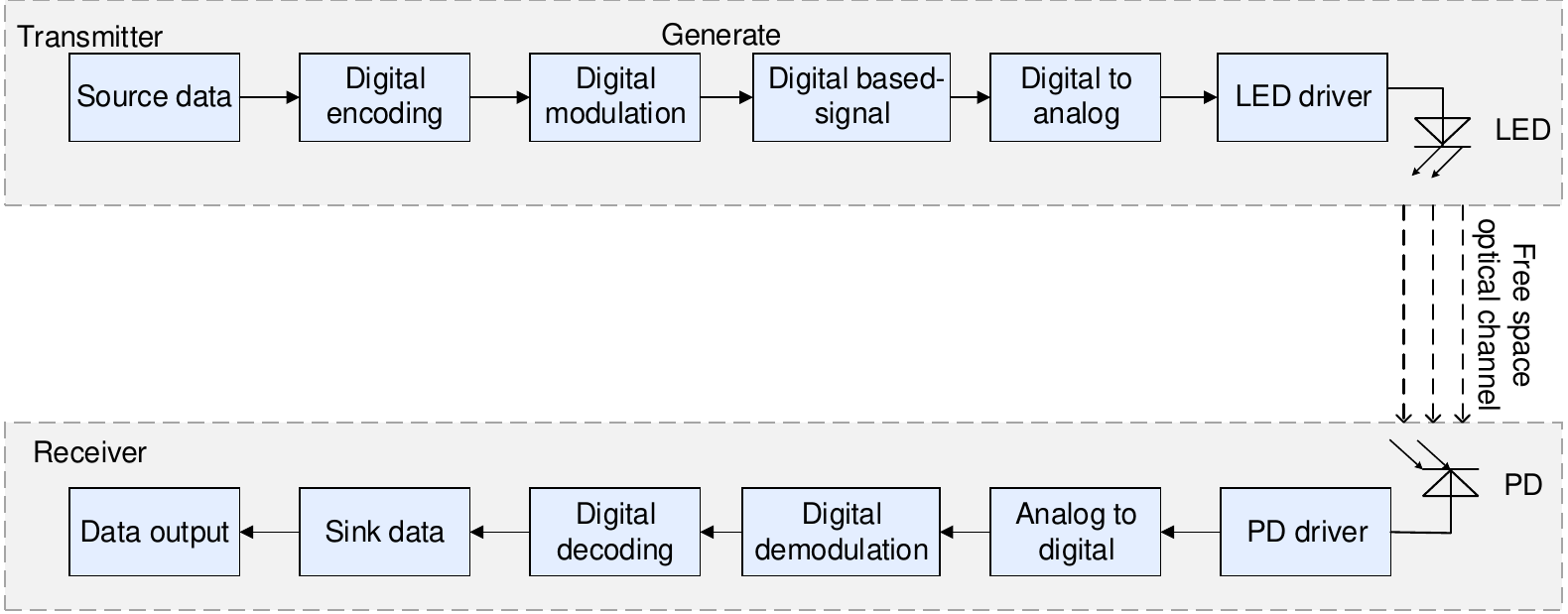}\\
  \vspace{-0.1in}
  \caption{The flow chart of the VLC communication system: the transmitter exploits the LED to transmit the non-negative amplitude and real-valued optical signals, and the receiver adopts the IM/DD method to transform the received optical signal into the electrical signal by the PD, and the LoS channel gain following the Lambert model.}
	\label{fig:IX.5.1}
  \end{center}
  \vspace{-0.25in}
\end{figure}

Optical communication systems comprise VLC~\cite{9756553,10174682,10043849}, ultraviolet (UV) communications~\cite{10462045,10305071,Ge:23}, etc. VLC utilizes EM waves in the visible light frequency band with wavelengths ranging from 380 nm to 780 nm for communication. In VLC communication systems, as shown in Fig. \ref{fig:IX.5.1}, the transmitter exploits the light emitting diode (LED) to transmit the non-negative amplitude and real-valued optical signals, and the receiver adopts intensity modulation/direction detection (IM/DD) method to transform the received optical signal into the electrical signal by the photo-detector (PD), and the LoS channel gain following the Lambert model \cite{9756553,9500409}.

Furthermore, the near-field condition is guaranteed in RIS-assisted VLC systems according to the nanoscale wavelength characteristics of visible light \cite{9756553}. Thus, the cascaded channel gain through the $m$-th RIS element follows the ``additive'' model \cite{9756553,10279487}. If there is a LoS link, the received signals will be the sum of LoS and cascaded links.
Due to the requirements of non-negative amplitude and real-valued optical signals in VLC systems, the achievable data rate cannot be exactly described by the typical Shannon capacity, and the tight lower bound of VLC channel capacity is given by \cite{9756553}
\begin{equation}
    C_{\mathrm{VLC},k}=\frac{1}{2}B\mathrm{log}_2\left(1+\frac{e}{2\pi}\gamma_k\right)
    \label{eq:16}
\end{equation}
where $C_{\mathrm{VLC},k}$ and $\gamma_k$ are the channel capacity and SINR of the receiver with $k\in\{\mathrm{Bob},\mathrm{Eve}\}$ in VLC systems, respectively, $B$ is the modulation bandwidth, and $e$ is the base of natural logarithms.

\subsubsection{RIS-Assisted Secure VLC System}\label{Sec.IX.5.B}
\begin{figure}
  \begin{center}
  \includegraphics[width=3.0in]{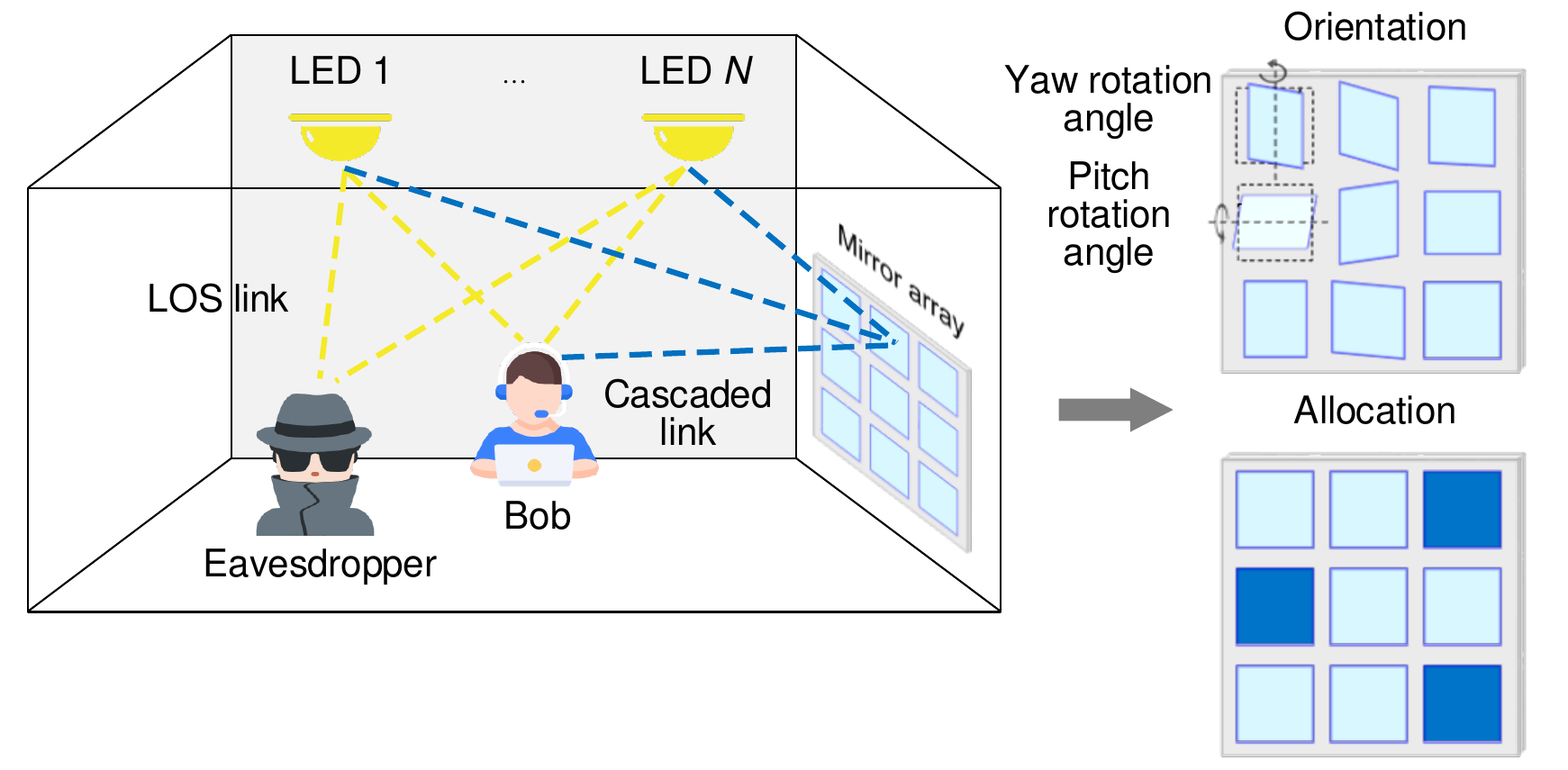}\\
  \vspace{-0.1in}
  \caption{RIS-assisted VLC system: the mirror array is utilized as the RIS to directly adjust the reflection direction of optical signals by optimizing mirror orientation or transforming into the assignment problem, and then improving the security performance of the VLC system.}
	\label{fig:IX.5.2}
  \end{center}
  \vspace{-0.25in}
\end{figure}

Beyond the advantages of low cost and power consumption, abundant spectrum resources, high transmission rate and efficiency, energy conservation and environmental protection, non-EM noise, safe and reliable, easy to reuse existing devices~\cite{10328848,electronics12245036,10.1117/1.OE.62.6.066103}, the VLC communication has greatly confidentiality advantage due to the weak diffraction ability of LED which is difficult to penetrate any non-transparent object. Consequently, the VLC communication is suitable for RF-sensitive areas such as hospitals, airports, and laboratories~\cite{9784887}. \par

\paragraph{Unique Security Challenges}\label{Sec.IX.5.B.1)}
The secrecy performance of RIS-assisted VLC systems has attracted extensive research attention due to their strong compatibility with indoor wireless security enhancement~\cite{10128157}. While VLC's inherent optical properties, particularly its negligible diffraction capability and inability to penetrate walls, naturally enhance physical layer security for indoor environments, eavesdroppers within the same coverage area can still intercept confidential optical signals. Furthermore, distinct from RF communications, VLC's IM/DD scheme eliminates phase information, restricting RIS implementations to mirror arrays that either manipulate reflection angles or solve binary assignment problems, according to Snell's law~\cite{9500409,10279487,9784887,9756553}, as shown in Fig.~\ref{fig:IX.5.2}. These fundamental characteristics introduce unique design constraints for security-oriented beamforming in optical domains. \par

\paragraph{For RIS's Defence in VLC System Mirror Orientation}\label{Sec.IX.5.B.2)}
In~\cite{9500409}, the mirror array is used as the RIS to improve the secrecy performance of the SISO VLC system monitored by an eavesdropper. Each mirror's yaw and pitch rotation angles can be independently controlled and optimized to tune mirror orientation and then adjust the direction of reflected optical signals as shown in the ``Orientation'' part in Fig.~\ref{fig:IX.5.2}. Furthermore, the mirrors' orientation optimization problem is transformed into the reflected spot finding (RSF) problem to reduce the complexity. It is solved by an improved heuristic algorithm named particle swarm optimization-initialization intervention (PSO-II). According to the proposed RSF method, unsafe areas in the VLC system can be further decreased, and then the system can remain secure all the time. The more practical dynamic system caused by mobile user movement is studied in~\cite{9784887}, and the beamforming weights at LEDs, mirror array sheet yaw angles, and individual mirror yaw and roll angle are jointly optimized by the DDPG algorithm. Compared with the PSO-based algorithm in~\cite{9500409}, the DDPG-based training can achieve better adaptability under the dynamic environment, and both of them show that SR drops to the lowest value when the eavesdropper is close to Bob. \par

\paragraph{For RIS's Defence in VLC System through Assignment}\label{Sec.IX.5.B.3)}
Instead of optimizing mirror orientation, the SR maximization process is transformed into the assignment problem of RIS in~\cite{10279487,9756553} as shown in the ``Allocation'' part in the lower right corner of Fig.~\ref{fig:IX.5.2} where the light-colored and blackish elements express the assigned and unassigned elements to the users respectively. A secure RIS-assisted MU-MISO VLC system with an eavesdropper is investigated in~\cite{9756553}, the binary RIS allocation matrix optimization is transformed into an assignment problem of a bipartite graph and then solved by the Kuhn-Munkres (KM) algorithm. The SR increases with the number of RIS units and reflectivity. Meanwhile, to satisfy the requirements of high reliability and low latency for massive communication at any time and anywhere for the arrival of the 6G era, non-orthogonal multiple access (NOMA) has become an appealing technology to achieve large-scale connectivity~\cite{9915477}. A NOMA-based RIS-assisted VLC system with Bob and an eavesdropper is proposed in~\cite{10279487}. The allocation of NOMA power and the binary RIS matrix are jointly optimized to boost the secrecy capacity. The adaptive-restart genetic algorithm (GA) can gain a computationally efficient solution under the constraints of users' rate requirements and the transmitter's power limitation. According to the simulation, the cascaded RIS channel can provide great flexibility and the highest DoF to control users' channel conditions.

\subsubsection{RIS-Assisted Secure VLC/RF Hybrid System}\label{Sec.IX.5.C}
\begin{figure}
  \begin{center}
  \includegraphics[width=3.0in]{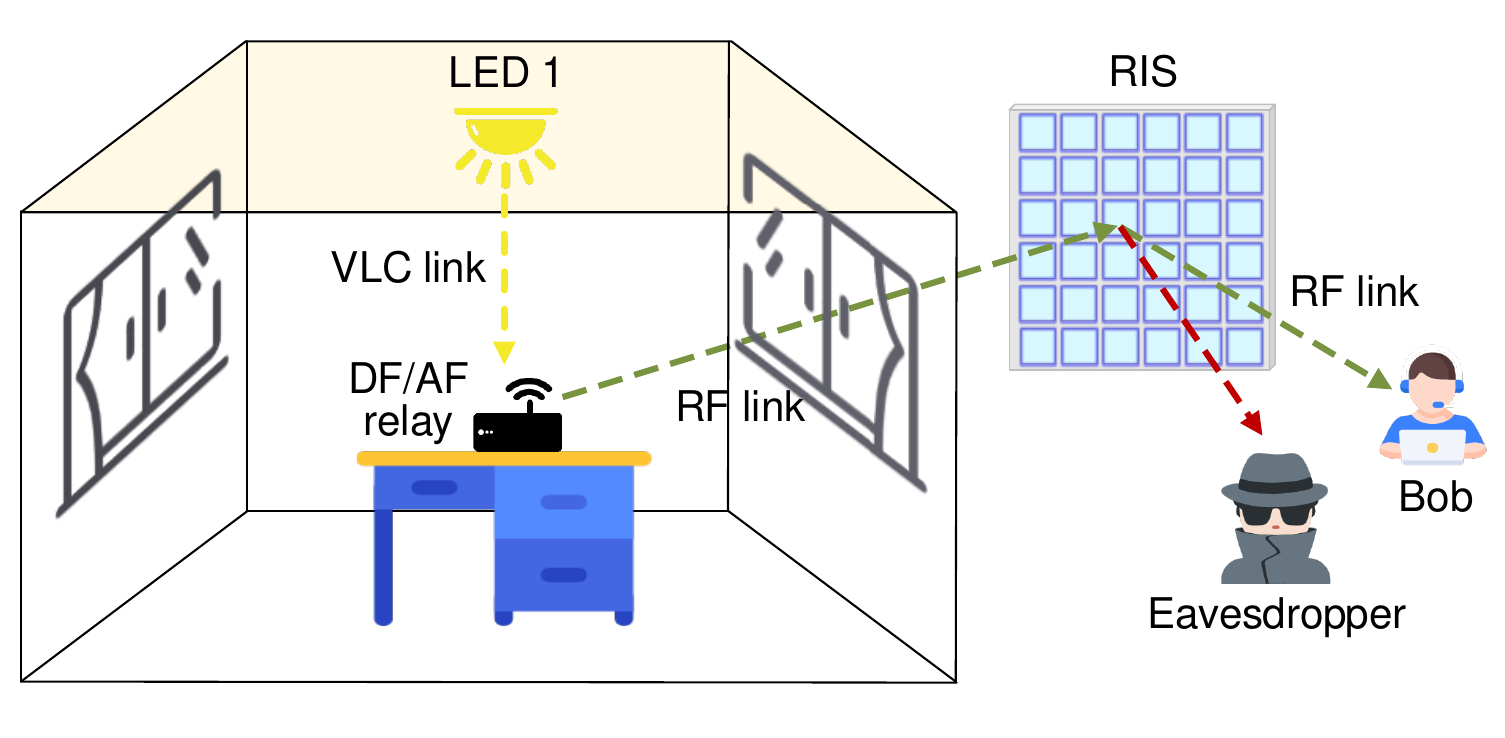}\\
  \vspace{-0.1in}
  \caption{RIS-assisted VLC/RF hybrid system: the optical signals firstly transmitted by the LEDs go through the VLC link indoors. Then,  a relay is adopted to transform the optical signals into the RF form and transmit the RF signals to Bob, assisted by the RIS in the presence of an eavesdropper.}
	\label{fig:IX.5.3}
  \end{center}
  \vspace{-0.25in}
\end{figure}

RIS-assisted VLC/RF hybrid secure networks have gained widespread attention in recent years to maximize the strengths of both the wide coverage of RF communication and high transmission rate of VLC system \cite{9817602,20233814752641} as shown in Fig.~\ref{fig:IX.5.3}. The hybrid system typically consists of two hops: firstly, the optical signals transmitted by the LEDs go through the VLC link indoors to bring the VLC unparalleled advantages into full play, such as high data rate and efficiency, no EM pollution; and then a relay is adopted to transform the received optical signal into RF signal by decode-and-forward (DF) or AF mode, and transmit the RF signal to users through the RF link assisted by the RIS to extend communication coverage and make up for the shortcoming of small coverage in VLC systems. \par

\paragraph{Unique Security Challenges}\label{Sec.IX.5.C.1)}
The RIS-assisted VLC/RF hybrid systems combine the broad coverage of RF with the high-speed transmission of VLC, yet this multi-hop architecture significantly expands potential eavesdropping opportunities. The dual-hop signal transmission from VLC to RF via relay nodes introduces vulnerabilities at multiple stages, as eavesdroppers may target either the indoor VLC link or the outdoor RIS-assisted RF link, each requiring distinct interception strategies. This heterogeneity creates challenges in maintaining consistent physical layer security across both optical and RF domains, particularly during signal conversion at the relay, where security policies may not align seamlessly. \par

\paragraph{RIS's Defensive Roles in VLC/RF Hybrid System}\label{Sec.IX.5.C.2)}
In~\cite{9817602} and~\cite{20233814752641}, the RIS-assisted SISO VLC/RF hybrid relaying system monitored by an eavesdropper is investigated, where the VLC link firstly transmits the signal and then converts into a RIS-assisted RF link via a relay to extend communication coverage of the system. There are a total of four combinations of the eavesdropper positions and relay modes, where the eavesdropper is located beside the relay or the RIS, and the relay adopts the DF or AF mode. The closed-form expressions of SOP and strictly positive secrecy capacity (SPSC) probability under the four situations are derived. Then, it is analyzed and verified by the asymptotic analysis and simulations, which show that the secrecy performance can be enhanced by increasing the SNR of the VLC link or decreasing the threshold of SOP, and is better in the AF relay mode than in the DF mode. \par

\subsubsection{Future Improvements}\label{Sec.IX.5.D}
In the optical communication system, RIS uses mirror arrays to adjust optical signal reflections based on Snell's law~\cite{10279487,9784887}, unlike the RF networks' impedance networks. Two main frameworks in RIS-assisted VLC enhance security by optimizing mirror orientation~\cite{9500409,9784887} and transforming SR maximization into a RIS assignment issue~\cite{10279487,9756553}. RIS-assisted VLC/RF networks are also studied to combine RF's coverage with VLC's high rate~\cite{9817602,20233814752641}. \par

However, the two-hop VLC/RF architecture introduces new security vulnerabilities that require comprehensive RIS solutions. Deploying RISs in both domains enables simultaneous optimization of mirror arrays for VLC and phase shift matrices for RF, ensuring robust security during mode transitions while addressing hybrid architecture threats. \par

\begin{table*}[!ht]
    \centering
    \caption{RIS-Assisted Security Enhancement and Privacy Protection With RF Scenarios: Summary Covers the Role of RIS, Paradigm, System Model, Channel Model, CSI, RIS Reflection Matrix, Optimization Objectives, Optimization Metrics, Optimization Method, a Form of Solution, Advantages, and Limitations of Various RF Scenarios Including UAV, SWIPT, D2D, and ISAC Systems}
    \vspace{-0.1in}
    \begin{adjustbox}{width=0.9\textwidth}
    \begin{tabular}{|m{1.2cm}<{\centering}|m{1.2cm}<{\centering}|m{1.2cm}<{\centering}|m{1.2cm}<{\centering}|m{1.2cm}<{\centering}|m{1.2cm}<{\centering}|m{1.2cm}<{\centering}|m{1.2cm}<{\centering}|m{1.2cm}<{\centering}|m{1.5cm}<{\centering}|m{1.2cm}<{\centering}|m{1.2cm}<{\centering}|m{1.2cm}<{\centering}|m{1.2cm}<{\centering}|m{2cm}<{\centering}|m{2cm}<{\centering}|m{0.5cm}<{\centering}|}
    \hline
\multirow{2}{*}[-0.5em]{\textbf{Scenario}} & \multirow{2}{*}[-0.5em]{\textbf{\begin{minipage}{1.2cm}
\centering
Role of RIS
\end{minipage}}} & \multirow{2}{*}[-0.5em]{\textbf{Paradigm}} & \multirow{2}{*}[-0.5em]{\textbf{System}} & \multicolumn{2}{c}{\textbf{Channel model}} & \multicolumn{2}{|c|}{\textbf{CSI}} & \multirow{2}{*}[-0.5em]{\textbf{\begin{minipage}{1.2cm}
\centering
RIS reflection matrix
\end{minipage}}} & \multirow{2}{*}[-0.5em]{\textbf{Opt. Obj.}} & \multirow{2}{*}[-0.5em]{\textbf{Metrics}} & \multirow{2}{*}[-0.5em]{\textbf{\begin{minipage}{1.2cm}
\centering
Global method
\end{minipage}}} & \multirow{2}{*}[-0.5em]{\textbf{\begin{minipage}{1.2cm}
\centering
Sub-problem method
\end{minipage}}} & \multirow{2}{*}[-0.5em]{\textbf{Solution}} & \multirow{2}{*}[-0.5em]{\textbf{Adv.}} & \multirow{2}{*}[-0.5em]{\textbf{Limits}} & \multirow{2}{*}[-0.5em]{\textbf{Ref.}} \\[5pt] \cline{5-8}
~ & ~ & ~ & ~ & \textbf{LoS link} & \textbf{Cascaded link} & \textbf{Legitimate channel} & \textbf{Wiretap channel} & ~ & ~ & ~ & ~ & ~ & ~ & ~ & ~ & ~\\ \hline
    \multirow{3}{*}[-6em]{\textbf{UAV}} & \multirow{3}{*}[-5em]{\textbf{\begin{minipage}{1.2cm}
    \centering
    Enhance coverage; SR; UAV system flexibility
    \end{minipage}}} & Constant UAV height & Single-Eve, single-user, SISO & Free-space path loss & Free-space path loss \& Rayleigh & Known & Unknown & Continuous & UAV: trajectory, power control; RIS: reflection matrix & Max. SR & AO algorithm & SCA \& phase alignment & Approx. & Provide a clear guideline for the RIS-assisted UAV system & Lack of UAV 3D trajectory design; single user & \cite{9209992}  \\ \cline{3-17}
    ~ & ~ & Constant UAV height & Single-Eve, MU-SISO & Free-space path loss & Free-space path loss \& Rayleigh & Known & Unknown & Continuous & UAV: trajectory, power control; RIS: reflection matrix; user scheduling & Max. SR & AO algorithm & SCA \& phase alignment & Approx. & Max. average SR for each user; more in
    line with the actual situation & Lack of UAV 3D trajectory and RIS placement design  & \cite{9880605} \\ \cline{3-17}
    ~ & ~ & Variable UAV height & Single-Eve, single-user, mmWave-MISO & Rician & Rician & Known & Known & Continuous & UAV \& RIS: position, beamforming; AN & Max. SR & AO algorithm & SDR \& derivation & Approxn. & Overcome mmWave communication blockages & Lack of analysis of imperfect CSI & \cite{9528924} \\ \hline
\multirow{4}{*}[-10em]{\textbf{SWIPT}} & \multirow{4}{*}[-10em]{\textbf{\begin{minipage}{1.2cm}
\centering
Enhance secure transmission and EH
\end{minipage}}} & Separate & Single-Eve, single-user, MIMO & Rayleigh & Rayleigh & Known & Known & Continuous & BS: precoding matrix, AN; RIS: reflection matrix & Max. SR & IBCD & MM algorithm & Approxn. & Give insights into the effectiveness of secure CR RIS-assisted, SWIPT MIMO systems & Cannot guarantee to converge to the global optimal solution & \cite{9852091}  \\ \cline{3-17}
~ & ~ & Unified & Single-Eve,  multi-user, MISO & Rician & Rician & Known & Unknown & Continuous & BS: precoding matrix; RIS: reflection matrix & Max. average SR & AO algorithm & MM algorithm  & Global optimum & Investigate RIS-assisted secrecy communication under passive Eve with unavailable CSI & Limited by strict computational time & \cite{9538921}  \\ \cline{3-17}
~ & ~ & Unified & Single-Eve, single-user, SISO & Rayleigh & Rician & Known & Known & Continuous & BS: transmit power; RIS: reflection matrix; Bob: PS & Max. SR & AO algorithm \& DL approach & FFP-SCA \& DNN structure & Global optimum & DL-based approach matches AO performance with notably less computation time & Can be extended to MU-MIMO / MISO system; integrated with other scenarios; adopt DRL with the real-time processing requirements & \cite{9734045} \\ \cline{3-17}
~ & ~ & Unified & Single-Eve, single-user, MISO & Rician & Rician & Known & Known & Discrete & BS: beamformer; RIS: reflection matrix; Bob: AN & Max. SR & AO algorithm & SDR \& CCP \& first-order derivative  & Approx. & Legitimate users can receive signal, harvest energy, and generate AN  simultaneously due to FD-CJ & May only ensure convergence to a local optimum & \cite{9766101} \\ \hline
    \multirow{3}{*}[-7em]{\textbf{D2D}} & \multirow{3}{*}[-3em]{\textbf{\begin{minipage}{1.2cm}
    \centering
    Suppress interference between cellular system and D2D link, the received signals at Eve; enhance desired signals for Bob, interference for Eve
    \end{minipage}}} & Downlink & Cellular system: Single-Eve, single-user, MISO; D2D: SISO & Cellular system: Rayleigh & D2D: Rayleigh & Cellular system: known; D2D: known & Cellular system: known & Continuous & RIS: reflection matrix & D2D outage probability \& SOP \& PNSC & CDF of SNR for D2D and cellular links & Phase alignment & Analytical expression & Drive analytical expressions for cellular system's SOP, PNSC and D2D's outage probability & Don't support more complex scenarios with cascaded and LoS link  in cellular network or D2D system simultaneously & \cite{9305710} \\ \cline{3-17}
    ~ & ~ & Down-link & Cellular system: Single-Eve, single-user, SISO; D2D: SISO & Cellular system: Rayleigh & Cellular system: Rayleigh; D2D: Rayleigh & Cellular system: known; D2D: known & Cellular system: unknown; D2D: unknown & Continuous & Cellular system: RIS: reflection matrix \& D2D: RIS: reflection matrix; DTX: PS; DRX: PS, beamformer & Achievable ergodic SR & Analytical framework & Phase alignment & Approxn. & Suggest an optimization of D2D power allocations for achievable ergodic SR & Lack consideration for bidirectional communication between the D2D link and EH & \cite{9478912} \\ \cline{3-17}
    ~ & ~ & Up-link & Cellular system: Single-Eve, single-user, SIMO; D2D: SISO & Cellular system: Rayleigh; D2D: Rayleigh & Cellular system: Rayleigh; D2D: Rayleigh & Cellular system: known; D2D: known & Cellular system: known & Continuous & BS: beamformer, PS; DTX: PS; RIS: reflection matrix & Max. SR & BCD algorithm & Rayleigh quotient Max. problem; SDR; linear program & Approxn. & Dramatically improve SE, security performance; suppress interference and Eve & Perfect CSI of wiretap channel may not be available as for the passive RIS & \cite{9625248} \\ \hline
\multirow{2}{*}[-4em]{\textbf{ISAC}} & \multirow{2}{*}[0em]{\textbf{\begin{minipage}{1.2cm}
\centering
Enhance radar functionality, expand coverage, and bolster ISAC communication security
\end{minipage}}} & Dual-function BS & Single-Eve, MU-MISO & Rayleigh & Rayleigh & Known & Known & Continuous & BS: beamformer; RIS: reflection matrix; radar; beamformer & Max. SR & AO algorithm & FP \& MM algorithm & Approx. & Active RIS-assisted ISAC system significantly outperforms passive RIS and non-RIS-assisted systems in terms of SR & Don't support situations with incomplete CSI or significant hardware impairments & \cite{9973319} \\ \cline{3-17}
 ~ & ~ & BS fully accessed by a wiretapper & Single-wiretapper, MU-MISO & / & Rician & Known & Known & Continuous & RIS reflection matrix & Max. SR \& Min. channel projection & Riemannian manifold & Riemannian manifold & Approx. & The scheme can effectively maintain high communication performance while ensuring user occultation & The RIS configuration guessed by the wiretapper is assumed as an identity matrix which requires further investigation &~\cite{2025arXiv250415033R} \\ \hline
    \end{tabular}
    \end{adjustbox}
    \label{Table:V.1}
    \vspace{-0.2in}
\end{table*}

\begin{table*}[ht]
    \centering
    \caption{RIS-Assisted Security Enhancement and Privacy Protection With VLC Scenarios: Summary Covers the Role of RIS, Paradigm, System Model, Channel Model, CSI, Optimization Objectives, Optimization Metrics, Optimization Method, a Form of Solution, Advantages and Limitations of VLC Scenarios Including Standalone VLC and VLC/RF Heterogeneous Systems}
    \vspace{-0.1in}
    \begin{adjustbox}{width=0.9\textwidth}
    \begin{tabular}{|m{1.2cm}<{\centering}|m{1.2cm}<{\centering}|m{1.2cm}<{\centering}|m{1.2cm}<{\centering}|m{1.2cm}<{\centering}|m{1.2cm}<{\centering}|m{1.2cm}<{\centering}|m{1.2cm}<{\centering}|m{1.2cm}<{\centering}|m{1.2cm}<{\centering}|m{1.5cm}<{\centering}|m{1.2cm}<{\centering}|m{1.2cm}<{\centering}|m{1.2cm}<{\centering}|m{1.2cm}<{\centering}|m{2cm}<{\centering}|m{2cm}<{\centering}|m{0.5cm}<{\centering}|}
    \hline
    \multirow{2}{*}[-0.5em]{\textbf{Scenario}} & \multirow{2}{*}[-0.5em]{\textbf{Network}} & \multirow{2}{*}[-0.5em]{\textbf{\begin{minipage}{1.2cm}
\centering
Role of RIS
\end{minipage}}} & \multirow{2}{*}[-0.5em]{\textbf{\begin{minipage}{1.2cm}
\centering
RIS Implementation
\end{minipage}}} & \multirow{2}{*}[-0.5em]{\textbf{Paradigm}} & \multirow{2}{*}[-0.5em]{\textbf{System}} & \multicolumn{2}{c}{\textbf{Channel model}} & \multicolumn{2}{|c|}{\textbf{CSI}} & \multirow{2}{*}[-0.5em]{\textbf{Opt. Obj.}} & \multirow{2}{*}[-0.5em]{\textbf{Metrics}} & \multirow{2}{*}[-0.5em]{\textbf{\begin{minipage}{1.2cm}
\centering
Global method
\end{minipage}}} & \multirow{2}{*}[-0.5em]{\textbf{\begin{minipage}{1.2cm}
\centering
Sub-problem method
\end{minipage}}} & \multirow{2}{*}[-0.5em]{\textbf{Solution}} & \multirow{2}{*}[-0.5em]{\textbf{Adv.}} & \multirow{2}{*}[-0.5em]{\textbf{Limits}} & \multirow{2}{*}[-0.5em]{\textbf{Ref.}} \\[5pt] \cline{7-10}
    ~ & ~ & ~ & ~ & ~ & ~ & \textbf{LoS link} & \textbf{Cascaded link} & \textbf{Legitimate channel} & \textbf{Wiretap channel} & ~ & ~ & ~ & ~ & ~ & ~ & ~ & ~\\ \hline
    \multirow{8}{*}[-18em]{\textbf{VLC}} & \multirow{4}{*}[-13em]{\begin{minipage}{1.2cm}
    \centering
    Standalone VLC
    \end{minipage}} & \multirow{4}{*}[-7em]{\begin{minipage}{1.2cm}
    \centering
    Directly adjust the reflection direction of optical signals, improve SR, and mitigate blockage problems in VLC systems
    \end{minipage}} & \multirow{4}{*}[-13em]{\begin{minipage}{1.2cm}
    \centering
    Mirror Array
    \end{minipage}} & \multirow{2}{*}[-5em]{\begin{minipage}{1.2cm}
    \centering
    Mirror array orientation
    \end{minipage}} & Single-Eve, single-user, SISO & Lambert model & ''Additive'' model & Known & Known & Each mirror: yaw and pitch rotation angles & Max. SR & PSO-II algorithm & / & Lower bound & Unsafe areas in the VLC system can be ulteriorly decreased, and then the system can remain secure all the time & Can be extended to a more practical dynamic system caused by mobile user movement & \cite{9784887} \\ \cline{6-18}
    ~ & ~ & ~ & ~ & ~ & Single-Eve, single-user, SISO dynamic system & Lambert model & ''Additive'' model & Known & Known & LEDs: beamformer; mirror array: yaw angle; each mirror: yaw and pitch rotation angles & Max. SR & DDPG algorithm & / & Lower bound & Can achieve better adaptability under the dynamic environment & Can be extended to complex scenarios, including multi-access points and multi-user & \cite{9500409} \\ \cline{5-18}
    ~ & ~ & ~ & ~ & \multirow{2}{*}[-5em]{\begin{minipage}{1.2cm}
    \centering
    Mirror array assignment
    \end{minipage}} & Single Eve, multi-user, MISO & Lambert model & ''Additive'' model & Known & Known & RIS: assignment & Max. SR & Assignment problem & Iterative KM algorithm & Approx. & Show enormous potentials of RIS for VLC security enhancement and future academic research &  Can be extended in the 6G era with the requirements of high reliability and low latency for massive communication at any time and anywhere & \cite{9756553} \\ \cline{6-18}
    ~ & ~ & ~ & ~ & ~ & Single-Eve, single-user, SISO & Lambert model & ''Additive'' model & Known & Known & LEDs: NOMA power allocation; RIS: assignment & Max. SR & AO algorithm & Linear program; adaptive-restart GA & Approx. & Examine security performance of NOMA-based RIS-assisted VLC system; gain computationally efficient solution; RIS path gives the highest DoF to manipulate Mus' channel conditions & Lack of analysis of imperfect CSI & \cite{10279487} \\ \cline{2-18}
    ~ & \multirow{2}{*}[-0.5em]{\textbf{Network}} & \multirow{2}{*}[-0.5em]{\textbf{\begin{minipage}{1.2cm}
    \centering
    Role of RIS
    \end{minipage}}} & \multirow{2}{*}[-0.5em]{\textbf{\begin{minipage}{1.2cm}
    \centering
    RIS Implementation
    \end{minipage}}} & \multirow{2}{*}[-0.5em]{\textbf{Paradigm}} & \multirow{2}{*}[-0.5em]{\textbf{System}} & \multicolumn{2}{c}{\textbf{Channel model}} & \multicolumn{2}{|c|}{\textbf{CSI}} & \multirow{2}{*}[-0.5em]{\textbf{Opt. Obj.}} & \multirow{2}{*}[-0.5em]{\textbf{Metrics}} & \multirow{2}{*}[-0.5em]{\textbf{\begin{minipage}{1.2cm}
    \centering
    Global method
    \end{minipage}}} & \multirow{2}{*}[-0.5em]{\textbf{\begin{minipage}{1.2cm}
    \centering
    Sub-problem method
    \end{minipage}}} & \multirow{2}{*}[-0.5em]{\textbf{Solution}} & \multirow{2}{*}[-0.5em]{\textbf{Adv.}} & \multirow{2}{*}[-0.5em]{\textbf{Limits}} & \multirow{2}{*}[-0.5em]{\textbf{Ref.}} \\[5pt] \cline{7-10}
    ~ & ~ & ~ & ~ & ~ & ~ & \textbf{VLC link} & \textbf{RF link} & \textbf{Legitimate channel} & \textbf{Wiretap channel} & ~ & ~ & ~ & ~ & ~ & ~ & ~ & ~\\ \cline{2-18}
    ~ & \multirow{2}{*}[-3em]{\begin{minipage}{1.2cm}
    \centering
    VLC/RF heterogeneous
    \end{minipage}} & \multirow{2}{*}[-3em]{\begin{minipage}{1.2cm}
    \centering
    Extend RF communication coverage
    \end{minipage}} & \multirow{2}{*}[-3em]{\begin{minipage}{1.2cm}
    \centering
    Programm-able metasurfaces
    \end{minipage}} & Eve with LoS link & Single-Eve, single-user, SISO & Lambert model & Rayleigh & Known & Unknown & RIS: reflection matrix & Derive SOP \& SPSC & CDF of SNR for VLC and RF links in different situations & Phase alignment & Closed-form expression & Derive closed-form expressions of SOP and SPSC with Eve besides relay & Can be extended in more complex situations & \cite{9817602} \\ \cline{5-18}
    ~ & ~ & ~ & ~ & Eve with LoS or cascaded link & Single-Eve, single-user, SISO & Lambert model & Rayleigh & Known & Unknown & RIS: reflection matrix & Derive SOP \& SPSC & CDF of SNR for VLC and RF links in different situations & Phase alignment & Closed-form expression & Derive closed-form expressions of SOP and SPSC within four situations & Can be extended in multi-eavesdroppers acting in collusion or non-collusion strategies & \cite{20233814752641} \\ \hline
    \end{tabular}
    \end{adjustbox}
    \label{Table:IX.2}
    \vspace{-0.25in}
\end{table*}

\subsection{Summary and Lessons Learned}\label{Sec.IX.6}
The RIS can intelligently reconstruct the wireless propagation environment by adjusting its reflection matrix to enhance transmission coverage and boost communication capacity by establishing virtual LoS links. According to the inherent characteristics of wireless channels, including openness, broadcast, and superposition, the RIS can considerably improve security performance. Specifically, the incident signals towards legitimate users and the eavesdropper can be boosted and suppressed to improve the system's security performance dramatically. As a result, a diverse array of RIS-enhanced wireless communication systems has been developed to optimize the benefits inherent to different RF environments. These include UAV SWIPT, D2D, and ISAC systems, as detailed in Table~\ref{Table:V.1}. Additionally, in the realm of VLC, both standalone VLC systems and VLC/RF heterogeneous systems assisted by RIS have also been implemented to exploit the unique advantages of these scenarios; see Table~\ref{Table:IX.2}. \par

The RIS technology not only maximizes the inherent advantages of diverse communication scenarios but also significantly enhances security performance by addressing each scenario's unique security challenges. As discussed in the preceding subsections, future research should enhance RIS-assisted scenarios, like UAV obstacle avoidance and VLC/RF defense schemes~\cite{10002850, 9922666, 9801642}, integrating them to improve SE, EE, and secrecy capacity. AI algorithms could address non-convex optimization issues to improve efficiency and reduce CPU time. Researchers may also consider imperfect or statistical CSI, which is common in real-world applications. These security improvements are further augmented through targeted technical approaches outlined in the ``Future Improvements'' analysis for each RIS-assisted secure scenario. \par

% ===X. Open Issues and Future Directions ========================================================================================================
\section{Open Issues and Future Research Directions}\label{Sec.X}
According to the sections above and plenty of recent research, incorporating RIS into wireless networks embodies a double-use nature. On one hand, RIS can significantly enhance security capabilities and fully leverage the benefits across a variety of wireless communication scenarios. On the other hand, it introduces new vulnerabilities and security challenges that require careful consideration to maintain the integrity of communication systems, as summarized in Table~\ref{Table:X.1}. \par

RIS technology can play a transformative role in enhancing the security and privacy of various wireless communication scenarios. The deployment of RIS has shown promising results in improving system security performance across diverse applications such as UAV systems, SWIPT, D2D, ISAC, and VLC systems, which not only fully exploit the benefits of various systems but also substantially boost security performance. Meanwhile, each application presents unique challenges that can be addressed to fully realize the potential of RIS-assisted communications. \par

The security and integrity of RIS microcontrollers are paramount, and they are susceptible to passive-active hybrid and AML attacks. 
The communication-enhancing capabilities of the RISs can be exploited by adversaries to compromise legitimate users, and can enable various RIS-assisted attacks, including eavesdropping, MITM, replay, reflection, jamming, and side-channel attacks. This poses severe threats to the integrity of communication systems, which should attract widespread attention. Furthermore, the AI-powered RIS-assisted secure wireless communication model, especially the DNN system, is susceptible to AML attacks, which can confuse the model to make incorrect classifications and predictions by adding elaborate and subtle perturbations to the input samples. Consequently, the corresponding defense and mitigation strategies should be investigated to prevent adversarial attacks and enhance the system's robustness. \par

According to the lessons learned and the summaries that lie ahead, this section addresses the open research issues and future research opportunities that will help shape RIS-assisted wireless communication in the future.

\begin{table*}[t]
    \centering
    \caption{Summary and Lessons Learned: Incorporating the RISs Into Wireless Networks Embodies a Double-Use Nature. Not Only Can the RISs Assist With Various Wireless Communication Scenarios in Boosting the Security Capacity and Maximizing Their Corresponding Advantages, but They Also Expose Users to Novel Vulnerabilities and Security Challenges That Demand Meticulous Attention to Preserve the Integrity of Communication Systems}
    \vspace{-0.1in}
    \begin{adjustbox}{width=0.83\textwidth}
    \begin{tabular}{|m{2cm}<{\centering}|m{2cm}<{\centering}|m{2cm}<{\centering}|m{2cm}<{\centering}|m{9cm}<{\centering}|}
    \hline
    \multicolumn{2}{|c|}{\textbf{Summary and lessons learned}} & \multicolumn{2}{c|}{\textbf{Scenarios}} & \textbf{Detailed description}\\ \hline
\multirow{12}{*}[-9em]{\textbf{\begin{minipage}{2cm}
\centering
The dual-use nature of RIS in a secure wireless communication system
\end{minipage}}} & \multirow{6}{*}[-5em]{\begin{minipage}{2cm}
\centering
RIS can assist in enhancing the security and privacy of various wireless communication scenarios
\end{minipage}} & \multirow{4}{*}[-3em]{\begin{minipage}{2cm}
\centering
RF scenarios
\end{minipage}} & UAV & Optimizing UAV trajectory and RIS position can enhance system security, but fixed flight altitude limits UAV flexibility to prevent collisions with buildings \\ \cline{4-5}
~ & ~ & ~ & SWIPT & RIS can assist in improving both the SR and EE, but the AO-based algorithms demand stringent computational timelines \\ \cline{4-5}
~ & ~ & ~ & D2D & RIS enhances SR and SE, boosts cellular coverage, and cuts delay, but complexities in cellular, D2D links, and D2D bidirectional communication are overlooked. \\ \cline{4-5}
~ & ~ & ~ & ISAC & RIS can assist in facilitating the concurrent execution of target sensing and user communication, but the security performance can be significantly affected by incomplete CSI or severe hardware impairments \\ \cline{3-5}
~ & ~ & \multirow{2}{*}[-1em]{\begin{minipage}{2cm}
\centering
VLC scenarios
\end{minipage}} & Pure VLC system & The mirror orientation and binary RIS allocation matrix are two primary frameworks to manipulate the reflection path of incoming signals \\ \cline{4-5}
~ & ~ & ~ & VLC/RF hybrid system & RIS-assisted VLC/RF system can integrate the advantages of the wide coverage of RF communication and high transmission rate of VLC system, but there will be more space and opportunities for the eavesdropper to choose \\ \cline{2-5}
    ~ & \multirow{6}{*}[-5em]{\begin{minipage}{2cm}
    \centering
    RIS also ushers in novel vulnerabilities and security
    \end{minipage}} & \multirow{4}{*}[-4.5em]{\begin{minipage}{2cm}
    \centering
    RIS can be exploited by adversaries to enable passive-active hybrid attacks
    \end{minipage}} & Eavesdropping attack & Eavesdropper can leverage RIS to enhance wiretap signal coverage and channel capacity while remaining undetectable to the BS without active connections, posing a significant security threat \\ \cline{4-5}
    ~ & ~ & ~ & MITM attack & Attackers use IRIS to intercept signals and attach malicious messages, complicating threat detection as passive IRISs remain undetectable to terminals without active connections. \\ \cline{4-5}
    ~ & ~ & ~ & Replay attack & Attackers exploit passive IRIS to capture and replay legitimate signals, remaining undetectable to terminals without active connections and complicating threat detection. \\ \cline{4-5}
    ~ & ~ & ~ & Reflection attack & Attackers exploit IRIS to reflect signals transmitted from legitimate users onto the victims and cause a reflective distributed denial of service attack, and their real IP addresses can not be exposed. \\ \cline{4-5}
    ~ & ~ & ~ & Jamming attack & Jammer exploit IRIS to forge virtual illegitimate links, transmitting interference to legitimate users or jamming signals to undermine the RIS's reflective capabilities. \\ \cline{4-5}
    ~ & ~ & ~ & Side-channel attack & Attackers exploit IRIS to degrade the secrecy capacity of legitimate users or destroy the channel reciprocity. Though they cannot directly decode the information, they can still capture signal characteristics and infer valuable intelligence. \\ \cline{3-5}
    ~ & ~ & \multirow{2}{*}[-1em]{\begin{minipage}{2cm}
    \centering
    Vulnerabilities of AI-enabled RIS
    \end{minipage}} & Adversarial attack & Adversarial attacks can add well-designed perturbations to the input samples and confuse the AI model, making incorrect reflection matrix predictions or incorrect signal classifications \\ \cline{4-5}
    ~ & ~ & ~ & Adversarial defense techniques & Most of the proposed adversarial defense techniques can often be bypassed by attackers, tweaking a small subset of the traffic characteristics based on prior 
     feedback, rendering them unable to function as expected \\ \hline
    \end{tabular}
    \end{adjustbox}
    \label{Table:X.1}
    \vspace{-0.25in}
\end{table*}

\begin{table*}[t]
    \centering
    \caption{Future Research Directions: Mainly Include RIS-Assisted Security Enhancement and Privacy Protection, and Exploiting RIS Vulnerabilities for Adversarial Attacks on Secure Wireless Networks. The Research Sub-Fields and Description of Research Details Are Discussed in Detail Subsequently}
    \vspace{-0.1in}
    \begin{adjustbox}{width=0.84\textwidth}
    \begin{tabular}{|m{3cm}<{\centering}|m{4cm}<{\centering}|m{9cm}<{\centering}|}
    \hline
    \textbf{Research field} & \textbf{Research direction} & \textbf{Brief description}\\ \hline
    \multirow{10}{*}[-4em]{\textbf{\begin{minipage}{3cm}
\centering
RIS-assisted security enhancement and privacy protection
\end{minipage}}} & \multirow{2}{*}[0em]{\begin{minipage}{4cm}
\centering
Introduce AI-based algorithms into RIS-assisted wireless communication scenarios
\end{minipage}} & Tackle challenges in traditional optimization, including coupled objectives \& constraints \\ \cline{3}
~ & ~ & DRL suits real-time processing, for example, in flexible UAV scenarios \\ \cline{2-3}
~ & \multirow{2}{*}[-1em]{\begin{minipage}{4cm}
\centering
Investigate the imperfect or statistical CSI
\end{minipage}} & CSI of wiretap channel is challenging to obtain for passive Eve \\ \cline{3}
~ & ~ & RIS-assisted systems' security and performance under imperfect or statistical CSI should be explored to enhance real-world compatibility and robustness \\ \cline{2-3}
~ & \multirow{2}{*}[0em]{\begin{minipage}{4cm}
\centering
Exploit more effectively DoFs offered by near-field channels to overcome security-blind zones and achieve better secrecy performance
\end{minipage}} & Security-blind zones in conventional schemes: passive eavesdroppers are located between the BS/RIS and the legitimate users \\ \cline{3}
~ & ~ & Increase of effective DoF in near-field channels: achieving more precise signal enhancement and allowing the beam to focus on legitimate users and be suppressed at the eavesdropper \\ \cline{2-3}
~ & \multirow{4}{*}[-1em]{\begin{minipage}{4cm}
\centering
Enhance adversarial defense techniques for AI-powered RIS-assisted communication networks
\end{minipage}} & Ordinary adversarial defense strategies may be bypassed by attackers \\ \cline{3}
~ & ~ & More robust, secure, and resilient adversarial defense techniques need to be proposed and developed \\ \cline{3}
~ & ~ & The differences between legitimate examples and adversarial examples can be further analyzed in high-level feature spaces \\ \cline{3}
~ & ~ & Adversarial data can be augmented using sophisticated learning models and algorithms \\ \cline{1-3}
\multirow{10}{*}[-4em]{\textbf{\begin{minipage}{3cm}
\centering
Exploiting RIS vulnerabilities for adversarial attacks on secure wireless networks
\end{minipage}}} & \multirow{4}{*}[-2em]{\begin{minipage}{4cm}
\centering
Channel attacks against large-dimensional RIS-assisted secure wireless networks
\end{minipage}} & Large pilot and training overhead is needed for BSs to observe all cascaded channels due to numerous coefficients and rapid channel changes \\ \cline{3}
~ & ~ & Codebooks are used to minimize pilot transmission overhead and achieve precise CSI estimation \\ \cline{3}
~ & ~ & Attackers can easily access the RIS-structured codebook and design interference signals \\ \cline{3}
~ & ~ & Inaccurate CSI disrupts orthogonality between precoder matrix and co-user channels, disrupting legitimate PKG \\ \cline{2-3}
~ & \multirow{3}{*}[0em]{\begin{minipage}{4cm}
\centering
Exploit the multi-sector RIS for strategic signal scattering and interference to advance adversarial interference capabilities
\end{minipage}} & RIS with multi-sector mode can be investigated and operated to scatter the impinging signals into several directions \\ \cline{3}
~ & ~ & Cause attenuation of useful signals at the legitimate receivers \\ \cline{3}
~ & ~ & Result in serious interference to uses in other directions \\ \cline{2-3}
~ & \multirow{3}{*}[-1.5em]{\begin{minipage}{4cm}
\centering
Intelligent hacking models against AI-powered RIS networks
\end{minipage}} & White-box attacks: require complete visibility of the target system \\ \cline{3}
~ & ~ & Black-box attacks: require plenty of queries to create the successful adversarial samples \\ \cline{3}
~ & ~ & Adversarial perturbations can be added in the most sensitive regions and be removed from the regions with less impact to confuse the AI-based model with minimum perturbations \\ \cline{1-3}
    \end{tabular}
    \end{adjustbox}
    \label{Table:X.2}
    \vspace{-0.25in}
\end{table*}

\subsection{Open Issues}\label{Sec.X.1}
According to the advantages above and disadvantages of various RIS-assisted wireless communication scenarios, future research directions can focus on:

\subsubsection{Cross-layer attacks exploiting network layer vulnerabilities in secure wireless communication systems}\label{Sec.X.1.A}
\textit{Attackers can exploit vulnerabilities in the network layer to seize control of RIS, carrying out attacks on the physical layer.} 
Specifically, attackers can discover network layer vulnerabilities, such as verification mechanisms and security protocols, to obtain control of the RIS microcontroller. After gaining control of RIS, attackers can manipulate signal propagation by adjusting the RIS reflection matrix, and cause reflection, jamming, MITM, replay, and side-channel attacks on the physical layer.
Such cross-layer attack methods expand the attack surface available to adversaries and may bypass security measures designed for single-layer protection.

Both legitimate and illegitimate users aim to gain control of RIS, and when each of them controls a RIS, a strategic competition unfolds \cite{9292435}. Concretely, the lawful terminals diligently work to augment the capacity of the legitimate channel while simultaneously striving to diminish the capacity of the wiretap channel by manipulating the reflection matrix of RIS and trying their best to maximize the BS gain and security performance of the system. However, attackers have purposes opposite those of legitimate users. Then, both will exert maximum effort to optimize their performance while concurrently attempting to degrade the adversary's performance.

Passive illegitimate users and their controlled IRIS are challenging to detect and locate due to their lack of active interaction with legitimate users, presenting a more serious risk to secure wireless communication. Consequently, to ensure the security of wireless communications, it is imperative to adopt comprehensive security measures that provide defense at each layer and address potential cross-layer attack strategies with appropriate protective measures.

\subsubsection{Exploring effective methods for detection and tracking of malicious RIS and attackers}\label{Sec.X.1.B}
Both passive attackers and illegally accessed RISs are rather difficult to identify by legitimate terminals because of their passive characteristics and the lack of active connections. The intercepted and reflected signals may come from legitimate terminals, and legitimate terminals may unknowingly participate in attacks, such as MITM attacks, replay attacks, and reflection attacks. The passive nature of malicious RIS and attackers, and the ``environment change attacks'' capability of RIS pose considerable security risks in wireless communication systems and increase attack complexity and stealth.

The practical prerequisites for launching these IRIS-enabled passive-active hybrid attacks vary, influencing their feasibility. For instance, as for the eavesdropping attack paradigm in Sec.~\ref{Sec.IV.3}, attackers need to obtain the CSI of the eavesdropping link, and as for jamming attacks in Sec.~\ref{Sec.VI.2}, attackers typically require accurate CSI of both the illegitimate and legitimate links to manipulate the IRIS reflection matrix effectively. This assumes that attackers may be former legitimate users or can exploit channel reciprocity in time-division duplex systems. Conversely, as for side-channel attacks by destroying channel reciprocity in Sec.~\ref{Sec.VII.2}, attackers only need to randomly reconfigure the IRIS reflection matrix to destroy channel reciprocity, which is a low-complexity tactic and requires minimal prior knowledge~\cite{10081025,10424421}.

Regarding the compromise of RIS microcontrollers, the practical difficulty spans a wide spectrum:
\begin{itemize}
    \item Commercial RISs: have weak physical security or default credentials and present a low barrier for attackers. Attackers can readily gain direct access to reprogram them, and orchestrate passive-active hybrid attacks.
    \item Hardened RISs: have robust authentication and encryption mechanisms and constitute a high-difficulty endeavor. Compromising such systems typically requires exploiting zero-day vulnerabilities in controller software~\cite{8991124}, raising the expertise, resource threshold and cost for attackers significantly.
    \item Dedicated RISs: are deployed by attackers, and the primary cost is the hardware itself which is low-cost. This makes the IRIS deployment a low-cost and high-concealment attack vector in practice. Specifically, attackers have full control by design, completely bypassing the need to compromise existing infrastructure. The inherent passivity of such unauthorized RISs reflecting signals without active transmission, coupled with their absence from the network's list of legitimate components, dramatically enhances their concealment and makes them exceptionally difficult to detect.
\end{itemize}

\textit{Therefore, how to promptly identify the type of attack endured and detect the location of malicious RIS and attackers has become exceptionally crucial for enhancing the security performance of wireless communication systems.} For instance, the ISAC technology can be utilized to enhance the capability of monitoring environmental changes and promptly detect alterations in physical indicators, such as EM fields and temperature surrounding legitimate users, by combining with various sensing devices. Furthermore, 
the monitoring results can be used to identify the type of attack and track the source of interference in a timely manner.

\subsubsection{RIS Deployment for Security and Performance Trade-offs}\label{Sec.X.1.C}
The deployment of RISs, including their geometric placement and orientation relative to transceivers, is a critical factor that fundamentally influences both communication performance and security dynamics. As established in~\cite{11007277}, optimal RIS deployment is governed by the need to mitigate the inherent double-hop path-loss and extend the coverage area, typically favoring locations near transmitters or receivers for passive RISs, or closer to receivers to balance amplification and path-loss for active RISs. However, this performance-centric deployment strategy has adverse effects on the security landscape of communication networks.

As discussed in Sec.~\ref{Sec.X.1.A}, a strategic competition unfolds when both legitimate users and attackers control RISs. The outcome of this competition is highly sensitive to relative deployment positions and orientations of the legal and illegal RISs. The non-ideal radiation patterns of practical RIS elements restrict beam steering capabilities to certain angular ranges~\cite{11007277}. This means that an IRIS will have the greatest attack efficacy when deployed within a specific angular range and efficiently manipulates signals toward attackers or target victims. In contrast, a legitimate RIS must be carefully positioned and oriented to maximize coverage for intended users while minimizing potential signal leakage to unauthorized areas.

\textit{Consequently, the deployment of RISs transitions from a pure performance optimization problem to a complex trade-off between enhancing legitimate link capacity and mitigating security threats.} A security-conscious deployment strategy may require sacrificing the performance-optimal deployment in favor of a more resilient location and orientation. This approach prioritizes minimizing unintended signal leakage from the RIS into areas potentially accessible to attackers, thereby enhancing the network's defenses against cross-layer and passive-active hybrid attacks.

\subsubsection{Trade-off between security performance and cost}\label{Sec.X.1.D}
\textit{Recent researchers have adopted and designed increasing RIS elements and joint optimization objectives to achieve better security performance. Though the security performance can be improved, the cost has increased dramatically.} Concretely, with the number of RIS elements and joint optimization objectives increasing,    
the computational complexity of multiple coupled objectives and constraints non-convex problems or the structures of the AI models will considerably increase, and the resultant computational complexity increases with the number of RIS elements on the order of $\mathcal{O}(N^3)$ \cite{10380596}.

For the large-dimensional RIS-assisted secure wireless networks, the radiative near-field regime will occupy a dominant position, and the distance variations have become a critical parameter that should be considered alongside the azimuth and elevation angle-of-arrivals (AoAs), and further increase the computational complexity.

With the number of RIS elements and bit resolution increasing, the power consumption of RIS becomes significant and cannot be ignored. Specifically, as investigated in \cite{10480438}, the power consumption of PIN-diode-based and continuous-type varactor-diode-based RISs increases with the number of RIS elements increasing with the slope of 0.01 W and 0.43 W, respectively, and the total power consumption of RIS can achieve approximately 8 W and 140 W, respectively, when the number of RIS elements is 300. Consequently, it is necessary to find the trade-off between security performance and power consumption with increasing RIS elements.

\subsubsection{Research on new security risks and countermeasures in emerging RISs and more subtle eavesdroppers}\label{Sec.X.1.E}
The deployment of enhanced RIS designs, as mentioned in Section~\ref{Sec.III.2}, introduces new security risks, though they can overcome the limitations of single-mode RISs, mainly including half-space coverage, double-fading attenuation, and energy dependency. For example, though the STAR-RIS can provide full-space coverage, the information can also leak to the eavesdropper distributed in both the reflection and transmission spaces \cite{10530140}.

The increase in the number of RIS components, although improving the ability to reflect signals and overcome multiplicative path loss by increasing DoFs, also increases the risk of being attacked. For instance, the pre-established codebook is adopted to reduce the overhead of polit transmission and achieve accurate CSI estimation. However, the pre-established codebook may be easily obtained by attackers because it is mainly designed according to the structure of RIS. Attackers can elaborately design interference signals based on the pre-established codebook to confuse the legitimate terminals to incorrectly estimate the CSI.

As RIS-assisted secure wireless communication networks evolve with new scenarios and schemes, the eavesdropper's tactics also become increasingly complex. The eavesdropper can enhance their eavesdropping capabilities by using multiple antennas and collaborating with other eavesdroppers to increase the overall eavesdropping rate. The passive eavesdropper can position themselves between the BS/RIS and legitimate users, remaining undetected, while the aerial eavesdropper can exploit their mobility to make their trajectories and positions difficult to track \cite{10478709}.

\textit{The deployment of enhanced RIS designs and the increase of RIS elements create new attack domains for attackers, and the eavesdropper has become increasingly cunning, which leads to the increasingly difficult situation of secure wireless communication.} Consequently, it is crucial to investigate and consider more complex and realistic secure wireless scenarios. New frameworks and schemes must be developed to counteract the increasing security risks posed by the deployment of enhanced RIS designs and cunning eavesdroppers and enhance the overall secrecy performance of secure wireless networks.

\subsubsection{Potential risks of RIS-assisted emerging secure wireless communication scenarios}\label{Sec.X.1.F}
\textit{There are various RIS-assisted emerging wireless communication scenarios to satisfy the increasing passenger demand for data-intensive services, and emerging technologies in 5G. However, the risks in these areas are still lacking in research.} For example, the RIS-assisted high-speed train (HST) communication \cite{10210625}. The transmission signals include user privacy information and train control messages. Consequently, the security and reliability performance must be guaranteed in HST communication scenarios. According to \cite{6849076}, the link SE should satisfy 0.25 bits/s/Hz in a 350 km/h vehicular environment.

Due to the high-speed movement of high-speed trains, the Doppler effect of signals is significant, and the time window for passing through BSs is very short, making it difficult for the eavesdropper to capture and process signals in an extremely short time, by lurking next to high-speed rail BSs. However, the eavesdropper can more easily capture confidential information inside the train. The refraction RIS, as mentioned in Section~\ref{Sec.III.1.2}, can be deployed on the train window, to reconfigure the propagation environment. The refraction signal can be dynamically constructively or destructively superimposed at the legitimate users or the eavesdropper, respectively, by intelligently controlling the phase shifts of the refraction RIS.

\subsection{Future Research Directions}\label{Sec.X.2}
Future research directions can be divided into RIS-assisted security enhancement and privacy protection, and RIS-based attacks on wireless networks, as shown in Sections~\ref{Sec.X.2.1} and \ref{Sec.X.2.2}, respectively, and summarized in Table~\ref{Table:X.2}.

\subsubsection{RIS-assisted security and privacy protection}\label{Sec.X.2.1}
\paragraph{Introduce AI-based algorithms into RIS-assisted secure wireless communication scenarios}
In RIS-assisted secure wireless communication scenarios, traditional optimization methods struggle with various challenges, including multiple coupled objectives and constraints non-convex problems, which lead to inefficiencies in computation, especially in systems with strict CPU running time requirements, and often result in solutions that fall into local optima.

AI-based algorithms, such as DL~\cite{10445177}, RL~\cite{10460315}, DRL \cite{10445710,10187161}, and FL~\cite{10478904, 10649032, 9829190}, can be introduced in RIS-assisted secure wireless communication system to address the challenges above, and have the potential to significantly improve computational efficiency and avoid local optima, thereby enhancing overall security performance. For example, as for the flexible UAV scenario, the algorithms known as DQN and DDPG~\cite{10021680} within the domain of DRL \cite{10380323, 10234427} are aptly suited for real-time processing requirements.

\paragraph{Investigate the imperfect or statistical CSI}
In various RIS-assisted wireless communication scenarios, the CSI of legitimate links, including the LoS and cascaded links, can be obtained due to the pilot transmission between the BS and legitimate users, however, the CSI of the wiretap channel is difficult to gain, especially for the passive the eavesdropper. Thus, RIS-assisted systems' security schemes and performances should be investigated under the imperfect or statistical CSI to further compatibility with real-world applications and improve the system's robustness.

\paragraph{Exploit more effectively DoF offered by near-field channels to overcome security-blind zones and achieve better secrecy performance}
The number of RIS elements has become more significant to compensate for the double-fading attenuation of the cascaded links and achieve better secrecy performance. The signal propagation of BS-RIS and RIS-Bob can be divided into near-field and far-field regions according to the ``Rayleigh distance'' expressed as $d_{\mathrm{F}}=2D^{2}/\lambda$ \cite{10436390}. Here, $D$ and $\lambda$ are the aperture of RIS and the length of carrier wavelength, respectively. When the distances of BS-RIS or/and RIS-Bob are shorter than the Rayleigh distance, the assumption of far-field propagation in most existing RIS-assisted secure wireless networks is no longer applicable. The radiative near-field regime will occupy a dominant position towards the large-dimensional RIS with large aperture \cite{10462912} in which the distance variations between each RIS element and the terminals become a critical parameter that should be considered alongside the azimuth and elevation AoAs.

As for conventional security schemes, if the eavesdroppers are located between the BS/RIS and the legitimate users, there will be security-blind zones, and it will be difficult for BS/RIS to suppress or interfere with the eavesdropper by optimizing the beamforming. Fortunately, with the consideration of the distance domain, the effective DoFs of near-field channels increase and achieve more precise signal enhancement. Additionally, the unique spherical wave-based near-field propagation can enable array radiation patterns to concentrate on a specific point~\cite{10436390}, and allow the beam to focus on legitimate users and be suppressed at the eavesdropper. Even if the eavesdropper is located between the BS/RIS and legitimate users, the increased DoFs in the near-field channel can still achieve excellent secrecy capacity and overcome the security-blind zones in secure wireless communication systems based on far-field effects.

\paragraph{Enhance adversarial defense techniques for AI-powered RIS-assisted communication networks}
More robust, secure, and resilient adversarial defense techniques must be proposed and developed to counter diverse and ever-evolving adversarial attacks on AI-powered RIS-assisted communication networks. Although some adversarial defense strategies have been developed to counteract adversarial attack methods and improve system robustness—such as adversarial attack detection, NIDs, gradient masking, adversarial training, and input transformation—they may still be bypassed by attackers to continue confusing the RIS microcontroller into making incorrect judgments and predictions, suffer from incompatibility with different adversarial robustness techniques, incur high computational costs, exhibit relatively weak generalization abilities, or encounter other challenges.

Fortunately, there have been recent breakthroughs in the domain of deep image classification~\cite{10521508}, and can be adopted to enhance the robustness of AI-powered RIS-assisted communication networks against adversarial attacks. Specifically, the differences between legitimate examples and adversarial examples can be further analyzed in high-level feature spaces, including the characteristics of manifolds, local intrinsic dimension (LID), and constellation diagrams~(CD)~\cite{10315205,10411959}. Input data can be preprocessed in high-level feature spaces \cite{wang2023intelligent}; for instance, adversarial samples can be projected from off-the-manifold into the native manifold and estimated in the class activation feature space by minimizing their distinctiveness from clean samples. Adversarial data can be augmented using sophisticated learning models and algorithms, such as hierarchical learning \cite{10061433}, GANs \cite{10239214}, and deep spiking neural networks (SNNs) \cite{10428029}, instead of directly employing gradient information to produce adversarial perturbations similar to the process of adversarial attacks \cite{tang2023divfusion}. 
The adversarially enhanced data can then be used in RIS microcontroller adversarial learning \cite{10386756,10171408} during the process of predicting RIS reflection matrix or reconstructing QPSs, and strengthen the robustness and security of AI-modeled RIS-assisted networks.

\subsubsection{Exploiting RIS vulnerabilities for attacks on secure wireless networks}\label{Sec.X.2.2}
This research direction investigates methods for leveraging RIS to undermine security performance in wireless communication systems. This includes designing sophisticated channel attacks, strategic signal scattering, and applying intelligent hacking models to compromise AI-powered RIS networks.

\paragraph{Channel attacks on large-dimensional RIS-assisted secure wireless networks}
With RIS elements increasing and near-field propagation gradually dominating, real-time and accurate channel estimation faces severe challenges due to the large number of coefficients and the high-frequency dynamic changes of the channel \cite{9875062}. The pilot and training overhead will become enormous for the BSs to observe all possible cascaded channels \cite{10380596}. Instead of relying on exhaustive search-based near-field beam training, the codebook is adopted to reduce the overhead of pilot transmission and achieve accurate CSI estimation.

However, the predefined codebook brings new risks for large-dimensional RIS-assisted secure wireless networks. The attackers may quickly obtain the codebook, which is mainly designed according to the structure of RIS, and then they can elaborately design interference signals, which can lead the BSs to incorrectly estimate the array response vector for the large-dimensional RIS and then incorrectly estimate the CSI. Due to the incorrect CSI, the orthogonality between MU's precoder matrix and co-user channels could be destroyed. Consequently, the communication performance of legitimate users will dramatically decrease due to the serious IUI, and the secrecy capacity will be considerably suppressed. Meanwhile, the incorrect CSI estimation will disrupt the PKG between legitimate users and BSs, mainly relying on accurate CSI estimation.

\paragraph{Exploit multi-sector RIS for strategic signal scattering and interference to advance adversarial interference capabilities}
Except for the aforementioned malicious operation mode targeting RIS, including eavesdropping attacks, MITM and replay attacks, and reflection and jamming attacks, the RIS with multi-sector mode can be investigated and operated to scatter the impinging signals into several directions. Though incident signals can be reflected into the $L$ sectors to avoid overlapping among sectors, the multi-sector RIS can also be exploited to not only suppress the LoS link effectively signals at the legitimate users, but also scatter the signal as interference to other users, causing interference to them.

Different from conventional RIS with a single connected reconfigurable impedance network, as for the RIS with $L$-sector mode, there are $L$ antennas placed on the corresponding fixed points of the $L$-side shape and connected by the $L$-port fully-connected reconfigurable impedance network \cite{10316535}. Consequently, if the attackers operate the RIS with $L$-sector mode, the incident signals can be scattered into multiple directions with higher gains, and not only cause attenuation of valuable signals at the legitimate receivers but also result in severe interference to users in other directions, which should be attended in the future research on RIS-assisted wireless communication system and corresponding potential threat.

\paragraph{Intelligent hacking models against AI-powered RIS networks}
Among the existing adversarial attacks, the white-box attacks require complete visibility of the target system, which is impractical in many real-world applications. As for the black-box attacks, they require plenty of queries to create successful adversarial samples. Some intelligent hackers have been successfully applied in the field of image classification, and they can be applied to attack AI-powered RIS-assisted communication networks. 

For example, the RL agent can learn an optimal policy to execute an adversarial attack with fewer queries while achieving a 100\% success rate in image classification \cite{sarkar2023robustness}. Similarly, in an AI-powered RIS-assisted network, each ``environment descriptor'' of the training model can be regarded as a ``patch'' in the image classification region. The adversarial perturbations can be added in the most sensitive areas and be removed from the regions with less impact to confuse the AI-based model to predict the reflection matrix with minimum perturbations incorrectly. According to the learning model, intelligent hackers can covertly attack networks with fewer queries to produce successful distortions.

% ================== Sec. XI Conclusion
\section{Concluding Remarks}\label{Sec.XI}
RIS technology has the potential to shape next-generation wireless communication networks. By enabling control over wireless propagation environments, RISs open new avenues for improving connectivity, efficiency, and security. However, this survey has demonstrated that the dual-use nature of RISs introduces significant security challenges that require immediate attention.
We identified ``\textit{passive-active hybrid attacks}'' as a new class of vulnerabilities, where adversaries exploit the passive nature of RISs to orchestrate malicious activities. Such attacks, combined with the inherent openness of wireless channels, amplify the risks of eavesdropping, jamming, and adversarial manipulations in AI-driven RIS networks. These findings highlight the critical need for advanced detection, tracking, and defense mechanisms to mitigate emerging threats.
To address these challenges, we encourage concerted efforts from both academia and industry toward standardized security frameworks for RIS-assisted networks, ensuring practical alignment between theoretical innovations and deployment needs.
Moreover, the survey has highlighted the importance of cross-layer collaboration between the physical and network layers to enhance security situational awareness. By integrating anomaly detection with EM signal analysis, holistic and robust security frameworks can be achieved. 
This survey has also shed light on the trade-offs between improving security and the associated costs, including computational complexity and hardware overhead. Balancing these aspects is pivotal for the practical deployment of RIS technology in secure communication networks.
Future research should focus on developing innovative countermeasures, exploring cross-layer integration, and advancing adversarial defense techniques to unlock the full promise of RIS technology in secure and resilient wireless networks.
\ifCLASSOPTIONcaptionsoff
  \newpage
\fi

\bibliographystyle{IEEEtran}
% \bibliography{IEEEabrv,Bibliography}
\begin{IEEEbiography}[{\includegraphics[width=1in,height=1.25in,clip,keepaspectratio]{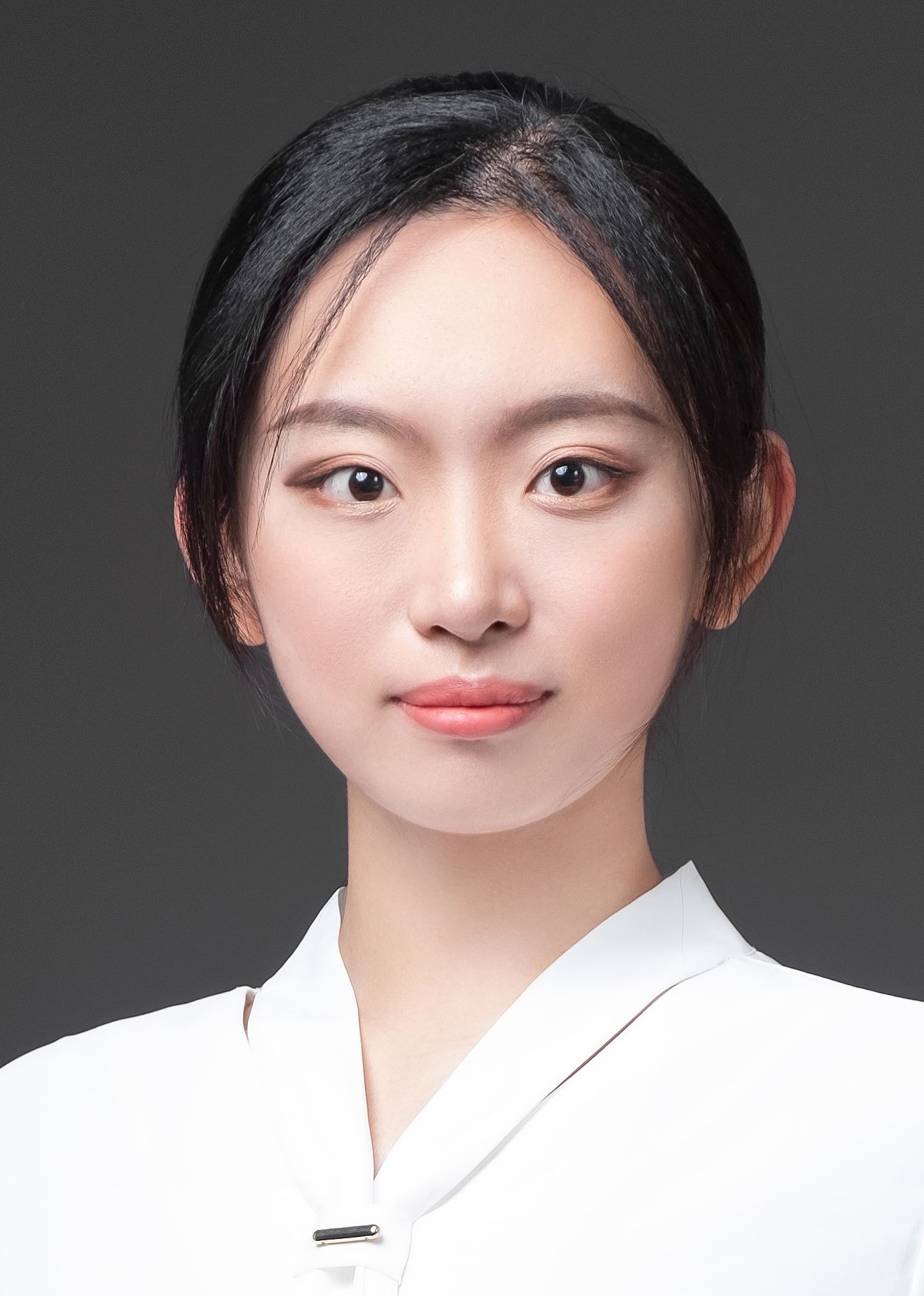}}]{Hetong Wang}
received the B.S. and M.S. degrees from the School of Telecommunications Engineering, Xidian University, Xi'an, China, in 2020 and 2023, respectively. She is currently pursuing the Ph.D. degree at the School of Information and Communication Engineering, Beijing University of Posts and Telecommunications (BUPT), Beijing, China. Her current research interests include Physical Layer Security, Reconfigurable Intelligent Surface, Stacked Intelligent Metasurface, and Machine Learning.
\end{IEEEbiography}

\begin{IEEEbiography}[{\includegraphics[width=1in,height=1.25in,clip,keepaspectratio]{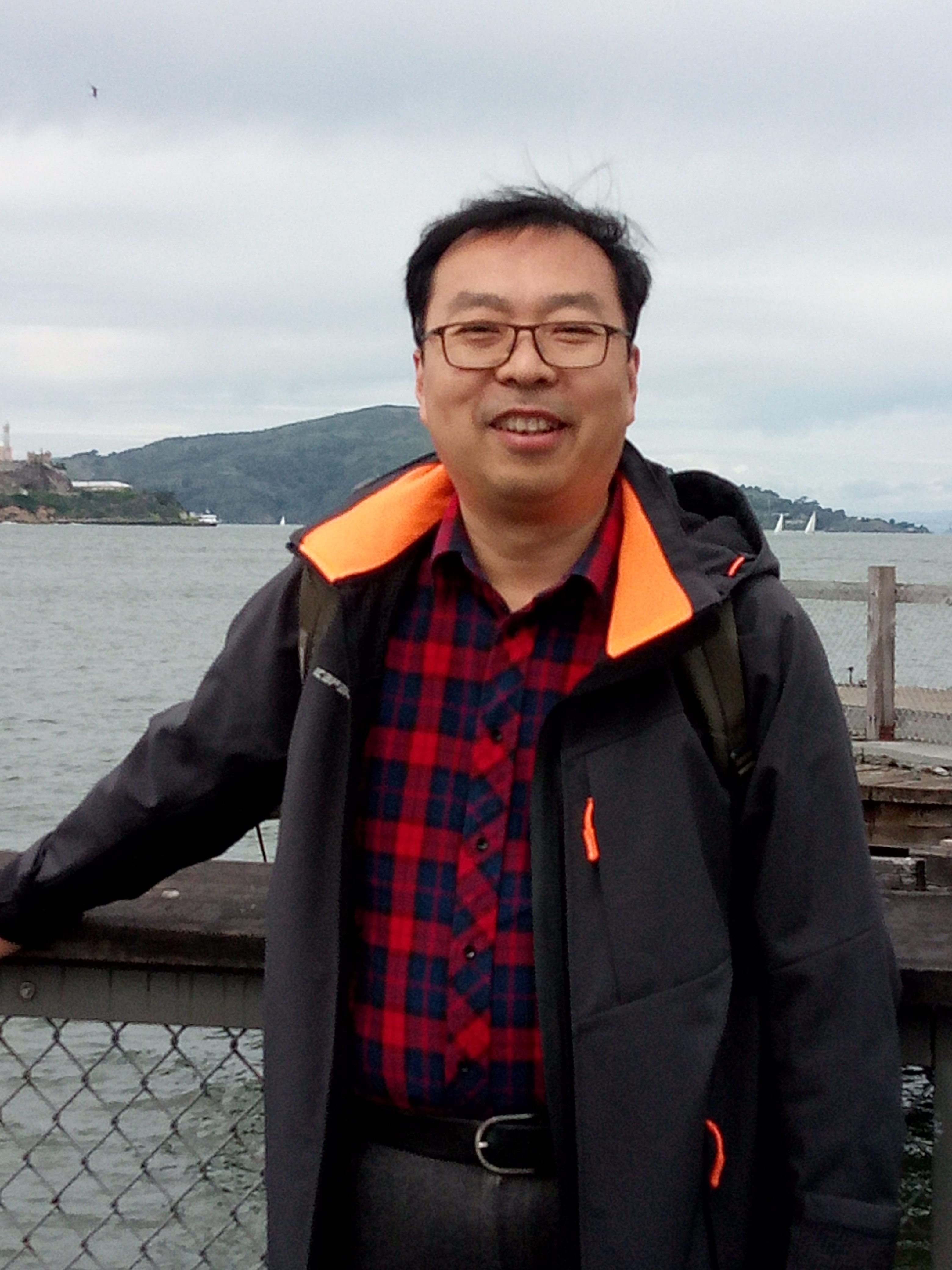}}]{Tiejun Lv}
received the M.S. and Ph.D. degrees in electronic engineering from the University of Electronic Science and Technology of China (UESTC), Chengdu, China, in 1997 and 2000, respectively. From January 2001 to January 2003, he was a Postdoctoral Fellow at Tsinghua University, Beijing, China. In 2005, he was promoted to Full Professor at the School of Information and Communication Engineering, Beijing University of Posts and Telecommunications (BUPT). From September 2008 to March 2009, he was a Visiting Professor with the Department of Electrical Engineering at Stanford University, Stanford, CA, USA. He is the author of four books, one book chapter, more than 160 published journal papers and 200 conference papers on the physical layer of wireless mobile communications. His current research interests include signal processing, communications theory and networking. He was the recipient of the Program for New Century Excellent Talents in University Award from the Ministry of Education, China, in 2006. He received the Nature Science Award from the Ministry of Education of China for the hierarchical cooperative communication theory and technologies in 2015 and the Shaanxi Higher Education Institutions Outstanding Scientific Research Achievement Award in 2025.
\end{IEEEbiography}

\begin{IEEEbiography}[{\includegraphics[width=1in,height=1.25in,clip,keepaspectratio]{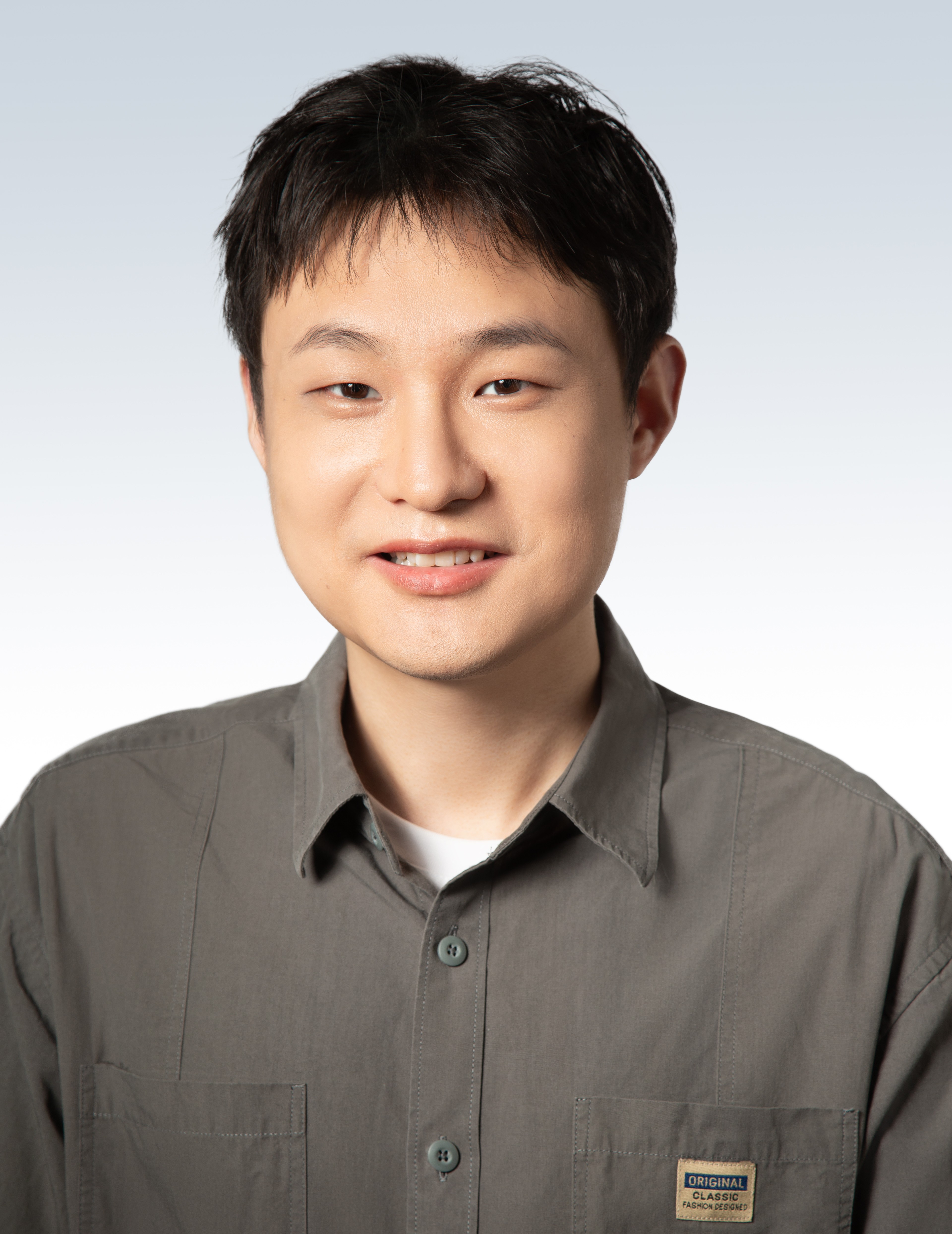}}]{Yashuai Cao}
received the B.E. and Ph.D. degrees in communication engineering from Chongqing University of Posts and Telecommunications (CQUPT) and Beijing University of Posts and Telecommunications (BUPT), China, in 2017 and 2022, respectively. From 2022 to 2023, he was a lecturer in the Department of Electronics and Communication Engineering, North China Electric Power University (NCEPU), Baoding. From 2023 to 2025, he was a Postdoctoral Research Fellow with the Department of Electronic Engineering, Tsinghua University, Beijing, China. He is currently a Distinguished Associate Professor with the School of Intelligence Science and Technology, University of Science and Technology Beijing, Beijing, China. His research interests include Intelligent Reflecting Surface, Stacked Intelligent Metasurface, Environment-Aware Communications, and Channel Twinning.
\end{IEEEbiography}

\begin{IEEEbiography}[{\includegraphics[width=1in,height=1.25in,clip,keepaspectratio]{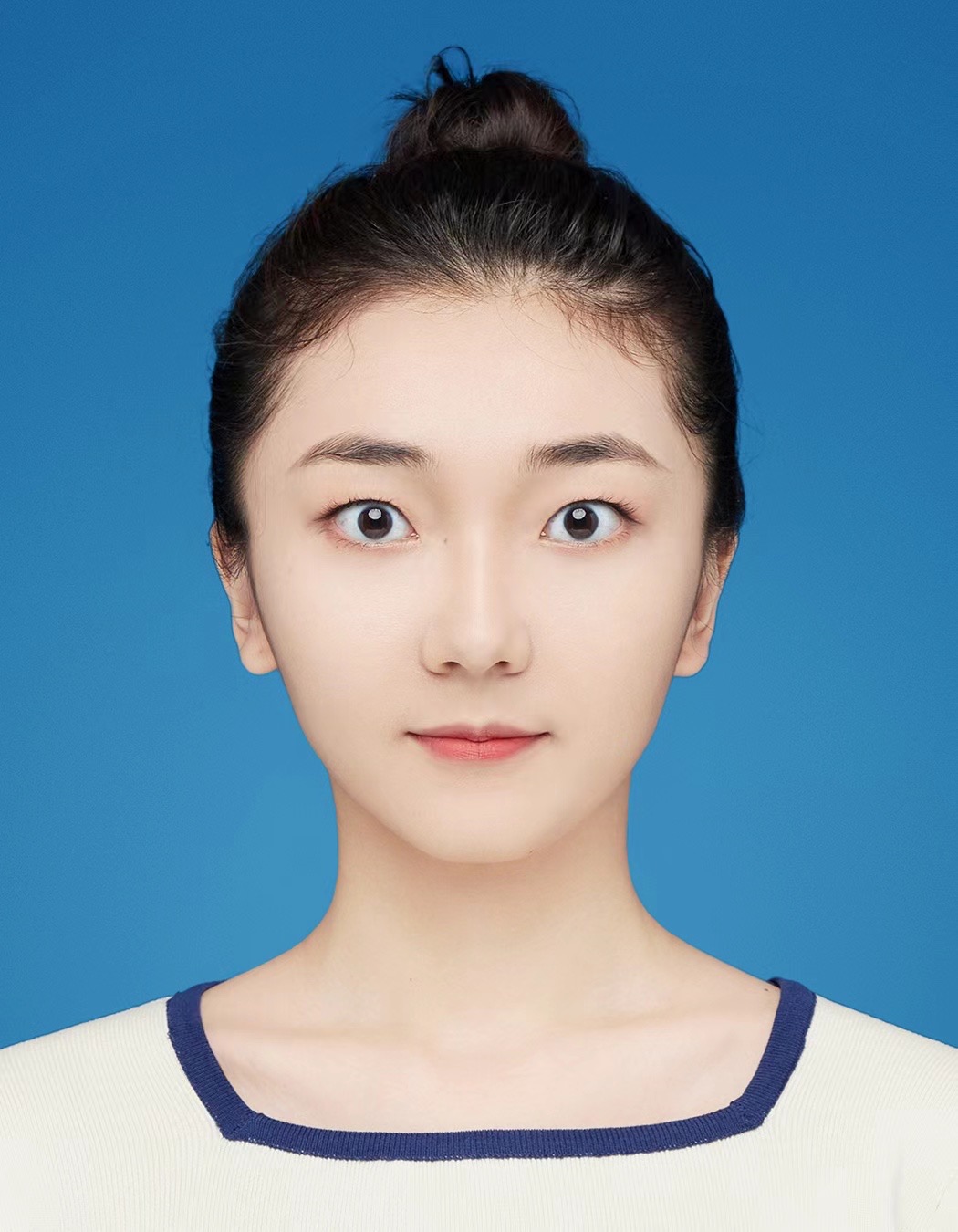}}]{Weicai Li}
received the B.E. and Ph.D. degrees from the School of Information and Communication Engineering, Beijing University of Posts and Telecommunications, China, in 2020 and 2025, respectively. From December 2022 to December 2023, she was a Visiting Student at the University of Technology Sydney, Australia. She is currently with the School of Information Communication Engineering, Beijing Information Science and Technology University, Beijing, China. Her research interests include integrated sensing and communications, wireless federated learning, distributed computing, and privacy preservation.
\end{IEEEbiography}

\begin{IEEEbiography}[{\includegraphics[width=1in,height=1.25in,clip,keepaspectratio]{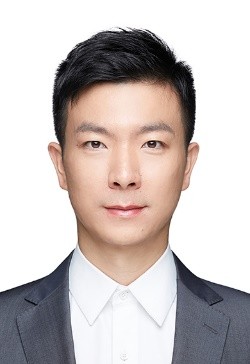}}]{Jie Zeng}
received the B.S. and M.S. degrees from Tsinghua University in 2006 and 2009, respectively, and received two Ph.D. degrees from Beijing University of Posts and Telecommunications in 2019 and the University of Technology Sydney in 2021, respectively.
From July 2009 to May 2020, he was with the Research Institute of Information Technology, Tsinghua University. From May 2020 to April 2022, he was a postdoctoral researcher with the Department of Electronic Engineering, Tsinghua University. Since May 2022, he has been an associate professor with the School of Cyberspace Science and Technology, Beijing Institute of Technology.
His research interests include 5G/6G, URLLC, satellite internet, and novel network architecture. He has published over 100 journal and conference papers, and holds more than 40 Chinese and international patents. He participated in drafting one national standard and one communication industry standard in China. 
He received Beijing’s science and technology award in 2015, the best cooperation award of Samsung Electronics in 2016, and Dolby Australia’s best scientific paper award in 2020.
\end{IEEEbiography}

\begin{IEEEbiography}[{\includegraphics[width=1in,height=1.25in,clip,keepaspectratio]{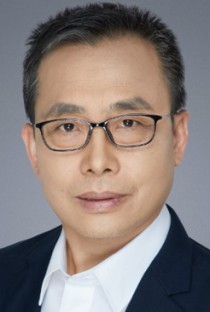}}]{Pingmu Huang}
Lecturer, School of Artificial Intelligence, Beijing University of Posts and Telecommunications. He received the M.S. degrees from Xi’an Jiaotong University, Xian, China, in 1996 and received Ph.D. degree of Signal and Information Processing from Beijing University of Posts and Telecommunications (BUPT), Beijing, China, in 2009. His current research interests include machine learning and signal processing. He published more than twenty journal papers and conference papers on signal processing and machine learning.
\end{IEEEbiography}

\begin{IEEEbiography}[{\includegraphics[width=1in,height=1.25in,clip,keepaspectratio]{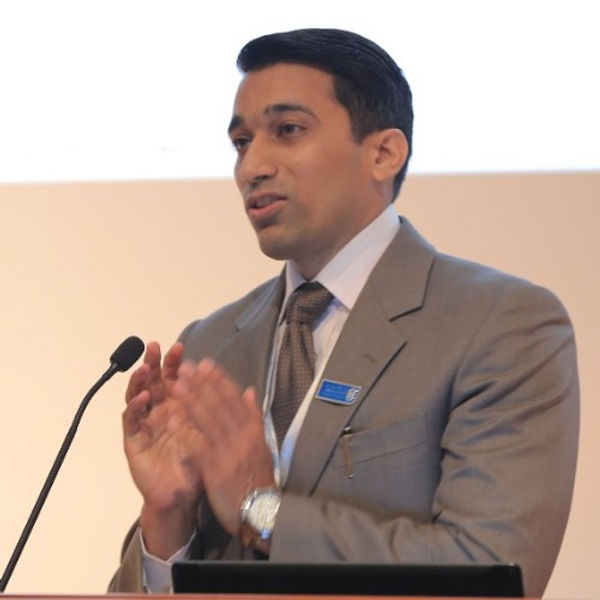}}]{Muhammad Khurram Khan}
(Senior Member, IEEE) is currently a Professor in cybersecurity with the Center of Excellence in Information Assurance, King Saud University, Saudi Arabia. He is the Founder and the CEO of the Global Foundation for Cyber Studies and Research (https://www.gfcyber.org), an independent and nonpartisan cybersecurity think-tank in Washington D.C, USA. He is also the Editor-in-Chief of Cyber Insights Magazine. He is on the editorial board of several journals including, IEEE Communications Surveys \& Tutorials, IEEE Communications Magazine, IEEE Internet of Things Journal, IEEE Transactions on Consumer Electronics, Journal of Network \& Computer Applications (Elsevier), IEEE Access, IEEE Consumer Electronics Magazine, and Electronic Commerce Research. He has published more than 450 papers in the journals and conferences of international repute. In addition, he is an inventor of ten U.S./PCT patents. He has edited ten books/proceedings published by Springer-Verlag, Taylor \& Francis, and IEEE. His research interests include cybersecurity, digital authentication, the IoT security, biometrics, multimedia security, cloud computing security, cyber policy, and technological innovation management. He is a fellow of the IET (U.K.), a fellow of the BCS (U.K.), and a fellow of the FTRA, South Korea. For more information visit the link (https://www.professorkhurram.com).

\end{IEEEbiography}

\vfill

\end{document}